\documentclass[twocolumn,english,prb, twocolumn,final,10pt]{revtex4-2}
\usepackage[T1]{fontenc}
\usepackage{textcomp}
\usepackage[utf8]{inputenc}
\usepackage{color}
\usepackage{babel}
\usepackage{array}
\usepackage{booktabs}
\usepackage{multirow}
\usepackage{varwidth}
\usepackage{amsmath}
\usepackage{amssymb}
\usepackage{graphicx}
\usepackage{wasysym}
\usepackage{microtype}
\usepackage[bookmarks=false,
 breaklinks=true,pdfborder={0 0 1},backref=false,colorlinks=true]
 {hyperref}
\hypersetup{pdftitle={Theory of Valley Splitting in Si/SiGe Spin-Qubits: Interplay of Strain, Resonances and Random Alloy Disorder},
 pdfauthor={Abel Thayil, Lasse Ermoneit, Markus Kantner},
 citecolor=purple,linkcolor=purple,urlcolor=black}

\makeatletter

\providecommand{\tabularnewline}{\\}

\usepackage{wasysym}
\usepackage{orcidlink}
\usepackage{todonotes}

\providecommand{\tabularnewline}{\\}

\def\fnum@figure{\textbf{Fig.~\thefigure}}
\def\fnum@table{\textbf{Tab.~\thetable}}

\usepackage{babel}

\makeatother

\begin{document}
\title{Theory of Valley Splitting in Si/SiGe Spin-Qubits:\\
 Interplay of Strain, Resonances and Random Alloy Disorder}
\author{Abel Thayil \orcidlink{0000-0002-3438-5901}}
\email{thayil@wias-berlin.de}

\address{Weierstrass Institute for Applied Analysis and Stochastics (WIAS),
Mohrenstr. 39, 10117 Berlin, Germany}
\author{Lasse Ermoneit \orcidlink{0009-0006-0329-0164}}
\address{Weierstrass Institute for Applied Analysis and Stochastics (WIAS),
Mohrenstr. 39, 10117 Berlin, Germany}
\author{Markus Kantner \orcidlink{0000-0003-4576-3135}}
\email{kantner@wias-berlin.de}

\address{Weierstrass Institute for Applied Analysis and Stochastics (WIAS),
Mohrenstr. 39, 10117 Berlin, Germany}
\begin{abstract}
Electron spin-qubits in silicon-germanium (SiGe) heterostructures
are a major candidate for the realization of scalable quantum computers.
A critical challenge in strained Si/SiGe quantum wells (QWs) is the
existence of two nearly degenerate valley states at the conduction
band minimum that can lead to leakage of quantum information. To address
this issue, various strategies have been explored to enhance the valley
splitting (\emph{i.e.}, the energy gap between the two low-energy
conduction band minima), such as sharp interfaces, oscillating germanium
concentrations in the QW (known as wiggle wells) and shear strain engineering.
In this work, we develop a comprehensive envelope-function theory
augmented by an empirical nonlocal pseudopotential model to incorporate
the effects of alloy disorder, strain, and non-trivial resonances
arising from interactions between valley states across neighboring
Brillouin zones. We apply our model to analyze common epitaxial profiles
studied in the literature with a focus on wiggle well type structures
and compare our results with previous work. Our framework provides
an efficient tool for quantifying the interplay of these effects on
the valley splitting, enabling complex epitaxial profile optimization
in future work.
\end{abstract}
\maketitle

\section{Introduction}

Electron spin-qubits in Si/SiGe quantum dots (QDs) are one of the
major candidates for the realization of fault-tolerant universal quantum
computers \cite{Loss1998,Zwanenburg2013,Burkard2023}. The material
platform has excellent scalability prospects because of the abundance
of nuclear spin free isotopes (\emph{e.g.}, \textsuperscript{28}Si
and \textsuperscript{76}Ge) required for long coherence times and
its compatibility with industrial fabrication technology \cite{Neyens2024,George2024,Koch2024}.
Experiments have demonstrated high-fidelity state initialization and
readout in combination with one and two-qubit gates exceeding the
fault-tolerance threshold \cite{Mills2022,Noiri2022,Xue2022}. Scalable
quantum computing architectures require coherent coupling of distant
qubits to overcome crosstalk and qubit wiring limitations \cite{Vandersypen2017,Kuenne2024}.
As a major step in this direction, coherent qubit transfer across
the chip was recently demonstrated using conveyor-mode electron spin-qubit
shuttles \cite{Seidler2022,Struck2024,Xue2024}.

\begin{figure}[t]
\includegraphics[width=1\columnwidth]{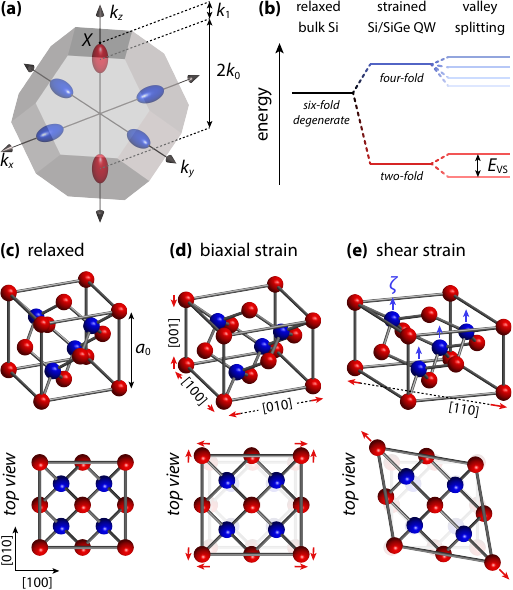} \caption{\textbf{(a)}~First Brillouin zone of the face-centered cubic (fcc)
lattice. The degeneracy between the six equivalent conduction band
minima in Si near the $X$-points is lifted by strain. \textbf{(b)}~Energy
diagram for the conduction band ground states in Si/SiGe quantum dots.
Biaxial strain due to lattice mismatch between Si and SiGe leads to
a separation of the two valleys oriented along $[001]$ and $[00\overline{1}]$
from the other four conduction band ground state valleys. The heterostructure
potential and alloy disorder finally lift the remaining degeneracies.
\textbf{(c)}~Cubic unit cell of relaxed bulk Si (diamond crystal)
composed of two interpenetrating fcc sub-lattices separated by a non-primitive
translation along a quarter of the face diagonal. Atoms of the first
and second fcc sub-lattice are shown in red and blue, respectively.
\textbf{(d)}~Biaxial strain along $[100]$ and $[010]$ yields a
tetragonal crystal with reduced symmetry. Similar to the relaxed crystal,
the two sub-lattices are interchangeable by a nonsymmorphic screw
symmetry \cite{Feng2022,Woods2024}. \textbf{(e)}~Additional shear
strain along $[110]$ further reduces the symmetry to an orthorhombic
system. The displacement between the two sub-lattices is controlled
by Kleinman's internal ionic displacement parameter $\zeta$ (blue
arrows). The nonsymmorphic screw symmetry, which maps the two sub-lattices
onto each other, is broken by the shear strain. }\label{fig: valley splitting}
\end{figure}

One of the key challenges in the design of reliable Si/SiGe qubits
is the enhancement of the energy splitting between the two nearly
degenerate valley states at the conduction band minimum of a biaxially
strained SiGe/Si/SiGe quantum well (QW), see Fig.~\ref{fig: valley splitting}.
The energy splitting between these states, called \emph{valley splitting},
is caused by the coupling of the two valley states by the heterostructure
potential \cite{Friesen2007,Saraiva2009,Saraiva2011}. Interface
roughness and random alloy disorder in the SiGe barrier \cite{PaqueletWuetz2022,Losert2023,Lima2023}
cause the valley splitting to fluctuate across the chip with typical
values ranging from several tens to hundreds of \textmu eV \cite{Borselli2011,Neyens2018,Hollmann2020,DegliEsposti2024}.
These statistical fluctuations of the valley splitting are notoriously
difficult to control and inevitably lead to spatial domains with very
low splitting, where it becomes comparable with the Zeeman-splitting
\cite{Yang2013,Borjans2019,Hollmann2020}. As a result, low valley
splittings lead to so-called \emph{spin-valley hotspots}, which are
a potential source for spin-dephasing and leakage of quantum information
\cite{Losert2024,David2024}. While these issues might be secondary
for stationary qubits, they are particularly critical in spin-qubit
shuttles, where the electron is conveyed over micrometer distances
across a disordered landscape \cite{Langrock2023,Volmer2024}. For
such applications a reliably large valley splitting is desirable to
avoid spin-valley hotspots \cite{Feng2022,Langrock2023,Klos2024}.

Recently, several heuristic strategies have been proposed to enhance
the valley splitting by engineering the Si/SiGe heterostructure \cite{McJunkin2022,Feng2022,Losert2023,Woods2024}.
These include sharp interfaces, narrow QWs and QWs with uniform low
Ge-concentration \cite{PaqueletWuetz2022} and more advanced epitaxial
profiles in the QW such as a Ge-spike \cite{McJunkin2021} or an
oscillating Ge concentration, known as the \emph{wiggle well} \cite{McJunkin2022,Feng2022,Woods2024,Gradwohl2025}.
In all these concepts, there is a complex interplay of resonances,
strain and disorder effects, which can lead to an enhancement of the
mean valley splitting. For practical applications, however, it is
important to assess the magnitude of the deterministic enhancement
along with the strength of the disorder-induced fluctuations.

In the literature, the theoretical description of valley splitting is primarily based
on envelope function theory and effective mass type models, partially augmented
by empirical tight-binding models \cite{Boykin2004,Chutia2008,Losert2023,PaqueletWuetz2022,Woods2024},
pseudopotential theory \cite{Zhang2013,Feng2022} or density functional
theory \cite{Saraiva2009,Saraiva2011,Cvitkovich2024}. Moreover,
these models have been combined with statistical models to incorporate
the effects of alloy disorder. For instance, in Refs.~\cite{PaqueletWuetz2022,Losert2023,Lima2023}
alloy fluctuations have been modeled by sampling of Ge atoms in the
primitive unit cells of the crystal from a (scaled) binomial distribution.
Alternatively, there are also approaches in which the alloy disorder
is already taken into account at the level of the electronic band
structure \cite{Feng2022}. Furthermore, the coupling between valley
states across different Brillouin zones is inconsistently accounted
for in the literature. These effects lead to non-trivial resonances,
that are addressed in several studies \cite{Feng2022,McJunkin2022,Woods2024},
whereas most of the existing literature employs the so-called ``$2k_{0}$-theory'',
where only the direct interaction of the two valley states within
the same Brillouin zone is taken into account \cite{Friesen2007,Hosseinkhani2020,Losert2023,Lima2023,Lima2024,Losert2024}.
Finally, only few works explicitly discuss the impact of strain \cite{Boykin2004,Sverdlov2010,Sverdlov2011,Woods2024}.
In particular, merely the recent paper by Woods \emph{et al.} \cite{Woods2024}
provides a combined analysis of non-trivial resonances and (shear)
strain based on an effective mass theory derived from a tight-binding
model. Their work, however, disregards the effects of alloy disorder.

In order to improve the understanding of the complex physics determining
the valley splitting in Si/SiGe qubits, it is pertinent to develop
a comprehensive theoretical model that combines all of the aforementioned
effects -- namely strain, random alloy disorder and non-trivial resonances
-- into a unified framework. This objective has been accomplished
in the present work by means of an envelope function model, which merges
several existing concepts consistently in a common framework. We demonstrate
that our model faithfully reproduces several known results on valley
splitting statistics, interface effects and the shear strain-dependency
of the long-period wiggle well, but also extends the state of the
art.

The paper is organized as follows: In Sec.~\ref{sec:Valley splitting-Theory}
, we provide the theoretical model for the valley-splitting in Si/SiGe
QDs which involves a multi-valley coupled envelope equation, a statistical
model of the random alloy disorder and the empirical pseudopotential
method to account for the electronic band structure, strain and crystal
symmetries. We provide expressions for the statistical properties
of the valley splitting. Numerical results are described in Sec.~\ref{sec: results}.
Major attention is devoted to strain-induced effects, which extend
previous findings on the QW interface-width dependency and wiggle
well-type heterostructures. Finally, Sec.~\ref{sec: discussion}
provides a thorough discussion of the results and a comparison with
similar models followed by an outlook in Sec.~\ref{sec: conclusions}.
Several technical considerations on the derivation of the multi-valley
coupled envelope equation model, the pseudopotential model, band structure coefficients and statistics can be found in the appendix.

\section{Valley Splitting Theory}\label{sec:Valley splitting-Theory}

\subsection{Coupled Envelope Equations}

The interaction of the two nearly degenerate low-energy valley states
at the conduction band minimum $\mathbf{k}=\pm\mathbf{k}_{0}\approx\left(0,0,\pm0.84\right)\times2\pi/a_{0}$
of a biaxially (tensile) strained QW grown in [001] direction
is described by the coupled envelope equation model \cite{Saraiva2009,Feng2022,Woods2024}
\begin{equation}
\left(\begin{array}{cc}
H_{0}\left(\mathbf{r}\right) & V_{c}\left(\mathbf{r}\right)\\
V_{c}^{*}\left(\mathbf{r}\right) & H_{0}\left(\mathbf{r}\right)
\end{array}\right)\left(\begin{array}{c}
\Psi_{+}\left(\mathbf{r}\right)\\
\Psi_{-}\left(\mathbf{r}\right)
\end{array}\right)=E\left(\begin{array}{c}
\Psi_{+}\left(\mathbf{r}\right)\\
\Psi_{-}\left(\mathbf{r}\right)
\end{array}\right),\label{eq: coupled envelope equation model}
\end{equation}
where $\Psi_{\pm}\left(\mathbf{r}\right)$ denotes the envelope wave
functions of the corresponding valley states. Here, the valley splitting
corresponds to the energy difference between the first excited state
and the ground state. The Hamiltonian takes the form 
\begin{equation}
H_{0}\left(\mathbf{r}\right)=-\frac{\hbar^{2}}{2m_{t}}\left(\frac{\partial^{2}}{\partial x^{2}}+\frac{\partial^{2}}{\partial y^{2}}\right)-\frac{\hbar^{2}}{2m_{l}}\frac{\partial^{2}}{\partial z^{2}}+U\left(\mathbf{r}\right),\label{eq: H_0}
\end{equation}
where $m_{l}$ and $m_{t}$ are the effective mass tensor components
at the silicon conduction band minimum and $U\left(\mathbf{r}\right)$
is the total confinement potential. Since the effective
mass parameters of conduction band electrons in Si$_{1-x}$Ge$_{x}$
alloys are practically constant for Ge content $x\apprle0.85$ \cite{Schaeffler1997},
we assume the same effective masses in the QW and in the barrier.
The intervalley coupling is described by 
\begin{align}
V_{c}\left(\mathbf{r}\right) & =\mathrm{e}^{-2i\mathbf{k}_{0}\cdot\mathbf{r}}u_{+}^{*}\left(\mathbf{r}\right)u_{-}\left(\mathbf{r}\right)U\left(\mathbf{r}\right)\label{eq: Vc}\\
 & =\sum_{\mathbf{G},\mathbf{G}'}\mathrm{e}^{-i\left(\mathbf{G}-\mathbf{G}'+2\mathbf{k}_{0}\right)\cdot\mathbf{r}}c_{+}^{*}\left(\mathbf{G}\right)c_{-}\left(\mathbf{G}'\right)U\left(\mathbf{r}\right),\nonumber 
\end{align}
which involves the plane wave expansion coefficients $c_{\pm}\left(\mathbf{G}\right)=c_{\pm\mathbf{k}_{0}}\left(\mathbf{G}\right)$
of the lattice-periodic part of the Bloch factors $u_{\pm}\left(\mathbf{r}\right)=\sum_{\mathbf{G}}\mathrm{e}^{i\mathbf{G}\cdot\mathbf{r}}c_{\pm}\left(\mathbf{G}\right)$
at the two valleys. The band index is suppressed throughout this paper,
as we are solely concerned with the (lowest energy) conduction band.
A detailed derivation of the coupled envelope equation model~\eqref{eq: coupled envelope equation model}
is given in Appendix~\ref{sec: Derivation of coupled envelope equations}.

The total confinement potential 
\begin{equation}
U\left(\mathbf{r}\right)=U_{\mathrm{het}}\left(\mathbf{r}\right)+U_{\mathrm{QD}}\left(x,y\right)+U_{F}\left(z\right)\label{eq: total potential}
\end{equation}
describes the effects of both the epitaxial heterostructure and the
electrostatic fields induced by the metal gates at the top surface
of the device. The heterostructure potential $U_{\mathrm{het}}\left(\mathbf{r}\right)$
models the potential induced by the Ge atoms in the SiGe alloy,
\emph{i.e.}, a type-II Si/SiGe QW with random alloy disorder, which
will be described in more detail in Sec.~\ref{sec:Heterostructure-Potential}
below. Note that in the present model, the effects of Ge atoms are
entirely described by the heterostructure potential, whereas the underlying
band structure coefficients (\emph{i.e.}, effective masses and Bloch factors) are
those of pure Si \footnote{In principle, one could also consider a (small) non-zero Ge concentration
in the bulk potential $V\left(\mathbf{r}\right)$, see Eq.~\eqref{eq: stationary Schroedinger equation},
and treat the resulting Si$_{1-x}$Ge$_{x}$ alloy using the virtual
crystal approximation (VCA). For consistency, the same Ge concentration
must then be removed from the mean of the heterostructure potential
$U\left(\mathbf{r}\right)$. While this approach could yield some enhancements for the band
structure coefficients and effective masses, the VCA will qualitatively
preserve the symmetries of the diamond crystal \cite{Feng2022},
leading to a similar shear strain dependency. For low Ge concentrations,
only minor changes in the quantitative results are expected since
the effective mass and band structure parameters of the Si$_{1-x}$Ge$_{x}$-alloy
(in VCA) are very close to the pure Si values \cite{Rieger1993,Schaeffler1997}.}. We assume a harmonic QD confinement potential induced by the gate
electrodes 
\begin{equation}
U_{\mathrm{QD}}\left(x,y\right)=\frac{m_{t}}{2}\left(\omega_{x}^{2}x^{2}+\omega_{y}^{2}y^{2}\right),\label{eq: QD potential}
\end{equation}
where $\omega_{x}$ and $\omega_{y}$ describe the lateral extension
of the QD an thus the orbital splitting $\Delta E_{\mathrm{orb}}=\min\left(\hbar\omega_{x},\hbar\omega_{y}\right)$.
In the limiting case of $\omega_{x}=\omega_{y}$, the QD takes a circular
shape. Finally, we assume a constant electric field $F$ along the
growth direction, which induces the potential 
\begin{equation}
U_{F}\left(z\right)=-e_{0}Fz,\label{eq: electric field potential}
\end{equation}
where $e_{0}$ is the elementary charge.

\subsection{Heterostructure Potential and Alloy Disorder}\label{sec:Heterostructure-Potential}

The heterostructure potential describes the built-in potential due
to the epitaxial profile. In order to account for disorder in the
Si$_{1-x}$Ge$_{x}$ alloy, we choose a statistical model similar
to that in Refs.~\cite{Lima2023,Lima2024,Pena2024}, where the heterostructure
potential is described as a random field 
\begin{equation}
U_{\mathrm{het}}\left(\mathbf{r}\right)=\Delta E_{c}\,\Omega_{a}\sum_{i}N_{i}\,\delta\left(\mathbf{r}-\mathbf{R}_{i}\right).\label{eq: heterostructure potential}
\end{equation}
Here, $\Delta E_{c}$ is the Si/Ge conduction band energy offset \cite{Schaeffler1997,VandeWalle1986},
$\Omega_{a}=\left(a_{0}/2\right)^{3}$ is the atomic volume (not to
be confused with the volume of the primitive unit cell) and $\mathbf{R}_{i}$
is a lattice vector of the (strained) diamond crystal, see Fig.~\ref{fig: valley splitting}\,(c)--(e).
The number of local Ge atoms at each lattice site is modeled as an
independent random variable $N_{i}$, which follows a Bernoulli distribution
depending on the local Ge concentration $X=X\left(\mathbf{R}_{i}\right)$
\begin{equation}
N_{i}\sim\mathrm{Bernoulli}\left(p=X\left(\mathbf{R}_{i}\right)\right).\label{eq: Bernoulli distribution}
\end{equation}
In the following, we assume a one-dimensional epitaxial profile characterized
by $X\left(\mathbf{r}\right)=X\left(z\right)$ describing the Ge concentration
in the QW 
\begin{equation}
X\left(z\right)=X_{b}\left(1-\Xi\left(z\right)\right),\label{eq: QW epitaxial profile X}
\end{equation}
where $X_{b}$ is the Ge concentration in the barrier (we assume $X_{b}=0.3$
throughout) and 
\begin{equation}
\Xi\left(z\right)=\frac{1}{2}\left(\tanh\left(\frac{h+z}{\sigma_{l}}\right)+\tanh\left(-\frac{z}{\sigma_{u}}\right)\right)\label{eq: QW indicator}
\end{equation}
is a smoothed indicator function that describes the shape of the QW.
Here, $h$ is the QW thickness and $\sigma_{u}$ and $\sigma_{l}$
describe the width of the upper and lower QW interfaces, respectively.
The interface widths can be obtained experimentally using scanning
transmission electron microscopy. Typical values for Si/SiGe QWs are
$\sigma_{u}\approx\sigma_{l}\approx0.5\,\mathrm{nm}$ \cite{PaqueletWuetz2022,Pena2024,Klos2024}.

The heterostructure potential is separated into a deterministic and
a random component 
\begin{equation}
U_{\mathrm{het}}\left(\mathbf{r}\right)=U_{\mathrm{QW}}\left(z\right)+\delta U_{\mathrm{het}}\left(\mathbf{r}\right),\label{eq: separation deterministic + random}
\end{equation}
where the deterministic component describes the nominal QW confinement
potential given by the expectation value 
\begin{align}
U_{\mathrm{QW}}\left(z\right)=\left\langle U_{\mathrm{het}}\left(\mathbf{r}\right)\right\rangle  & =\Delta E_{c}\Omega_{a}\sum_{i}X\left(\mathbf{R}_{i}\right)\delta\left(\mathbf{r}-\mathbf{R}_{i}\right)\nonumber \\
 & \approx\Delta E_{c}X\left(z\right).\label{eq: QW potential}
\end{align}
Here we used the mean value of the Bernoulli-distributed random Ge
number at each lattice site $\langle N_{i}\rangle=X\left(\mathbf{R}_{i}\right)$.
The random component has zero mean 
\begin{equation}
\left\langle \delta U_{\mathrm{\mathrm{het}}}\left(\mathbf{r}\right)\right\rangle =0\label{eq: zero mean deltaU}
\end{equation}
by construction. Using the covariance of the Bernoulli distribution
\[
\langle\left(N_{i}-\langle N_{i}\rangle\right)\left(N_{j}-\langle N_{j}\rangle\right)\rangle=\delta_{i,j}\,X\left(\mathbf{R}_{i}\right)\left(1-X\left(\mathbf{R}_{i}\right)\right),
\]
the covariance function of the heterostructure potential is obtained
as 
\begin{align}
\langle\delta U_{\mathrm{\mathrm{het}}} & \left(\mathbf{r}\right)\delta U_{\mathrm{\mathrm{het}}}\left(\mathbf{r}'\right)\rangle=\label{eq: covariance deltaU}\\
 & =\left(\Delta E_{c}\right)^{2}\Omega_{a}\delta\left(\mathbf{r}-\mathbf{r}'\right)\times\nonumber \\
 & \phantom{=}\times\Omega_{a}\sum_{i}X\left(\mathbf{R}_{i}\right)\left(1-X\left(\mathbf{R}_{i}\right)\right)\delta\left(\mathbf{r}-\mathbf{R}_{i}\right)\nonumber \\
 & \approx\left(\Delta E_{c}\right)^{2}\Omega_{a}\,X\left(z\right)\left(1-X\left(z\right)\right)\delta\left(\mathbf{r}-\mathbf{r}'\right).\nonumber 
\end{align}
The covariance function reflects the assumption of locally independent
distribution of Ge atoms stated in Eq.~\eqref{eq: Bernoulli distribution}
(\emph{i.e.}, no clustering of Ge atoms) and determines the statistical
properties of the intervalley-coupling parameter, see Appendix~\ref{sec: statistics}
for details.

\subsection{Empirical Pseudopotential Theory and Strain }\label{sec:Empirical-Pseudopotentials}

The empirical pseudopotential method (EPM) provides an accurate description
of the electronic band structure with only a few parameters fitted
to experimental data \cite{Cohen1970,Chelikowsky1974,Chelikowsky1976}.
A major advantage of the EPM is that it can naturally account for
strain effects arising from a displacement of the crystal ions $\mathbf{R}_{i}'=\left(I+\varepsilon\right)\mathbf{R}_{i}$,
where $\varepsilon$ is the strain tensor. For strained SiGe alloys,
numerous empirical pseudopotential models are available in the literature
\cite{Fischetti1991,Fischetti1996,Rieger1993,Ungersboeck2007b,Kim2010a,Sverdlov2011,Sant2013}. The key steps for the inclusion of strain in EPMs are \cite{Ungersboeck2007b}:
\begin{enumerate}
\item computation of strained reciprocal lattice vectors $\mathbf{G}_{i}'\approx\left(I-\varepsilon\right)\mathbf{G}_{i}$
(assuming small strain to linear order), where $\textbf{G}_{i}$ denotes
the reciprocal lattice vectors of the relaxed crystal 
\item strain-induced modification of the primitive unit cell volume $\Omega_{p}'\approx\left(1+\mathrm{tr}\left(\varepsilon\right)\right)\Omega_{p}$
\item interpolation of the Fourier coefficients of the pseudopotential at
strained reciprocal lattice vectors 
\item consideration of internal ionic displacement, see Fig.~\ref{fig: valley splitting}\,(e). 
\end{enumerate}
In this paper, we employ the nonlocal empirical pseudopotential
model described by Rieger \& Vogl \cite{Rieger1993}. From this model,
we have obtained the plane wave expansion coefficients of the Bloch
factors $c_{\mathbf{k}}\left(\mathbf{G}\right)$, the wave numbers
of the conduction band minima $\pm\mathbf{k}_{0}=\left(0,0,\pm k_{0}\right)$
and the corresponding effective mass tensor components $m_{l}$ and
$m_{t}$, see Tab.~\ref{tab: band parameters}. We have used 181 plane waves corresponding
to a cutoff energy of $12\,\mathrm{Ry}$ for convergence \cite{Rieger1993}. Details on the EPM
are given in Appendix~\ref{sec: EPM}.

The theory presented in this paper is limited to spatially
homogeneous strain distributions. If the strain field is only slowly varying (on a length scale that is large compared to the characteristic
size of the QD envelope wave function), the model can be applied locally as is. For spatially rapidly varying strain
fields, a generalization of the model might be required. We assume
biaxial (tensile) strain due to the lattice mismatch between the $\mathrm{Si}_{0.7}\mathrm{Ge}_{0.3}$
substrate and the Si QW \cite{Yu2010} 
\begin{equation}
\varepsilon_{\mathrm{QW}}=\left(\begin{array}{ccc}
\varepsilon_{\parallel} & 0 & 0\\
0 & \varepsilon_{\parallel} & 0\\
0 & 0 & \varepsilon_{\perp}
\end{array}\right)\label{eq: biaxial strain QW}
\end{equation}
with $\varepsilon_{\parallel}=\varepsilon_{x,x}=\varepsilon_{y,y}=a_{0}^{\mathrm{SiGe}}/a_{0}^{\mathrm{Si}}-1\approx1.14\,\%$
and $\varepsilon_{\perp}=\varepsilon_{z,z}=-2C_{1,2}/C_{1,1}\varepsilon_{\parallel}\approx-0.88\,\%$,
where $C_{1,1}$ and $C_{1,2}$ are elastic constants of Si, see Tab.~\ref{tab:parameter-pseudopotential}.
Below, we will consider additional shear strain along the [110]
crystallographic direction. Our analysis of shear strain induced effects
will be purely phenomenological, \emph{i.e.},
we will not discuss how it might be generated. We remark that the
shear strain can be engineered by design of the macroscopic device
geometry \cite{Sverdlov2011,CorleyWiciak2023,Adelsberger2024,Woods2024},
but it might also originate from crystal defects \cite{Mooney1996,Gradwohl2023} or alloy disorder \cite{Zaiser2022,Geslin2021a}.
Effective masses and the conduction
band minimum wave numbers are evaluated specifically for the strain tensors considered
in the simulations.

\begin{table}[t]
\begin{tabular*}{1\columnwidth}[t]{@{\extracolsep{\fill}}@{\extracolsep{\fill}}cll}
\toprule 
\textbf{symbol} & \textbf{description} & \textbf{value}\tabularnewline
\midrule 
$k_{0}$ & conduction band min. wave number & $0.839\times2\pi/a_{0}$\tabularnewline
$m_{t}$ & transverse effective mass & $0.202\times m_{0}$\tabularnewline
$m_{l}$ & longitudinal effective mass & $0.920\times m_{0}$\tabularnewline
\bottomrule
\end{tabular*}\caption{Band parameters for the conduction band minimum computed
from the pseudopotential model for relaxed bulk Si. The wave number
$k_{0}$ has been obtained from minimization of the conduction band
energy. The effective masses were computed using a finite difference
approximation of the band curvature at $\mathbf{k}_{0}$. Parameter
values used in the simulations were specifically evaluated for the
respective strain tensors.
\label{tab: band parameters}
}
\end{table}

\subsection{Perturbation Theory }

The valley splitting can be approximated using first-order degenerate
perturbation theory by assuming that both the intervalley-coupling
term $V_{c}\left(\mathbf{r}\right)$ and the disorder
potential $\delta U_{\mathrm{\mathrm{het}}}\left(\mathbf{r}\right)$
can be treated as small perturbations.

\subsubsection{Unperturbed Problem}

The unperturbed problem corresponding to Eq.~\eqref{eq: coupled envelope equation model}
is the single-valley Schrödinger equation 
\begin{align}
E_{0}\Psi_{0}\left(\mathbf{r}\right) & =-\frac{\hbar^{2}}{2}\nabla\cdot\left(m^{-1}\nabla\Psi_{0}\left(\mathbf{r}\right)\right)\label{eq: zeroth order problem}\\
 & \phantom{=}+\left(U_{\mathrm{QD}}\left(x,y\right)+U_{\mathrm{QW}}\left(z\right)+U_{F}\left(z\right)\right)\Psi_{0}\left(\mathbf{r}\right)\nonumber 
\end{align}
where $\left\langle U\left(\mathbf{r}\right)\right\rangle =U_{\mathrm{QD}}\left(x,y\right)+U_{\mathrm{QW}}\left(z\right)+U_{F}\left(z\right)$
is the mean potential energy and $m=\mathrm{diag}\left(m_{t},m_{t},m_{l}\right)$
is the effective mass tensor. The unperturbed problem \eqref{eq: zeroth order problem}
describes two energetically degenerate (decoupled) valley states with
identical orbital wave function $\Psi_{0}\left(\mathbf{r}\right)$.
Using the separation ansatz $\Psi_{0}\left(\mathbf{r}\right)=\phi_{0}\left(x,y\right)\psi_{0}\left(z\right)$
and $E_{0}=E_{t,0}+E_{l,0}$, the problem separates into two scalar
effective mass-type Schrödinger equations, \emph{i.e.}, the transverse
problem 
\begin{align}
\left(-\frac{\hbar^{2}}{2m_{t}}\left(\frac{\partial^{2}}{\partial x^{2}}+\frac{\partial^{2}}{\partial y^{2}}\right)+U_{\mathrm{QD}}\left(x,y\right)\right) & \phi_{n}\left(x,y\right)=\label{eq: 1D effective mass eq QD}\\
 & =E_{t,n}\phi_{n}\left(x,y\right)\nonumber 
\end{align}
and the longitudinal problem 
\begin{equation}
\left(-\frac{\hbar^{2}}{2m_{l}}\frac{\partial^{2}}{\partial z^{2}}+U_{\mathrm{QW}}\left(z\right)+U_{F}\left(z\right)\right)\psi_{n}\left(z\right)=E_{l,n}\psi_{n}\left(z\right).\label{eq: 1D effective mass eq QW}
\end{equation}
The exact ground state wave function of the transverse problem reads
\begin{equation}
\phi_{0}\left(x,y\right)=\left(\pi l_{x}l_{y}\right)^{-1/2}\mathrm{e}^{-\frac{1}{2}\left(\frac{x}{l_{x}}\right)^{2}}\mathrm{e}^{-\frac{1}{2}\left(\frac{y}{l_{y}}\right)^{2}}\label{eq: QD wave function}
\end{equation}
with ground state energy $E_{t,0}=\left(\hbar\text{\ensuremath{\omega}}_{x}+\hbar\text{\ensuremath{\omega}}_{y}\right)/2$
and QD width $l_{j}=\sqrt{\hbar/\left(m_{t}\omega_{j}\right)}$, $j\in\left\{ x,y\right\} $.
The ground state $\left\{ E_{l,0},\psi_{0}\left(z\right)\right\} $
of the longitudinal problem is computed numerically using a finite
difference\textcolor{red}{{} }approximation. As the unperturbed single-valley
problem \eqref{eq: zeroth order problem} is identical for both valleys,
the ground state is twofold degenerate. 

\subsubsection{First-Order Degenerate Perturbation Theory}

A perturbative expression for the valley splitting is obtained using
first-order degenerate perturbation theory \cite{Friesen2007}. The
degenerate ground state subspace is spanned by the two independent
basis states $\boldsymbol{\Psi}_{+}\left(\mathbf{r}\right)=\left(\Psi_{0}\left(\mathbf{r}\right),0\right)^{T}$
and $\boldsymbol{\Psi}_{-}\left(\mathbf{r}\right)=\left(0,\Psi_{0}\left(\mathbf{r}\right)\right)^{T}$.
Substitution of the linear combination $\eta_{+}\boldsymbol{\Psi}_{+}\left(\mathbf{r}\right)+\eta_{-}\boldsymbol{\Psi}_{-}\left(\mathbf{r}\right)=\left(\eta_{+},\eta_{-}\right)^{T}\Psi_{0}\left(\mathbf{r}\right)$
into the equation for the first-order correction yields---after projection
on the basis states---an eigenvalue equation for the first-order energy
correction $E_{1}$
\begin{equation*}
\left(\begin{array}{cc}
\delta E & \Delta\\
\Delta^{*} & \delta E
\end{array}\right)\left(\begin{array}{c}
\eta_{+}\\
\eta_{-}
\end{array}\right)=E_{1}\left(\begin{array}{c}
\eta_{+}\\
\eta_{-}
\end{array}\right).
\end{equation*}
Here we introduced the complex-valued intervalley-coupling parameter
\begin{align}
\Delta & =\int\mathrm{d}^{3}r\,\Psi_{0}^{*}\left(\mathbf{r}\right)V_{c}\left(\mathbf{r}\right)\Psi_{0}\left(\mathbf{r}\right)\label{eq: Delta}\\
 &=\sum_{\mathbf{G},\mathbf{G}'}
 c_{+}^{*}\left(\mathbf{G}\right)
 c_{-}\left(\mathbf{G}'\right) \times
 \nonumber
 \\
 &
 \hphantom{=\sum_{\mathbf{G},\mathbf{G}'}}
 \times
 \int\mathrm{d}^{3}r\,
 \mathrm{e}^{
 -i\left(\mathbf{G}-\mathbf{G}'+2\mathbf{k}_{0}\right)\cdot\mathbf{r}}\,
U\left(\mathbf{r}\right)\left|\Psi_{0}\left(\mathbf{r}\right)\right|^{2} \nonumber
\end{align}
and the disorder-induced energy shift
\begin{equation}
\delta E=\int\mathrm{d}^{3}r\,\Psi_{0}^{*}\left(\mathbf{r}\right)\delta U_{\mathrm{het}}\left(\mathbf{r}\right)\Psi_{0}\left(\mathbf{r}\right).\label{eq: deltaE}
\end{equation}
We note that Eq.~\eqref{eq: Delta} involves all of the aforementioned
effects, \emph{i.e.}, strain (via modification of the reciprocal lattice
vectors and the Bloch factor expansion coefficients), alloy disorder
and non-trivial resonances due to coupling of valley states within
different Brillouin zones. Finally, the energy correction $E_{1,\pm}=\delta E\pm\left|\Delta\right|$
yields the valley splitting
\begin{equation}
E_{\mathrm{VS}}=2\left|\Delta\right|.\label{eq: valley splitting}
\end{equation}
An accurate description of the intervalley coupling
parameter provides the key to the engineering of deterministic enhancements
of the valley splitting in Si/SiGe qubits.

\subsection{Intervalley Coupling Parameter}

The intervalley coupling parameter $\Delta$ has a deterministic and
a random component 
\begin{equation}
\Delta=\Delta_{\mathrm{det}}+\Delta_{\mathrm{rand}},\label{eq: Delta deterministic and random}
\end{equation}
reflecting the deterministic and the stochastic components of the
total confinement potential \eqref{eq: total potential}: 
\begin{align*}
\Delta_{\mathrm{det}} & =\int_{V}\mathrm{d}^{3}r\,\mathrm{e}^{-2i\mathbf{k}_{0}\cdot\mathbf{r}}u_{+}^{*}\left(\mathbf{r}\right)u_{-}\left(\mathbf{r}\right)\left\langle U\left(\mathbf{r}\right)\right\rangle \left|\Psi_{0}\left(\mathbf{r}\right)\right|^{2},\\
\Delta_{\mathrm{rand}} & =\int_{V}\mathrm{d}^{3}r\,\mathrm{e}^{-2i\mathbf{k}_{0}\cdot\mathbf{r}}u_{+}^{*}\left(\mathbf{r}\right)u_{-}\left(\mathbf{r}\right)\delta U_{\mathrm{het}}\left(\mathbf{r}\right)\left|\Psi_{0}\left(\mathbf{r}\right)\right|^{2}.
\end{align*}
In the following, we will evaluate both components.

\subsubsection{Deterministic Component }\label{sec: deterministic component}

Using the exact ground state wave function \eqref{eq: QD wave function},
the deterministic component of the intervalley coupling parameter
is obtained after in-plane integration as 
\begin{align}
\Delta_{\mathrm{det}} & =\sum_{\mathbf{G},\mathbf{G}'}c_{+}^{*}\left(\mathbf{G}\right)c_{-}\left(\mathbf{G}'\right)\times\label{eq: Delta det}\\
 & \phantom{=}\times\mathrm{e}^{-\frac{1}{4}\left(G_{x}-G_{x}'\right)^{2}l_{x}^{2}}\mathrm{e}^{-\frac{1}{4}\left(G_{y}-G_{y}'\right)^{2}l_{y}^{2}}\times\nonumber \\
 & \phantom{=}\times\int\mathrm{d}z\,\mathrm{e}^{-i\left(G_{z}-G_{z}'+2k_{0}\right)z}\,\bigg[U_{\mathrm{QW}}\left(z\right)+U_{F}\left(z\right)\nonumber \\
 & \phantom{\times=}+\frac{\hbar\omega_{x}}{2}\left(\frac{1}{2}-\left(\frac{\left(G_{x}-G_{x}'\right)l_{x}}{2}\right)^{2}\right)\nonumber \\
 & \phantom{\times=}+\frac{\hbar\omega_{y}}{2}\left(\frac{1}{2}-\left(\frac{\left(G_{y}-G_{y}'\right)l_{y}}{2}\right)^{2}\right)\bigg]\left|\psi_{0}\left(z\right)\right|^{2}.\nonumber 
\end{align}
The Gaussian damping terms in the second line result from the shape
of the in-plane wave function~\eqref{eq: QD wave function}. The
contributions from the in-plane QD potential (last two lines) are
typically small.

Deterministic enhancements of the valley splitting can be achieved,
when the product of the confinement potential and the longitudinal
wave function component resonate with the complex exponential in Eq.~\eqref{eq: Delta det},
cf. Ref.~\cite{Feng2022}. This means, that the Fourier spectrum
of 
\[
S\left(z\right)=\left(U_{\mathrm{QW}}\left(z\right)+U_{F}\left(z\right)\right)\left|\psi_{0}\left(z\right)\right|^{2}
\]
must provide large amplitudes $\tilde{S}\left(q\right)$ at wave numbers
$q$ that obey the resonance condition 
\begin{equation}
G_{z}-G_{z}'+2k_{0}-q=0\label{eq: resonance condition}
\end{equation}
for any possible combination of $G_{z}$ and $G_{z}'$. In this case,
\[
\int\mathrm{d}z\,\mathrm{e}^{-i\left(G_{z}-G_{z}'+2k_{0}\right)z}S\left(z\right)=2\pi\,\tilde{S}\left(G_{z}-G_{z}'+2k_{0}\right)
\]
gives a strong contribution via an enhancement of the coupling strength
between certain valley states, possibly across different Brillouin
zones. This concept of confinement potential engineering to achieve
large Fourier amplitudes $\tilde{S}\left(q\right)$ is explicitly
addressed in the wiggle well \cite{McJunkin2022,Feng2022,Woods2024},
but is implicitly employed also in other approaches (\emph{e.g.},
sharp interfaces \cite{DegliEsposti2024}, Ge-spike \cite{McJunkin2021}).

The expression \eqref{eq: Delta det} can be considerably simplified
by exploiting the fact that the in-plane extension of the QD wave
function is much larger than the lattice constant $l_{x},l_{y}\gg a_{0}$.
With this, the Gaussians in the second line of Eq.~\eqref{eq: Delta det}
effectively reduce to a Kronecker-Delta reproducing the selection
rule for quantum wells~\cite{Saraiva2009}, see Appendix~\ref{sec: eval of valley splitting and bandstuct coeff}
for details. This approximation allows for a very compact notation
\begin{equation}
\Delta_{\mathrm{det}}=\sum_{n=-\infty}^{\infty}\Delta_{\mathrm{det},n}=\sum_{n=-\infty}^{\infty}C_{n}^{\left(2\right)}J_{n}^{\mathrm{det}},\label{eq: Delta det compact}
\end{equation}
where we have introduced the coefficients $C_{n}^{\left(2\right)}$
described in Eq.~\eqref{eq: bandstructure coefficient C2} and the
integrals $J_{n}^{\mathrm{det}}$ defined in Eq.~\eqref{eq: integral J det}.

\begin{figure}[t]
\includegraphics[width=1\columnwidth]{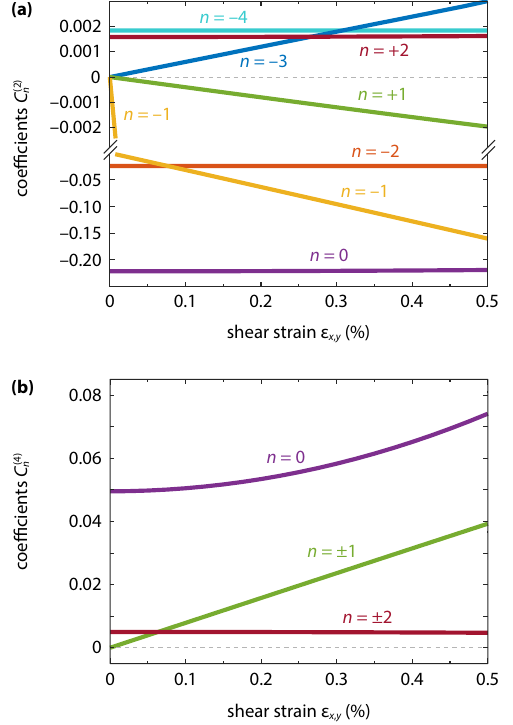}\caption{\textbf{(a)}~Plot of the coefficients $C_{n}^{\left(2\right)}$ governing
the magnitude of the deterministic contribution to the valley splitting
as a function of shear strain $\varepsilon_{x,y}$. The coefficients
have been computed from Eq.~\eqref{eq: bandstructure coefficient C2}
using the plane wave expansion coefficients of the Bloch factors for
strained silicon at the conduction band minimum. In addition to shear
strain, biaxial tensile strain arising from the $\mathrm{Si}/\mathrm{Si}_{0.7}\mathrm{Ge}_{0.3}$
heterostructure was assumed, see Eq.~\eqref{eq: biaxial strain QW}.
Coefficients with odd $n$ show a linear dependence on shear strain,
while coefficients with even $n$ are practically constant. \textbf{(b)}~Same
as (a) for the coefficients $C_{n}^{\left(4\right)}$ given by Eq.~\eqref{eq: coefficients C4}
governing the magnitude of the disorder-induced contribution to the
valley splitting. }\label{fig:Bandstructure-coefficients}
\end{figure}

The above expressions fully account for strain (to linear order) and
non-trivial resonances due to coupling with valley states in neighboring
Brillouin zones. Hence, the present model goes beyond the commonly
employed ``$2k_{0}$-theory'', which accounts only for the $n=0$
contribution in Eq.~\eqref{eq: Delta det compact}. This corresponds
to the trivial resonance condition $\mathbf{G}=\mathbf{G}'$, cf.~Ref.~\cite{Saraiva2009},
where only the spectral component at 
\begin{equation}
\left.q\right\vert _{n=0}=2k_{0}\label{eq: 2k0 resonance}
\end{equation}
is taken into account. The plot of the coefficients $C_{n}^{\left(2\right)}$
in Fig.~\ref{fig:Bandstructure-coefficients}\,(a) shows, that this
approximation is indeed justified in many scenarios, but might fail
when the integrand of Eq.~\eqref{eq: Delta det} exhibits strong
non-trivial resonances with valley states outside of the first Brillouin
zone, \emph{i.e.}, with $\mathbf{G}-\mathbf{G}'=n\mathbf{G}_{0}$
for $n\neq0$ and $\mathbf{G}_{0}$ given in Eq.~\eqref{eq: G0 vector}.
In these cases, contributions proportional to other coefficients $C_{n\neq0}^{\left(2\right)}$
become relevant. A particularly important resonance is the one at
\begin{equation}
\left.q\right\vert _{n=-1}=-2k_{1},\label{eq: modified 2k1 resonance}
\end{equation}
that is associated with the long-period wiggle-well \cite{McJunkin2022,Feng2022,Woods2024}.
Here we introduced 
\begin{equation}
k_{1}=\frac{2\pi}{a_{0}}\left(1-\varepsilon_{z,z}\right)-k_{0},\label{eq:definition k1}
\end{equation}
which is the reciprocal-space distance of the conduction band minimum
from the Brillouin zone boundary, see Fig.~\ref{fig: valley splitting}\,(a).
We note that uniaxial strain $\varepsilon_{z,z}$ (which is typically
compressive) leads to a slight modification of this distance. As shown
in Fig.~\ref{fig:Bandstructure-coefficients}, the coefficient $C_{n=-1}^{\left(2\right)}$
obtained from the EPM is linearly dependent on the shear strain component
as $C_{n=-1}^{\left(2\right)}\approx-32.0\times\varepsilon_{x,y}$.
This result is in good agreement with the theory recently presented
by Woods \emph{et al.} \cite{Woods2024}, which predicts a linear
dependency of the long-period wiggle-well on shear strain following
a $\mathrm{s}\mathrm{p}^{3}\mathrm{d}^{5}\mathrm{s}^{*}$ tight-binding
model. One can clearly see that the resonance is suppressed in the
absence of shear strain $\varepsilon_{x,y}=0$, which is explained
by a nonsymmorphic symmetry of the (either relaxed or biaxially strained)
silicon crystal structure \cite{Feng2022,Woods2024}. In the presence
of shear strain, however, this symmetry is broken such that the resonance
at $n=-1$ also yields enhancements to structures such as sharp interfaces or
QWs with uniform Ge concentrations, in addition to the long period wiggle well. This is discussed in more detail
along with numerical results in Sec.~\ref{sec: results} below. Finally,
in Sec.~\ref{sec: results} we will also observe small contributions
from the $n=-2$ resonance, in particular at sharp interfaces.
This corresponds to 
\begin{equation}
\left.q\right\vert _{n=-2}=-2k_{0}-4k_{1}\label{eq: n=00003D00003D-2 resonance}
\end{equation}
\emph{i.e.}, a coupling between distant valley states separated by
an intermediate Brillouin zone. The impact of further non-trivial
resonances was found to be negligible.

\subsubsection{Random Component}

The random contribution to the valley splitting in Eq.~\eqref{eq: Delta deterministic and random}
results from alloy disorder and is described by 
\begin{align}
\Delta_{\mathrm{rand}} & =\sum_{\mathbf{G},\mathbf{G}'}c_{+}^{*}\left(\mathbf{G}\right)c_{-}\left(\mathbf{G}'\right)\times\label{eq: Delta rand}\\
 & \phantom{=}\times\int\mathrm{d}^{3}r\,\mathrm{e}^{-i\left(\mathbf{G}-\mathbf{G}'+2\mathbf{k}_{0}\right)\cdot\mathbf{r}}\,\delta U_{\mathrm{het}}\left(\mathbf{r}\right)\left|\Psi_{0}\left(\mathbf{r}\right)\right|^{2}.\nonumber 
\end{align}
As shown in Appendix~\ref{sec: statistics - Intervalley-Coupling-Parameter},
$\Delta_{\mathrm{rand}}$ obeys a complex normal distribution 
\begin{equation}
\Delta_{\mathrm{rand}}\sim\mathrm{ComplexNormal}\left(\mu=0,\Gamma,C\right),\label{eq: complex normal}
\end{equation}
with zero mean $\mu=0$, covariance $\Gamma=\langle\left|\Delta_{\mathrm{rand}}\right|^{2}\rangle$
and pseudo-covariance $C=\langle\Delta_{\mathrm{rand}}^{2}\rangle$.
As the pseudo-covariance is typically negligible in comparison to
the covariance, the random contribution is well approximated by a
circular symmetric normal distribution in the complex plane with independent
and identically distributed real and imaginary parts 
\begin{align*}
\mathrm{Re}\left(\Delta_{\mathrm{rand}}\right) & \sim\mathrm{Normal}\left(\mu=0,\sigma^{2}=\frac{1}{2}\Gamma\right),\\
\mathrm{Im}\left(\Delta_{\mathrm{rand}}\right) & \sim\mathrm{Normal}\left(\mu=0,\sigma^{2}=\frac{1}{2}\Gamma\right).
\end{align*}
We refer to Appendix~\ref{sec: statistics - Intervalley-Coupling-Parameter}
for a detailed derivation. Consequently, the characterization of the
disorder-induced contribution to the valley splitting requires solely
the computation of the covariance $\Gamma$. Using the covariance
function of the random potential \eqref{eq: covariance deltaU} and
the in-plane wave functions \eqref{eq: QD wave function}, one obtains
\begin{align}
 & \Gamma=\langle\left|\Delta_{\mathrm{rand}}\right|^{2}\rangle=\frac{1}{2\pi l_{x}l_{y}}\left(\Delta E_{c}\right)^{2}\Omega_{a}\label{eq: Delta rand abs}\\
 & \phantom{=}\times\sum_{\mathbf{G},\mathbf{G}',\mathbf{G}'',\mathbf{G}'''}c_{+}^{*}\left(\mathbf{G}\right)c_{-}\left(\mathbf{G}'\right)c_{+}\left(\mathbf{G}''\right)c_{-}^{*}\left(\mathbf{G}'''\right)\nonumber \\
 & \phantom{=}\times\mathrm{e}^{-\frac{1}{2}\big(\frac{(G_{x}-G_{x}'-G_{x}''+G_{x}''')l_{x}}{2}\big)^{2}}\mathrm{e}^{-\frac{1}{2}\big(\frac{(G_{y}-G_{y}'-G_{y}''+G_{y}''')l_{y}}{2}\big)^{2}}\nonumber \\
 & \phantom{=}\times\int\mathrm{d}z\,\mathrm{e}^{-i\left(G_{z}-G_{z}'-G_{z}''+G_{z}'''\right)z}X(z)\left(1-X(z)\right)\left|\psi_{0}(z)\right|^{4}.\nonumber 
\end{align}
Following the same steps as in the deterministic part above, the fourfold
summation over reciprocal lattice vectors can be reduced to a single
summation 
\begin{equation}
\langle\left|\Delta_{\mathrm{rand}}\right|^{2}\rangle=\sum_{n=-\infty}^{\infty}C_{n}^{\left(4\right)}J_{n}^{\mathrm{rand}},\label{eq: Delta rand abs compact}
\end{equation}
where the integrals $J_{n}^{\mathrm{rand}}$ are given in Eq.~(\ref{eq: integral J rand})
and the coefficients $C_{n}^{\left(4\right)}$ are defined in Eq.~\eqref{eq: coefficients C4}.
We refer to Appendix~\ref{sec: eval of valley splitting and bandstuct coeff}
for details. The coefficients $C_{n}^{\left(4\right)}$ are plotted
as a function of shear strain in Fig.~\ref{fig:Bandstructure-coefficients}\,(b).

\subsubsection{Valley Splitting Statistics}

From the normal distribution of $\Delta_{\mathrm{rand}}$ it follows
that the valley splitting obeys a Rice distribution 
\begin{equation}
E_{\mathrm{VS}}\sim\mathrm{Rice}\left(\nu=2\left|\Delta_{\mathrm{det}}\right|,\sigma^{2}=2\,\Gamma\right),\label{eq: E_VS distribution law}
\end{equation}
see Appendix~\ref{sec: statistics - valley splitting} for a derivation.
This result has been obtained previously using the $2k_{0}$-theory
in Refs.~\cite{Losert2023,PaqueletWuetz2022} with a very similar
model for the alloy disorder. Here, it has been extended to a more
complex case that includes non-trivial resonances and strain. We find
that the qualitative result of a Rice distribution for $E_{\mathrm{VS}}$
is unchanged but the shape parameters of the
distribution $\nu=2\left|\Delta_{\mathrm{det}}\right|$ and $\sigma^{2}=2\Gamma=2\,\langle\left|\Delta_{\mathrm{rand}}\right|^{2}\rangle$
are modified to account for the more complex physics, see Eqs.~\eqref{eq: Delta det}
and \eqref{eq: Delta rand abs}. The Rice distribution was found to
be in good agreement with fully atomistic tight-binding simulations~\cite{PaqueletWuetz2022,Klos2024},
density functional theory~\cite{Pena2024} and experimental results
from conveyor-mode shuttling tomography of the valley splitting~\cite{Volmer2024}.

In the following, we will frequently consider the mean and the variance
of the Rice distribution given by 
\begin{align}
\left\langle E_{\mathrm{VS}}\right\rangle  & =\sqrt{\frac{\pi}{2}}\sigma\,f\left(\left(\frac{\nu}{2\sigma}\right)^{2}\right)\label{eq: Rice mean}\\
\mathrm{Var}\left(E_{\mathrm{VS}}\right) & =2\sigma^{2}+\nu^{2}-\left\langle E_{\mathrm{VS}}\right\rangle ^{2}\label{eq: Rice variance}
\end{align}
with the shape parameters $\nu$ and $\sigma$ specified above. The
expression for the mean involves the function 
\[
f\left(x\right)=\mathrm{e}^{-x}\left(\left(1+2x\right)I_{0}\left(x\right)+2xI_{1}\left(x\right)\right),
\]
where $I_{n}\left(x\right)$ denotes the modified Bessel functions
of the first kind. The asymptotics 
\begin{align*}
f\left(x\right) & \sim\begin{cases}
1+x & x\ll1\\
2\sqrt{2x/\pi} & x\gg1
\end{cases}
\end{align*}
indicate that the expected valley splitting is dominated by the deterministic
part only if $\nu\gg2\sigma$, but is disorder-dominated otherwise:
\begin{align*}
\left\langle E_{\mathrm{VS}}\right\rangle  & \sim\begin{cases}
\sqrt{\frac{\pi}{2}}\sigma & \nu\ll2\sigma,\\
\nu & \nu\gg2\sigma.
\end{cases}
\end{align*}
In order to distinguish between regimes with primarily deterministic
or disorder-dominated contributions to the mean valley splitting,
we introduce the separatrix defined by the condition 
\begin{align}
\left.\frac{\nu}{\left\langle E_{\mathrm{VS}}\right\rangle }\right|_{\mathrm{separatrix}} & =\frac{1}{2},\label{eq: separatrix condition}
\end{align}
where both components contribute with equal weight. The separatrix
condition can be solved explicitly for the shape parameters as 
\[
\left.\frac{\nu}{2\sigma}\right|_{\mathrm{separatrix}}=\left.\frac{\left|\Delta_{\mathrm{det}}\right|}{\sqrt{2\Gamma}}\right|_{\mathrm{separatrix}}\approx0.3507
\]
where $x_{0}\approx0.3507$ satisfies $f\left(x_{0}^{2}\right)=4\sqrt{2/\pi}\,x_{0}$,
which reflects Eq.~\eqref{eq: separatrix condition}.

\begin{table}[b]
\begin{tabular*}{1\columnwidth}[t]{@{\extracolsep{\fill}}@{\extracolsep{\fill}}cll}
\toprule 
\textbf{symbol} & \textbf{description} & \textbf{value}\tabularnewline
\midrule 
$\Delta E_{c}$ & Si/Ge conduction band offset & $0.5\,\mathrm{eV}$\tabularnewline
$X_{b}$ & mean barrier Ge concentration & $0.3$\tabularnewline
$\hbar\omega_{x},\hbar\omega_{y}$ & orbital splitting energy (circular QD) & $3.0\,\mathrm{meV}$\tabularnewline
$F$ & electric field strength & $5\,\mathrm{mV}/\mathrm{nm}$\tabularnewline
$\sigma_{u}$, $\sigma_{l}$ & upper and lower QW interface width & $0.5\,\mathrm{nm}$\tabularnewline
$h$ & quantum well thickness & $75\,\mathrm{ML}$\tabularnewline
$C_{1,1}$ & longitudinal stiffness constant of Si & $167.5\,\mathrm{GPa}$ \cite{Rieger1993}\tabularnewline
$C_{1,2}$ & transverse stiffness constant of Si & $65.0\,\mathrm{GPa}$ \cite{Rieger1993}\tabularnewline
\bottomrule
\end{tabular*}\caption{Parameter values used in the computations, if not stated otherwise.
The QW thickness is given in units of monolayers $\mathrm{ML}=a_{0}/4$
of the relaxed Si crystal.}
\label{tab: parameters}
\end{table}

\section{Results }\label{sec: results}

In this section, we compute the valley splitting for different types of
engineered heterostructures
using the model described in Sec.~\ref{sec:Valley splitting-Theory}. We will specifically highlight the effects
of shear strain and non-trivial resonances, that go beyond previously
reported results. Parameter values used in the simulations are given
in Tab.~\ref{tab: band parameters} and \ref{tab: parameters}.

\begin{figure}[t]
\includegraphics[width=1\columnwidth]{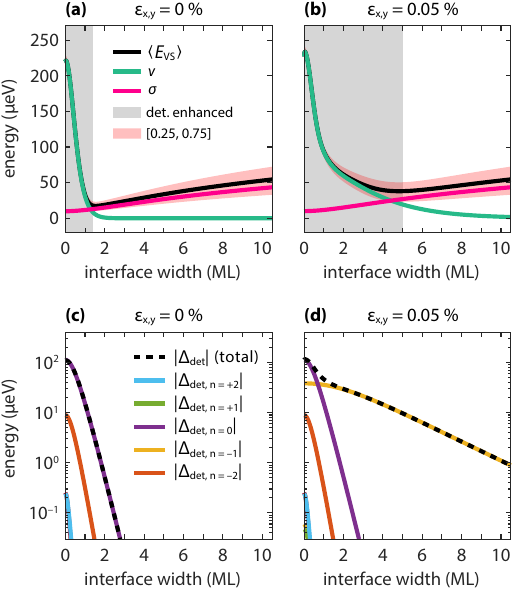}
\caption{Impact of interface width on valley splitting. \textbf{(a)}~Mean
valley splitting $\langle E_{\mathrm{VS}}\rangle$ and shape parameters
$\nu=2\vert\Delta_{\mathrm{det}}\vert$ and $\sigma=\sqrt{2\Gamma}$
of the Rice distribution in the absence of shear strain $\varepsilon_{x,y}=0$
as a function of the QW interface width.
The interface width is given in units of monolayers $\mathrm{ML}=a_{0}/4\approx0.1357\,\mathrm{nm}$
of relaxed Si. The gray shaded region indicates the deterministically
enhanced regime according to Eq.~\eqref{eq: separatrix condition}
and the red shaded region shows the {[}25\%, 75\%{]} quantile of the
Rice distribution. Deterministic enhancements are observed for sharp
interfaces with up to one ML width. \textbf{(b)}~Same as (a), but
for small shear strain $\varepsilon_{x,y}=0.05\%$. The deterministically
enhanced regime is extended to about five MLs. \textbf{(c)}~Absolute
values of the individual components $\Delta_{\mathrm{det},n}$ contributing
to the deterministic valley splitting, see Eq.~\eqref{eq: Delta det compact},
as a function of the interface width. In the absence of shear strain,
the resulting valley slitting is dominated by the $n=0$ resonance
with a small correction induced by the $n=-2$ resonance. \textbf{(d)}~In
the case of small shear strain $\varepsilon_{x,y}=0.05\%$, the $n=-1$
resonance provides the dominant contribution over almost the entire
parameter domain. In the simulation, an electric field of $F=10\,\mathrm{mV}/\mathrm{nm}$
was assumed. }\label{fig: interface width}
\end{figure}

\subsection{Conventional Heterostructure: Dependence on Quantum Well Interface
Width }\label{sec:Interface-width}

Sharp interfaces are a common strategy to enhance the valley splitting
\cite{Saraiva2009,Culcer2010,DegliEsposti2024}. From the perturbative
expression for the intervalley coupling parameter in Eq.~\eqref{eq: Delta},
it is clear that any heterostructure which contains sufficiently sharp
spatial features overlapping with the wave function will lead to an
enhanced valley splitting. This is because the corresponding broadband
Fourier spectrum still has comparatively large amplitudes even at
high wave numbers, especially close to $2k_{0}$. Consequently, there
is a great potential for further enhancement of the valley splitting
in the presence of shear strain $\varepsilon_{x,y}\neq0$, which unlocks
the low-frequency resonance at $2k_{1}$ that is typically supported
by much larger Fourier amplitudes.

The results in Fig.~\ref{fig: interface width} show that the influence
of shear strain on sharp interfaces is indeed significant. Without
shear strain, extremely sharp interfaces with a width $\sigma_{u,l}\approx a_{0}/4$
of about one monolayer (ML) are required to achieve a deterministic
enhancement of the valley splitting, see Fig.~\ref{fig: interface width}\,(a).
This result is consistent with previous findings based on the $2k_{0}$-theory
\cite{Losert2023,Lima2023}. If shear strain is applied, the range
in which a deterministic enhancement occurs is significantly
increased. In Fig.~\ref{fig: interface width}\,(b) it is shown
that already moderate shear strain of $\varepsilon_{x,y}=0.05\%$
yields deterministic enhancements for rather broad interfaces with
a width of up to five MLs. In any case, sharp interfaces must be
combined with a strong vertical electric field to enhance
the overlap of the wave function with the interface.

\subsection{Unconventional Heterostructures }\label{sec: unconventional heterostructures}

A very promising approach to achieve deterministic enhancements of
the valley splitting is the \emph{wiggle well} heterostructure, which
employs an oscillating Ge concentration within the QW \cite{McJunkin2022,Feng2022,Woods2024}.
In the following, we assume an epitaxial profile of the form
\begin{equation*}
X\left(z\right)=X_{\mathrm{QW}}\left(z\right)+\Xi\left(z\right)x\left(z\right),
\end{equation*}
where $X_{\mathrm{QW}}=X_{b}\left(1-\Xi\left(z\right)\right)$ is
the profile of the conventional SiGe/Si/SiGe QW, see Eq.~\eqref{eq: QW epitaxial profile X},
and 
\begin{equation}
x\left(z\right)=\frac{1}{2}X_{\mathrm{ww}}\left(1+\cos\left(qz\right)\right)\label{eq: wiggle well profile}
\end{equation}
describes the oscillating Ge concentration in the QW with amplitude
$X_{\mathrm{ww}}$ and wave number $q$. The smoothed QW indicator
function $\Xi\left(z\right)$ is given in Eq.~\eqref{eq: QW indicator}.

In the following, we will discuss a number of special cases of Eq.~(\ref{eq: wiggle well profile})
that are each characterized by a particular wave number $q$.

\begin{figure}[t]
\includegraphics[width=1\columnwidth]{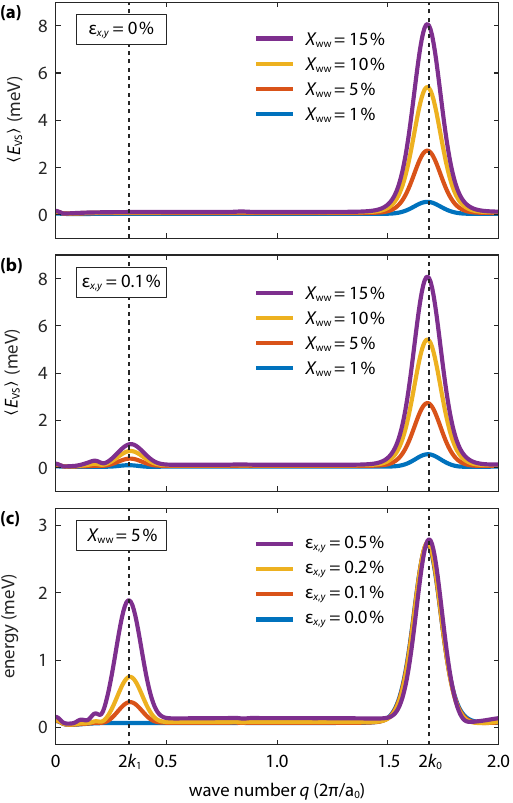}
\caption{Mean valley splitting in the wiggle well heterostructure as a function
of wave number. \textbf{(a)}~In the absence of shear strain $\varepsilon_{x,y}=0\%$
deterministic enhancements are observed only at $q=2k_{0}$ (short-period
wiggle-well), which increases with increasing amplitude $X_{\mathrm{ww}}$.
Away from the resonance, the non-zero valley splittings are disorder-dominated.
\textbf{(b)~}A small amount of shear strain $\varepsilon_{x,y}=0.1\,\%$
unlocks a new resonance at $q=2k_{1}$ (long-period wiggle-well).
\textbf{(c)}~Mean valley splitting for fixed amplitude $X_{\mathrm{ww}}=5\%$
and different values of shear strain $\varepsilon_{x,y}$. A constant
electric field $F=5\,\mathrm{mV}/\mathrm{nm}$ and biaxial tensile
strain, see Eq.~\eqref{eq: biaxial strain QW}, was assumed in the
simulations.}
\label{fig: ww line scans}
\end{figure}

\subsubsection{Uniform Germanium Concentration}\label{sec:Uniform-Germanium}

For $q=0$, the profile in Eq.~\eqref{eq: wiggle well profile},
reduces to a uniform Ge concentration in the QW. In this configuration,
which was first proposed in Ref.~\cite{PaqueletWuetz2022}, the
mean valley splitting is enhanced because the random Ge concentration
corresponds to a flat (white noise) power spectrum in reciprocal space,
contributing even at large wave numbers. The corresponding enhancement
of the valley splitting is, however, entirely disorder-dominated.
Therefore, numerous spin-valley hotspots must be expected \cite{Losert2024}.

\begin{figure*}[t]
\includegraphics[width=1\textwidth]{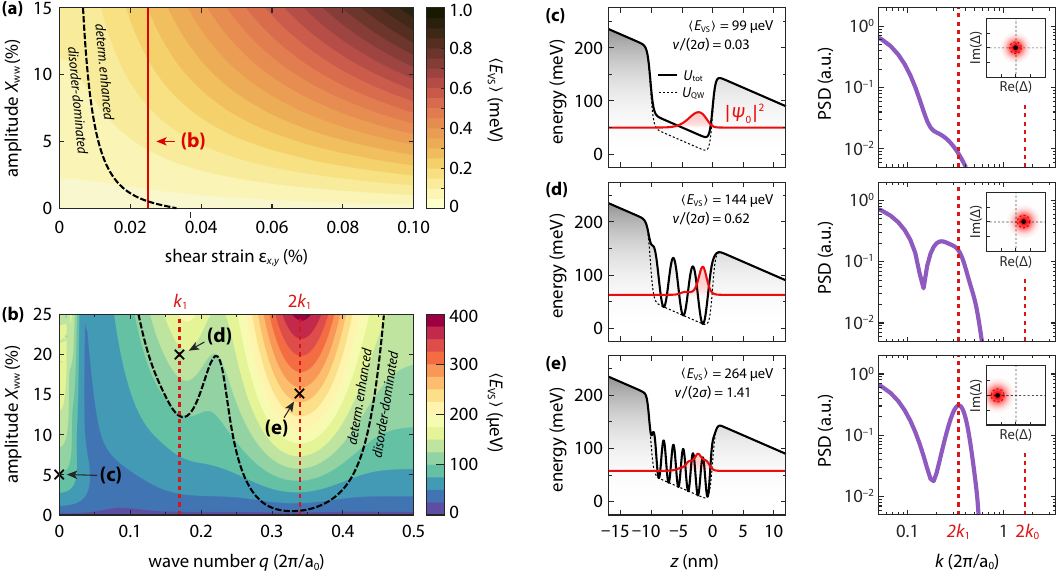}
\caption{\textbf{(a)}~Mean valley splitting of the long-period wiggle-well
with fixed wave number $q=0.32\times2\pi/a_{0}\approx2k_{1}$ as a
function of shear strain $\varepsilon_{x,y}$ and Ge amplitude $X_{\mathrm{ww}}$.
The dashed line is the separatrix defined in Eq.~\eqref{eq: separatrix condition}
that separates disorder-dominated and deterministically enhanced regimes.
For large shear strains, a deterministic enhancement is achieved already
for very small Ge amplitudes. The red line indicates the constant
shear strain value considered in panel (b). \textbf{(b)}~Mean valley
splitting as a function of wave number $q$ and amplitude $X_{\mathrm{ww}}$
for fixed shear strain $\varepsilon_{x,y}=0.025\,\%$. Three domains
with valley splitting enhancements can be observed, namely, at very
low $q\approx0$ (neary constant uniform Ge concentration in the QW),
near $q\approx k_{1}$ (lower harmonic/ Ge spike, only for sufficiently
high $X_{\mathrm{ww}}$) and near $q\approx2k_{1}$ (long-period wiggle-well).
The dashed line is again the separatrix. \textbf{(c)}~Left: Epitaxial
profile and ground state electron density distribution for the uniform
Ge concentration in the well ($q=0$, $X_{\mathrm{ww}}=5\%$) at shear
strain $\varepsilon_{x,y}=0.025\,\%$. The low value of the ratio
$\nu/\left(2\sigma\right)$ indicates that this configuration falls
into the disorder-dominated regime. Right: Power spectral density
(PSD) of the product $\left(U_{\mathrm{QW}}\left(z\right)+U_{F}\left(z\right)\right)\left|\psi_{0}\left(z\right)\right|^{2}$
as a function of the wave number $k$. The contribution to the valley
splitting is dominated by the $2k_{1}$-resonance and is thus larger
than predicted by the $2k_{0}$-theory. The inset shows the statistical
distribution of the intervalley coupling parameter in the complex
plane in the range of $-200\,\text{\ensuremath{\mathrm{\text{\textmu}eV}}}\protect\leq\mathrm{Re}\left(\Delta\right),\mathrm{Im}\left(\Delta\right)\protect\leq200\,\mathrm{\text{\textmu}eV}$.
\textbf{(d)}~Same as (c), but at ($q=k_{1}$, $X_{\mathrm{ww}}=20\%$).
The product of the potential and the wave function yields a strong
Fourier component at $2k_{1}$, although the potential has only
half of the frequency $q=k_{1}$. The configuration has a moderate
deterministic enhancement $\nu/\left(2\sigma\right)\approx0.62$.
\textbf{(e)}~Same as (c), but for a long-period wiggle-well at ($q=2k_{1}$,
$X_{\mathrm{ww}}=15\%$). The $q=2k_{1}$-periodic modulation of the
potential provides a significant deterministic enhancement of the
valley splitting $\nu/\left(2\sigma\right)\approx1.41$. A constant
electric field $F=5\,\mathrm{mV}/\mathrm{nm}$ and biaxial (tensile)
strain was assumed in all simulations. }\label{fig: wiggle well maps and wave functions}
\end{figure*}

For a sufficient amount of shear strain, when the deterministic component
starts to be dominated by the $n=-1$ resonance, cf.~Fig.~\ref{fig: interface width}\,(d),
slight enhancements of the deterministic contribution can be expected.
These enhancements are, however, much smaller than the disorder-induced
component for typical parameters. This is shown in Fig.~\ref{fig: ww line scans}\,(a)--(b),
where the valley splittings at $q=0$ are practically unchanged even
in the presence of shear strain. Moreover, Fig.~\ref{fig: wiggle well maps and wave functions}\,(b)
shows a regime with enhanced mean valley splitting at very low wave
numbers, which is clearly outside of the deterministically enhanced
regime (indicated by the separatrix condition~\eqref{eq: separatrix condition},
dashed line). Finally, the typical epitaxial profile and electron
density distribution is shown in Fig.~\ref{fig: wiggle well maps and wave functions}\,(c)
along with the power spectral density (PSD) of the product $\left(U_{\mathrm{QW}}\left(z\right)+U_{F}\left(z\right)\right)\left|\psi_{0}\left(z\right)\right|^{2}$
and the complex-plane distribution of the intervalley coupling parameter
$\Delta$, showing that the valley splitting is indeed disorder-dominated.

\subsubsection{Short-Period Wiggle-Well}\label{sec:short-period-wiggle-well}

The results shown in Fig.~\ref{fig: ww line scans} feature a strong
resonance at $q=2k_{0}$, which grows with increasing Ge amplitude
$X_{\mathrm{ww}}$, but is practically independent of shear strain.
This configuration is called the \emph{short-period wiggle-well}.
The effect is based on a direct enhancement of the coupling between
the valley states within the same Brillouin zone triggered by the
periodicity of the confinement potential. The short-period wiggle-well
requires a high-frequency Ge modulation with a very short wave length
$\lambda=\pi/k_{0}\approx2.4\,\mathrm{ML}\approx0.32\,\mathrm{nm}$.
The epitaxial growth of such rapidly modulated structures is out of
reach with current growth technology.

\subsubsection{Long-Period Wiggle-Well}\label{sec:long-period-wiggle-well}

In the presence of shear strain, a second resonance emerges at $q=2k_{1}$,
see Fig.~\ref{fig: ww line scans}\,(b). The peak grows almost linearly
with increasing amplitude $X_{\mathrm{ww}}$. This resonance is associated
with the \emph{long-period wiggle-well} \cite{McJunkin2022,Feng2022,Woods2024},
which enhances the coupling of valley states in neighboring Brillouin
zones separated by $2k_{1}$. The corresponding wave length is $\lambda\approx11.8\,\mathrm{ML}\approx1.60\,\mathrm{nm}$
(for a biaxially strained QW). Such structures can be grown using
molecular beam epitaxy \cite{Gradwohl2025}.

The dependency on shear strain is illustrated in Fig.~\ref{fig: ww line scans}\,(c),
showing a linear relation between the mean valley splitting and shear
strain at $q=2k_1$.
The two-parametric plot in Fig.~\ref{fig: wiggle well maps and wave functions}\,(a)
shows the dependency of the mean valley splitting in a long-period
wiggle-well on shear strain and the Ge amplitude. The separatrix (dashed
line) indicates that a minimum shear strain of about $\varepsilon_{x,y}\approx0.006\%$
is required to enter the deterministically enhanced regime even for
high Ge amplitudes (where disorder effects are strong). For shear strain above $\varepsilon_{x,y}\gtrsim0.035\%$,
the deterministic enhancement sets in already for tiny Ge amplitudes.
In Fig.~\ref{fig: wiggle well maps and wave functions}\,(b), the
mean valley splitting $\langle E_{\mathrm{VS}}\rangle$ is shown for
fixed shear strain $\varepsilon_{x,y}=0.025\%$ as a function of the
wave number $q$ and the Ge oscillation amplitude $X_{\mathrm{ww}}$.
The broad resonance around $q\approx2k_{1}$ falls clearly in the
deterministically enhanced regime and is relatively robust against
deviations from the ideal wave number. The typical epitaxial profile
and the corresponding electron density distribution are shown in Fig.~\ref{fig: wiggle well maps and wave functions}\,(e)
along with the PSD and the complex-plane distribution of the intervalley
coupling parameter (inset). The strong spectral contribution at $2k_{1}$
leads to a considerable deterministic enhancement of the valley splitting
such that the complex normal distribution governing the statistical
properties is steered away from the origin. Hence, a significant suppression
of spin-valley hotspots can be expected. The long-period wiggle-well
is thus a very promising approach to design qubits with deterministically
large valley splittings.

\subsubsection{Lower Harmonic Wiggle-Well / Ge-Spike }\label{sec:lower-harmonic-wiggle-well}

Finally, we note the emergence of a new resonance in Fig.~\ref{fig: ww line scans}\,(b)
at approximately $q\approx k_{1}$, \emph{i.e.}, about half the wave
number of the long-period wiggle-well. This lower harmonic resonance
is suppressed in the absence of shear strain and sets in only at fairly
high Ge amplitudes. The resonance is also clearly visible in Fig.~\ref{fig: wiggle well maps and wave functions}\,(b)
for Ge amplitudes above $X_{\mathrm{ww}}\gtrsim12\%$ and falls clearly
within the deterministically enhanced regime. The underlying mechanism
becomes clear in Fig.~\ref{fig: wiggle well maps and wave functions}\,(d),
which shows an intricate interplay of the periodic heterostructure
potential and the electron ground state wave function. While both
of them individually have a frequency spectrum dominated by the $k_{1}$-component,
their combination effectively triggers the $2k_{1}$-resonance mechanism
that underlies the long-period wiggle-well. From the plot it becomes also
 clear why a large Ge amplitude is required, since this is crucial
to achieve the strong electronic confinement.

While at first glance the longer wave length might be favorable due
to reduced demands on the epitaxial growth process, the required high
Ge content causes severe drawbacks. In fact, we expect a strong enhancement
of the spin-orbit interaction in such a structure \cite{Woods2023},
which would significantly complicate the control of the qubit. The
deterministic enhancements achieved with such a structure are lower
than that of a corresponding long-period wiggle-well.

Finally, we remark that the second Ge-peak near the lower interface
is not essential for the observed effect. It is simply enforced by
the parametrization in Eq.~\eqref{eq: wiggle well profile}. Without
the second Ge-peak, the structure has similarities with the
Ge-spike described in Ref.~\cite{McJunkin2021}.

\section{Discussion}\label{sec: discussion}

The theoretical model developed in this work combines a number of previously existing modeling concepts
into a unified framework. In this way, a comprehensive description
of a number of important effects contributing to the valley splitting
in engineered heterostructures, \emph{i.e.}, alloy-disorder, strain
and resonances, has been achieved at the level of envelope function
theory.

The numerical results in Sec.~\ref{sec: results} are consistent
with previously reported findings, but also extend them by systematically investigating the effects of strain. For instance, significant
shear strain-induced enhancements of the valley splitting even at
wide interfaces were predicted. In addition, the model includes all
non-trivial resonances that mediate intervalley coupling across neighboring
Brillouin zones. Most notably, this includes the important resonance
at $2k_{1}$, which triggers the mechanism behind the long-period
wiggle well. This is the main difference with the prevailing $2k_{0}$-theory,
which does not explain the long-period wiggle-well. The key accomplishments
of the $2k_{0}$-theory as described in Refs.~\cite{Losert2023,Lima2023},
lies in the description of disorder-induced effects and the distinction
between deterministic and disorder-dominated enhancements of valley
splitting. This part of the theory has been adopted practically identically
in the present model.

The results reported in this paper depend to a minor
degree on the chosen pseudopotential model. In addition to the model
by Rieger \& Vogl~\cite{Rieger1993}, we have also tested the model
by Fischetti \& Laux~\cite{Fischetti1996} and obtained qualitatively
similar results with slight quantitative differences. A detailed comparison
is beyond the scope of this paper.

The present model differs in a number of aspects from the model by
Feng \& Joynt~\cite{Feng2022}, which has been the first to provide
an explanation for the mechanism behind the long-period wiggle-well
in the framework of envelope function theory. The mechanism proposed
there to induce a violation of the nonsymmorphic screw symmetry---which otherwise suppresses the resonance at $2k_{1}$---is based
on a randomization of the plane wave expansion coefficients of the
Bloch factors. The latter are determined for statistical ensembles
of disordered SiGe alloys using a combination of an empirical pseudopotential
model and an extended virtual crystal approximation, whereby the alloy
disorder induces the desired symmetry breaking. This approach is in
principle very plausible, but a subsequent analysis shows that the
corresponding statistical distribution of the valley splitting implied
by this method does not lead to a Rice distribution (as observed in
experiments and atomistic simulations). Instead, a significantly more
skewed distribution is found, which has an increased probability
density at low valley splittings compared to the Rice distribution.
Furthermore, in the model by Feng \& Joynt, the distribution of the
intervalley coupling parameter in the complex plane takes a strongly
elliptical shape, which is very different from the nearly circular
distribution obtained here (see Appendix~\ref{sec: statistics - Intervalley-Coupling-Parameter}).
Another consequence of the symmetry-breaking mechanism proposed there
is that the variance of the valley splitting in the long-period wiggle-well
becomes very large. Thus, the valley splitting enhancement due to
the long-period wiggle-well is found to be primarily disorder-dominated
and is by no means deterministic. This is reasonable when one assumes
that the symmetry breaking is disorder-induced. These predictions
on the characteristics of the long-period wiggle-well are in sharp
contrast to those obtained in the present work, where the nonsymmorphic
symmetry is broken via shear strain. As a consequence, the present
model predicts a strong deterministic enhancement of the valley splitting
in the long-period wiggle-well. The actual microscopic symmetry-breaking
mechanism, however, is very likely a combination of both effects.
The impact of disorder on the long-period wiggle-well might therefore
require further investigation. Finally, the model by Feng \& Joynt
shows another pronounced resonance for a wiggle well with intermediate
wave number $q=k_{0}$. In our model, we found that this is caused
by a harmonic similar to the configuration described in Sec.~\ref{sec:lower-harmonic-wiggle-well}.
For typical SiGe parameters (recall that Feng \& Joynt used a conduction
band offset more typical for Si-MOS), however, the small deterministic
resonance is entirely obscured by the disorder-induced component.
Thus, no peak at $q=k_{0}$ can be observed in the mean valley splitting
in Fig.~\ref{fig: ww line scans}.

Finally, we discuss the relation between the present work and the
paper by Woods \emph{et al.} \cite{Woods2024}. In Ref.~\cite{Woods2024},
the shear strain induced symmetry-breaking mechanism to unlock the
resonance behind the long-period wiggle-well has been studied for
the first time. The envelope function model presented there, which
is derived from a one-dimensional tight-binding model, features an
intervalley coupling matrix element with unconventional functional
form that is notably different from other expressions in the literature.
In our model, however, the standard form of the intervalley matrix
element is preserved even in the presence of strain effects. The numerical
results obtained here are qualitatively in excellent agreement
with those reported by Woods \emph{et al.}. In particular the joint
linear dependence of the long-period wiggle-well on shear strain and
Ge concentration is recovered using the EPM. Furthermore, we also
find good quantitative agreements in the magnitude of the obtained
valley splittings. The quantitative comparison, however, requires
caution, as the present model also contains disorder-induced contributions,
which are omitted in Ref.~\cite{Woods2024}.

\section{Conclusions and Outlook}

\label{sec: conclusions} We have developed a comprehensive theoretical
model that describes a broad range of physical mechanisms determining
the valley splitting in Si/SiGe heterostructures, namely alloy-disorder,
strain and resonances between valley states across neighboring Brillouin
zones. The numerical results are consistent with previously known
results but also offer new insights extending the state of the art.
The key accomplishment of our model is the explanation of the long-period wiggle-well in the framework of envelope function theory as a result of shear strain-induced symmetry
breaking, which unlocks the corresponding resonance mechanism.
Due to the critical role of shear strain, we conclude
that strain engineering \cite{Sverdlov2011,Frink2023,Adelsberger2024,Woods2024}
must be further advanced to enhance the design of Si/SiGe spin-qubits.

In contrast to detailed atomistic models, the present model offers
a comprehensive description of multiple physical phenomena at very
low computational cost. Therefore, it is well suited to rapidly characterize
the valley splitting in different epitaxial designs.
In a subsequent work, the present model will be employed for free-shape
optimization of the epitaxial profile to further enhance the valley
splitting beyond the heuristic strategies considered here.

\section*{Code Availability}

MATLAB code to reproduce the simulation results and figures of this
work is available on GitHub~\cite{Thayil2025b}.

\begin{acknowledgments}
This work was funded by the Deutsche Forschungsgemeinschaft (DFG,
German Research Foundation) under Germany's Excellence Strategy --
The Berlin Mathematics Research Center MATH+ (EXC-2046/1, project
ID: 390685689, project AA2--17). M.~K. is grateful for valuable
discussions with Mark Friesen, Merritt P. Losert, Lars R. Schreiber
and Thomas Koprucki. 
\end{acknowledgments}

\appendix

\section{Multi-Valley Coupled Envelope Equation Model  }\label{sec: Derivation of coupled envelope equations}

In this section, we provide a derivation of the coupled envelope equations
\eqref{eq: coupled envelope equation model} based on a Burt--Foreman
type envelope function theory for multi-valley semiconductors.

\subsection{Microscopic Schrödinger Equation}

We consider the stationary Schrödinger equation for an electronic
state in the Si/SiGe heterostructure 
\begin{equation}
\left(-\frac{\hbar^{2}}{2m_{0}}\nabla^{2}+V\left(\mathbf{r}\right)+U\left(\mathbf{r}\right)\right)\Psi\left(\mathbf{r}\right)=E\Psi\left(\mathbf{r}\right),\label{eq: stationary Schroedinger equation}
\end{equation}
where $m_{0}$ is the vacuum electron mass, $V\left(\mathbf{r}\right)$
is a lattice-periodic potential and $U\left(\mathbf{r}\right)$ is
a mesoscopic confinement potential. The crystal potential is invariant
under translations with lattice vectors $V\left(\mathbf{r}\right)\equiv V\left(\mathbf{r}+\mathbf{R}\right)$.
Throughout this paper we assume that $V_{0}$ describes solely the
perfectly periodic Si crystal, whereas the effects of Ge in both the
QW and the barrier are described by the heterostructure potential
$U_{\mathrm{het}}\left(\mathbf{r}\right)$ that is included in $U\left(\mathbf{r}\right)$,
see Eq.~\eqref{eq: total potential}. This way, double-counting of
Ge atoms is avoided and alloy disorder effects are separated from
the idealized band structure computation.

\subsection{Multi-Valley Envelope Wave Function}

\begin{figure}
\includegraphics[width=1\columnwidth]{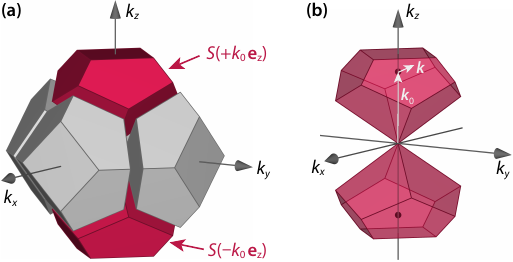}

\caption{\textbf{(a)}~Decomposition of the first Brillouin zone of the fcc
lattice into non-overlapping valley-specific sectors $S\left(\mathbf{k}_{0}\right)$.
The sectors of the two low-energy valleys are highlighted.
\textbf{(b)}~The wave vectors within the sectors are decomposed into
a valley vector $\mathbf{k}_{0}$ and a deviation $\mathbf{k}$ such
that $\left(\mathbf{k}_{0}+\mathbf{k}\right)\in S\left(\mathbf{k}_{0}\right)$.
}\label{fig:Brillouin-zone-decomposition}

\end{figure}

The single-particle wave function for electrons in semiconductor nanostructures
shows rapid spatial oscillations on the atomistic scale of the underlying
crystal as well as a slowly varying envelope on the mesoscopic scale
of the confinement potential. In order to obtain an effective description
of the envelope wave function, the rapid atomistic features must be
eliminated. There is extensive literature on the
(heuristic) derivation of envelope function theories for electrons
in Si-based nanostructures \cite{Kohn1955,Fritzsche1962,Ning1971,Shindo1976,Hui2013}
that employ a number of approximations, most notably the assumption
of a slowly varying potential. In the context of the present work,
the assumption of a slowly varying potential is questionable, as we
are particularly interested in the investigation of QWs with sharp
interfaces and highly oscillatory wiggle wells. Burt \cite{Burt1988,Burt1992}
and Foreman \cite{Foreman1995,Foreman1996} have developed an \emph{exact}
envelope function theory that avoids these assumptions, but leads
to a system of coupled integro-differential equations with a nonlocal
potential. The approach has been extended to multi-valley semiconductors
by Klymenko \emph{et al.}~\cite{Klymenko2014,Klymenko2015} based
on a decomposition of the first Brillouin zone into non-overlapping
valley-specific sectors, see Fig.~\ref{fig:Brillouin-zone-decomposition}\,(a).
In this section we provide a derivation of the coupled envelope equations
\eqref{eq: coupled envelope equation model} based on multi-valley
Burt--Foreman type envelope function theory.

Following \cite{Klymenko2014,Klymenko2015}, the wave function
is expanded as
\begin{equation}
\Psi\left(\mathbf{r}\right)=\sum_{n}\sum_{\mathbf{k}_{0}}u_{n,\mathbf{k}_{0}}\left(\mathbf{r}\right)\mathrm{e}^{i\mathbf{k}_{0}\cdot\mathbf{r}}f_{n,\mathbf{k}_{0}}\left(\mathbf{r}\right),\label{eq: expansion ansatz}
\end{equation}
where $n$ labels the band index, $\mathbf{k}_{0}$ are the wave vectors
of the conduction band valleys, $u_{n,\mathbf{k}_{0}}\left(\mathbf{r}\right)$
is the lattice-periodic Bloch factor and $f_{n,\mathbf{k}_{0}}\left(\mathbf{r}\right)$
is the (slowly varying) envelope wave function. The Fourier expansion
of the wave function reads
\[
\Psi\left(\mathbf{r}\right)=\sum_{\mathbf{G}}\sum_{\mathbf{k}_{0}}\sum_{\left(\mathbf{k}+\mathbf{k}_{0}\right)\in S\left(\mathbf{k}_{0}\right)}\mathrm{e}^{i\left(\mathbf{G}+\mathbf{k}_{0}+\mathbf{k}\right)\cdot\mathbf{r}}\Psi\left(\mathbf{G}+\mathbf{k}_{0}+\mathbf{k}\right),
\]
where each possible wave number is uniquely described by a reciprocal
lattice vector $\mathbf{G}$, a valley vector $\mathbf{k}_{0}$ and
a small wave number $\mathbf{k}$ such that the sum $\mathbf{k}+\mathbf{k}_{0}$
is restricted to the sector $S\left(\mathbf{k}_{0}\right)$ of the
Brillouin zone, see Fig.~\ref{fig:Brillouin-zone-decomposition}\,(b).
The exponential factor $\exp{\left(i\mathbf{G}\cdot\mathbf{r}\right)}$
can be expanded in terms of lattice-periodic Bloch functions
\[
\mathrm{e}^{i\mathbf{G}\cdot\mathbf{r}}=\sum_{n}u_{n,\mathbf{k}_{0}}\left(\mathbf{r}\right)c_{n,\mathbf{k}_{0}}^{*}\left(\mathbf{G}\right),
\]
such that we obtain from \eqref{eq: expansion ansatz} and its Fourier
transform an expression for the envelope
\begin{align}
 & f_{n,\mathbf{k}_{0}}\left(\mathbf{r}\right)=\label{eq: envelope wave function}\\
 & =\sum_{\left(\mathbf{k}+\mathbf{k}_{0}\right)\in S\left(\mathbf{k}_{0}\right)}\mathrm{e}^{i\mathbf{k}\cdot\mathbf{r}}\sum_{\mathbf{G}}c_{n,\mathbf{k}_{0}}^{*}\left(\mathbf{G}\right)\Psi\left(\mathbf{G}+\mathbf{k}_{0}+\mathbf{k}\right).\nonumber 
\end{align}
This function is slowly varying as its Fourier expansion is restricted
to a limited set of (small) wave vectors $\mathbf{k}$ relative to
the valley vector such that $\left(\mathbf{k}+\mathbf{k}_{0}\right)\in S\left(\mathbf{k}_{0}\right)$.

Following \cite{Burt1994}, the envelope equation is derived starting
out from the Fourier domain representation of Schrödinger's equation
\eqref{eq: stationary Schroedinger equation}. Substitution of the
ansatz \eqref{eq: expansion ansatz} yields
\begin{align*}
Ef_{n,\mathbf{k}_{0}}\left(\mathbf{r}\right) & =\left(-\frac{\hbar^{2}}{2m_{0}}\nabla^{2}-\frac{i\hbar^{2}}{m_{0}}\mathbf{k}_{0}\cdot\nabla+\frac{\hbar^{2}k_{0}^{2}}{2m_{0}}\right)f_{n,\mathbf{k}_{0}}\left(\mathbf{r}\right)\\
 & \phantom{=}+\frac{1}{m_{0}}\sum_{n'}\mathbf{p}_{n,n'}\left(\mathbf{k}_{0}\right)\cdot\left(\hbar\mathbf{k}_{0}-i\hbar\nabla\right)f_{n,'\mathbf{k}_{0}}\left(\mathbf{r}\right)\\
 & \phantom{=}+\sum_{n'}\left(T_{n,n'}\left(\mathbf{k}_{0}\right)+V_{n,n'}\left(\mathbf{k}_{0}\right)\right)f_{n,'\mathbf{k}_{0}}\left(\mathbf{r}\right)\\
 & \phantom{=}+\sum_{n'}\sum_{\mathbf{k}_{0}'}\int_{V}\mathrm{d}^{3}r'\,U_{\mathbf{k}_{0},\mathbf{k}_{0}'}^{n,n'}\left(\mathbf{r},\mathbf{r}'\right)f_{n',\mathbf{k}_{0}'}\left(\mathbf{r}'\right)
\end{align*}
where the nonlocal potential energy kernel reads
\begin{align}
U_{\mathbf{k}_{0},\mathbf{k}_{0}'}^{n,n'}\left(\mathbf{r},\mathbf{r}'\right) & =\mathrm{e}^{-i\left(\mathbf{k}_{0}\cdot\mathbf{r}-\mathbf{k}_{0}'\cdot\mathbf{r}'\right)}\times\label{eq: nonlocal potential}\\
 & \phantom{=}\times\int_{V}\mathrm{d}^{3}r''\,\Delta_{\mathbf{k}_{0}}\left(\mathbf{r}-\mathbf{r}''\right)u_{n,\mathbf{k}_{0}}^{*}\left(\mathbf{r}''\right)\times\nonumber \\
 &\hphantom{=\int_{V}\mathrm{d}^{3}r''}\times U\left(\mathbf{r}''\right)u_{n',\mathbf{k}_{0}'}\left(\mathbf{r}''\right)\Delta_{\mathbf{k}_{0}'}\left(\mathbf{r}''-\mathbf{r}'\right).\nonumber 
\end{align}
The matrix elements of the momentum, kinetic and potential energy
operator of the bulk crystal are
\begin{align*}
\mathbf{p}_{n,n'}\left(\mathbf{k}_{0}\right) & =\frac{1}{\Omega_{p}}\int_{\Omega_{p}}\mathrm{d}^{3}r\,u_{n,\mathbf{k}_{0}}^{*}\left(\mathbf{r}\right)\left(-i\hbar\nabla u_{n',\mathbf{k}_{0}}\left(\mathbf{r}\right)\right),\\
T_{n,n'}\left(\mathbf{k}_{0}\right) & =\frac{1}{\Omega_{p}}\int_{\Omega_{p}}\mathrm{d}^{3}r\,u_{n,\mathbf{k}_{0}}^{*}\left(\mathbf{r}\right)\left(-\frac{\hbar^{2}}{2m_{0}}\nabla^{2}u_{n',\mathbf{k}_{0}}\left(\mathbf{r}\right)\right),\\
V_{n,n'}\left(\mathbf{k}_{0}\right) & =\frac{1}{\Omega_{p}}\int_{\Omega_{p}}\mathrm{d}^{3}r\,u_{n,\mathbf{k}_{0}}^{*}\left(\mathbf{r}\right)V\left(\mathbf{r}\right)u_{n',\mathbf{k}_{0}}\left(\mathbf{r}\right).
\end{align*}
The nonlocal term \eqref{eq: nonlocal potential} for the mesoscopic
confinement potential involves the low-pass filtered delta function
\begin{equation}
\Delta_{\mathbf{k}_{0}}\left(\mathbf{r}-\mathbf{r}'\right)=\frac{1}{V}\sum_{\left(\mathbf{k}+\mathbf{k}_{0}\right)\in S\left(\mathbf{k}_{0}\right)}\mathrm{e}^{i\left(\mathbf{k}_{0}+\mathbf{k}\right)\cdot\left(\mathbf{r}-\mathbf{r}'\right)}\label{eq: low-pass Delta}
\end{equation}
that leads to smoothing of the effective potential and ensures proper
restriction to small wave vectors within the respective Brillouin
zone sectors.

Using the identity for the lattice-periodic problem 
\begin{align*}
T_{n,n'}\left(\mathbf{k}_{0}\right)+V_{n,n'}\left(\mathbf{k}_{0}\right)+ & \frac{\hbar}{m_{0}}\mathbf{k}_{0}\cdot\mathbf{p}_{n,n'}\left(\mathbf{k}_{0}\right)=\\
 & =\left(E_{n,\mathbf{k}_{0}}-\frac{\hbar^{2}k_{0}^{2}}{2m_{0}}\right)\delta_{n,n'},
\end{align*}
the system of coupled envelope equations takes the form
\begin{align}
Ef_{n,\mathbf{k}_{0}}\left(\mathbf{r}\right) & =\left(-\frac{\hbar^{2}}{2m_{0}}\nabla^{2}+E_{n,\mathbf{k}_{0}}\right)f_{n,\mathbf{k}_{0}}\left(\mathbf{r}\right)\label{eq: envelope function (3D)}\\
 &-\frac{i\hbar}{m_{0}}\sum_{n'}\left(\hbar\mathbf{k}_{0}\delta_{n,n'}+\mathbf{p}_{n,n'}\left(\mathbf{k}_{0}\right)\right)\cdot\nabla f_{n',\mathbf{k}_{0}}\left(\mathbf{r}\right)\nonumber \\
 &+\sum_{n'}\sum_{\mathbf{k}_{0}'}\int_{V}\mathrm{d}^{3}r'\,U_{\mathbf{k}_{0},\mathbf{k}_{0}'}^{n,n'}\left(\mathbf{r},\mathbf{r}'\right)f_{n',\mathbf{k}_{0}'}\left(\mathbf{r}'\right).\nonumber 
\end{align}

\subsection{Elimination of Remote Bands}

The multi-band system \eqref{eq: envelope function (3D)} can be
reduced to an effective equation for the conduction band $n=c$ by
eliminating remote bands $n\neq c$ (Löwdin renormalization). We approximate the remote band envelopes as \cite{Burt1988}
\begin{align*}
f_{n,\mathbf{k}_{0}}\left(\mathbf{r}\right) & \approx-\frac{i\hbar}{m_{0}}\frac{\mathbf{p}_{n,c}\left(\mathbf{k}_{0}\right)}{E_{c,\mathbf{k}_{0}}-E_{n,\mathbf{k}_{0}}}\cdot\nabla f_{c,\mathbf{k}_{0}}\left(\mathbf{r}\right)
\end{align*}
and neglect interaction between remote bands. Moreover, we ignore
the renormalization of the confinement potential term for the conduction
band. With this, we arrive at the single-band envelope equation in
effective mass approximation
\begin{align}
Ef_{c,\mathbf{k}_{0}}\left(\mathbf{r}\right) & =-\frac{\hbar^{2}}{2}\nabla\cdot\left(m_{c,\mathbf{k}_{0}}^{-1}\nabla f_{c,\mathbf{k}_{0}}\left(\mathbf{r}\right)\right)+E_{c,\mathbf{k}_{0}}f_{c,\mathbf{k}_{0}}\left(\mathbf{r}\right)\nonumber \\
 & \phantom{=}+\sum_{\mathbf{k}_{0}'}\int_{V}\mathrm{d}^{3}r'\,U_{\mathbf{k}_{0},\mathbf{k}_{0}'}^{c,c}\left(\mathbf{r},\mathbf{r}'\right)f_{c,\mathbf{k}_{0}'}\left(\mathbf{r}'\right)\label{eq: single band equation}
\end{align}
with the effective mass tensor
\begin{equation}
m_{c,\mathbf{k}_{0}}^{-1}=\frac{1}{m_{0}}I+\frac{2}{m_{0}^{2}}\sum_{n\neq c}\frac{\mathbf{p}_{c,n}\left(\mathbf{k}_{0}\right)\otimes\mathbf{p}_{c,n}^{\dagger}\left(\mathbf{k}_{0}\right)}{E_{c,\mathbf{k}_{0}}-E_{n,\mathbf{k}_{0}}}.\label{eq: effective mass}
\end{equation}
Here we have used the condition $\mathbf{p}_{c,c}\left(\mathbf{k}_{0}\right)=-\hbar\mathbf{k}_{0}$,
which ensures a minimum of the conduction band at the position of
the valley wave vector $\mathbf{k}_{0}$.

\subsection{Local Approximation}

If the momentum space envelopes are sufficiently localized near the
valley states \cite{Burt1992,Foreman1996}, the low-pass filtered
delta function \eqref{eq: low-pass Delta} can be replaced by a Dirac
delta leading to a local approximation of the envelope equation
\begin{align*}
E & f_{c,\mathbf{k}_{0}}\left(\mathbf{r}\right)\approx\\
\approx & -\frac{\hbar^{2}}{2}\nabla\cdot\left(m_{c,\mathbf{k}_{0}}^{-1}\nabla f_{c,\mathbf{k}_{0}}\left(\mathbf{r}\right)\right)+\left(E_{c,\mathbf{k}_{0}}+U\left(\mathbf{r}\right)\right)f_{c,\mathbf{k}_{0}}\left(\mathbf{r}\right)\\
 & +\sum_{\mathbf{k}_{0}'\neq\mathbf{k}_{0}}\mathrm{e}^{-i\left(\mathbf{k}_{0}-\mathbf{k}_{0}'\right)\mathbf{r}}u_{\mathbf{k}_{0}}^{*}\left(\mathbf{r}\right)u_{\mathbf{k}_{0}'}\left(\mathbf{r}\right)U\left(\mathbf{r}\right)f_{c,\mathbf{k}_{0}'}\left(\mathbf{r}\right).
\end{align*}
Restriction to the two low-energy valleys $\mathbf{k}_{0}=\pm k_{0}\mathbf{e}_{z}$
in a biaxially strained [001] QW and renaming $f_{c,\pm\mathbf{k}_{0}}\left(\mathbf{r}\right)\to\Psi_{\pm}\left(\mathbf{r}\right)$
yields the coupled envelope model~\eqref{eq: coupled envelope equation model}.

\section{Empirical Pseudopotential Model }\label{sec: EPM}

The lattice-periodic part of the problem is modeled using an empirical
pseudopotential model 
\begin{align*}
&\frac{\hbar^{2}}{2m_{0}}\left\Vert \mathbf{G}+\mathbf{k}\right\Vert ^{2}c_{n,\mathbf{k}}\left(\mathbf{G}\right)+\\
 &\hphantom{=}+\sum_{\mathbf{G}'}V_{\mathrm{ps}}\left(\mathbf{G}+\mathbf{k},\mathbf{G}'+\mathbf{k}\right)c_{n,\mathbf{k}}\left(\mathbf{G}'\right)=E_{n,\mathbf{k}}c_{n,\mathbf{k}}\left(\mathbf{G}\right).
\end{align*}
The pseudopotential has a local and a nonlocal component 
\[
V_{\mathrm{ps}}\left(\mathbf{K},\mathbf{K}'\right)=V_{\mathrm{loc}}\left(\mathbf{K}-\mathbf{K}'\right)+V_{\mathrm{nloc}}\left(\mathbf{K},\mathbf{K}'\right),
\]
where $\mathbf{K}=\mathbf{G}+\mathbf{k}$ and $\mathbf{K}'=\mathbf{G}+\mathbf{k}'$.
The local pseudopotential reads
\begin{align*}
V_{\mathrm{loc}}\left(\mathbf{G}\right) & =S\left(\mathbf{G}\right)V_{\mathrm{atom}}\left(\left\Vert \mathbf{G}\right\Vert \right),
\end{align*}
where $V_{\mathrm{atom}}\left(\left\Vert \mathbf{G}\right\Vert \right)$
is the radially symmetric atomic pseudopotential for silicon and $\mathbf{G}$
is a reciprocal lattice vector of the fcc lattice. The diamond
crystal structure is described via the structure factor
\begin{align*}
S\left(\mathbf{G}\right) & =\cos\left(\frac{\mathbf{G}\cdot\boldsymbol{\mathbf{\tau}}}{2}\right),
\end{align*}
where $\boldsymbol{\tau}=a_{0}/4\times\left(1,1,1\right)^{T}$ is
the offset between the two fcc sub-lattices. The nonlocal
pseudopotential reads
\[
V_{\mathrm{nloc}}\left(\mathbf{K},\mathbf{K}'\right)=\frac{4\pi R_{0}^{3}}{\Omega_{a}}A_{0}S\left(\mathbf{K}-\mathbf{K}'\right)f_{0}\left(R_{0}K,R_{0}K'\right),
\]
which describes a spherically symmetric $s$-well with depth $A_{0}$
and radius $R_{0}$. For a square well we use \cite{Fischetti1991,Hamaguchi2010}
\begin{align*}
f_{0}\left(x,x'\right) & =\frac{xj_{1}\left(x\right)j_{0}\left(x'\right)-x'j_{1}\left(x'\right)j_{0}\left(x\right)}{\left(x+x'\right)\left(x-x'\right)},
\end{align*}
where $j_{\nu}\left(x\right)$ are the spherical Bessel functions.
Following Ref.~\cite{Rieger1993}, spin-orbit interaction is ignored
for silicon.

\begin{table}[t]
\begin{tabular*}{1\columnwidth}
{@{\extracolsep{\fill}}c@{\extracolsep{\fill}}lc}
\toprule 
\textbf{symbol} & \textbf{description} & \textbf{value}\tabularnewline
\midrule 
$E_{\mathrm{cutoff}}$ & cutoff energy & $12.0\,\mathrm{Ry}$\tabularnewline
$V_{\mathrm{atom}}\left(\sqrt{3}\right)$ &
\multirow{3}{3.5cm}{atomic pseudopotential\\
shape parameters\\\ldots
}
& $-0.2241\;\mathrm{Ry}$ \cite{Rieger1993}\tabularnewline
$V_{\mathrm{atom}}\left(\sqrt{8}\right)$ &  & $\phantom{+}0.0520\;\mathrm{Ry}$ \cite{Rieger1993}\tabularnewline
$V_{\mathrm{atom}}\left(\sqrt{11}\right)$ &  & $\phantom{+}0.0724\;\mathrm{Ry}$ \cite{Rieger1993}\tabularnewline
$A_{0}$ & s-well depth & $0.03\;\mathrm{Ry}$ \cite{Rieger1993}\tabularnewline
$R_{0}$ & s-well radius & $1.06\,\mathring{\mathrm{A}}$ \cite{Rieger1993}\tabularnewline
$k_{F}$ & Fermi wave number & $1.66\times2\pi/a_{0}$ \cite{Ungersboeck2007b}\tabularnewline
$\zeta$ & internal strain parameter & $0.53$ \cite{Rieger1993}\tabularnewline
$a_{0}$ & lattice constant & $0.543\,\mathrm{nm}$\tabularnewline
\bottomrule
\end{tabular*}\caption{Parameters of the nonlocal empirical pseudopotential model for silicon.}
\label{tab:parameter-pseudopotential}
\end{table}

\subsubsection{Strain}

Strain leads to a deformation of the crystal and thereby to a modification
of the primitive lattice vectors
\begin{align*}
\mathbf{a}_{i}' & =\left(I+\varepsilon\right)\mathbf{a}_{i},
\end{align*}
where the relaxed basis vectors of the fcc lattice are
\begin{align*}
\mathbf{a}_{1} & =\frac{a_{0}}{2}\left(\begin{array}{c}
0\\
1\\
1
\end{array}\right), & \mathbf{a}_{2} & =\frac{a_{0}}{2}\left(\begin{array}{c}
1\\
0\\
1
\end{array}\right), & \mathbf{a}_{3} & =\frac{a_{0}}{2}\left(\begin{array}{c}
1\\
1\\
0
\end{array}\right).
\end{align*}
As a consequence, the basis vectors of the reciprocal lattice are
modified according to
\[
\mathbf{b}_{i}'=\frac{\pi}{\Omega_{p}'}\sum_{j,k}\epsilon_{i,j,k}\left(\mathbf{a}_{j}'\times\mathbf{a}_{k}'\right)
\]
where $\Omega_{p}'=\mathbf{a}_{1}'\cdot\left(\mathbf{a}_{2}'\times\mathbf{a}_{3}'\right)$
is the volume of the strained primitive unit cell and $\epsilon_{i,j,k}$
is the totally antisymmetric Levi--Civita tensor. In the small strain
limit, the modified reciprocal lattice vectors are well approximated
by $\mathbf{b}_{i}'\approx\left(I-\varepsilon\right)\mathbf{b}_{i}$.
The reciprocal lattice vectors of the strained crystal are integer
multiples of the basis vectors $\mathbf{G}'=\sum_{i=1}^{3}n_{i}\mathbf{b}_{i}'$.
The set of reciprocal lattice vectors is truncated at $\left\Vert \mathbf{G}\right\Vert ^{2}\leq2m_{0}E_{\mathrm{cutoff}}/\hbar^{2}$.
In the main text, the prime to indicate strain is omitted.

\subsubsection{Interpolation of the Atomic Pseudopotential}

Because of the strain-induced modification of the reciprocal lattice
vectors, the atomic pseudopotential $V_{\mathrm{atom}}\left(G\right)$
is required at all possible values of $G$. We follow the usual approach
\cite{Rieger1993,Ungersboeck2007b,Sant2013} and employ a cubic spline
interpolation using the form factors at $G=\left\{ \text{\ensuremath{\sqrt{3}}},\text{\ensuremath{\sqrt{8}}},\text{\ensuremath{\sqrt{11}}}\right\} \,2\pi/a_{0}$
and boundary conditions $V_{\mathrm{atom}}\left(0\right)=-\frac{2}{3}E_{F}$,
$V_{\mathrm{atom}}'\left(0\right)=0$, $V_{\mathrm{atom}}\left(3k_{F}\right)=0$
and $V_{\mathrm{atom}}'\left(3k_{F}\right)=0$. 
The Fermi energy is
taken as $E_{F}=\hbar^{2}k_{F}^{2}/\left(2m_{0}\right).$ Parameter
values are listed in Tab.~\ref{tab:parameter-pseudopotential}.

\subsubsection{Internal Ionic Displacement}

In addition to the deformation of the unit cell, we consider an additional
displacement of the atoms arising from bond-length distortion and
bond-bending \cite{Ungersboeck2007b,Sant2013}. Bond-length distortion
results from the (macroscopic) deformation of the regular tetrahedral
structure, where ions tend to shift towards the barycenter of the
distorted tetrahedron
\begin{align*}
\boldsymbol{\tau}_{0} & =\left(I+\varepsilon\right)\boldsymbol{\tau}.
\end{align*}
This ionic displacement is accompanied by bond-bending and electrostatic
repulsion from nearest neighbors which tend to restore its original
position at the circumcenter of the tetrahedron. For small strain
this is
\begin{align*}
\boldsymbol{\tau}_{1} & \approx\left(I+\varepsilon\right)\boldsymbol{\tau}-\frac{a_{0}}{2}\left(\begin{array}{c}
\varepsilon_{y,z}\\
\varepsilon_{z,x}\\
\varepsilon_{x,y}
\end{array}\right),
\end{align*}
which deviates from the barycenter in the presence of shear strain.
As a consequence, the ions take a position between the circumcenter
$\boldsymbol{\tau}_{0}$ and the barycenter $\boldsymbol{\tau}_{1}$
of the distorted tetrahedron. The ionic displacement is modeled phenomenologically
as
\begin{align*}
\boldsymbol{\tau}' & =\left(1-\zeta\right)\boldsymbol{\tau}_{0}+\zeta\boldsymbol{\tau}_{1}\approx\left(I+\varepsilon\right)\boldsymbol{\tau}-\frac{a_{0}}{2}\left(\begin{array}{c}
\varepsilon_{y,z}\\
\varepsilon_{z,x}\\
\varepsilon_{x,y}
\end{array}\right)\zeta,
\end{align*}
where $\zeta$ is the internal strain parameter, cf.~Fig.~\ref{fig: valley splitting}\,(e).

\section{Evaluation of Valley Splitting Components and Bloch Factor Coefficients
}\label{sec: eval of valley splitting and bandstuct coeff}

In the small strain limit, the reciprocal lattice vectors are written
as $\mathbf{G}_{n}\simeq\sum_{j=1}^{3}n_{j}\left(I-\varepsilon\right)\mathbf{b}_{j}$,
where the $\mathbf{b}_{j}$ are the primitive reciprocal lattice vectors
of the relaxed fcc lattice \cite{Yu2010} 
\begin{equation*}
\mathbf{b}_{1}=\frac{2\pi}{a_{0}}\left(\begin{array}{c}
-1\\
\phantom{+}1\\
\phantom{+}1
\end{array}\right),\,
\mathbf{b}_{2}=\frac{2\pi}{a_{0}}\left(\begin{array}{c}
\phantom{+}1\\
-1\\
\phantom{+}1
\end{array}\right),\,
\mathbf{b}_{3}=\frac{2\pi}{a_{0}}\left(\begin{array}{c}
\phantom{+}1\\
\phantom{+}1\\
-1
\end{array}\right).
\end{equation*}
Therefore, the difference between two (strained) reciprocal lattice
vectors reads 
\[
\mathbf{G}_{n}-\mathbf{G}_{n'}=\sum_{j=1}^{3}\Delta n_{j}\mathbf{b}_{j}^{\varepsilon}
\]
where $\Delta n_{j}=n_{j}-n_{j}'\in\mathbb{Z}$ is an integer and
$\mathbf{b}_{j}^{\varepsilon}=\left(I-\varepsilon\right)\mathbf{b}_{j}$
is a compact notation for the strained primitive reciprocal lattice
vector.

We will now evaluate the contribution to the valley splitting from
Eq.~\eqref{eq: Delta det}. The summation in Eq.~\eqref{eq: Delta det}
can be written as
\begin{align*}
\Delta_{\mathrm{det}} & =\sum_{\Delta n_{1}}\sum_{\Delta n_{2}}\sum_{\Delta n_{3}}\sum_{\mathbf{G},\mathbf{G}'}c_{+}^{*}\left(\mathbf{G}\right)c_{-}\left(\mathbf{G}'\right)\delta_{\mathbf{G}-\mathbf{G}',\sum_{j}\Delta n_{j}\mathbf{b}_{j}^{\varepsilon}}\\
 & \phantom{=}\times\mathrm{e}^{-\left(\frac{1}{2}l_{x}\sum_{j}\Delta n_{j}\mathbf{e}_{x}^{T}\mathbf{b}_{j}^{\varepsilon}\right)^{2}}\mathrm{e}^{-\left(\frac{1}{2}l_{y}\sum_{j}\Delta n_{j}\mathbf{e}_{y}^{T}\mathbf{b}_{j}^{\varepsilon}\right)^{2}}\\
 & \phantom{=}\times\int\mathrm{d}z\,\mathrm{e}^{-i\left(\sum_{j}\Delta n_{j}\mathbf{e}_{z}^{T}\mathbf{b}_{j}^{\varepsilon}+2k_{0}\right)z}\,\bigg[U_{\mathrm{QW}}\left(z\right)+U_{F}\left(z\right)\\
 & \phantom{=}+\frac{\hbar\omega_{x}}{2}\left(\frac{1}{2}-\left(\frac{l_{x}}{2}\sum_{j}\Delta n_{j}\mathbf{e}_{x}^{T}\mathbf{b}_{j}^{\varepsilon}\right)^{2}\right)\\
 & \phantom{=}+\frac{\hbar\omega_{y}}{2}\left(\frac{1}{2}-\left(\frac{l_{y}}{2}\sum_{j}\Delta n_{j}\mathbf{e}_{x}^{T}\mathbf{b}_{j}^{\varepsilon}\right)^{2}\right)\bigg]\left|\psi_{0}\left(z\right)\right|^{2},
\end{align*}
where we have inserted a three-dimensional Kronecker delta and summation
with $\Delta n_{i=1,2,3}$ running over all integers. As the in-plane
extension of the QD is much larger than the lattice constant $l_{x},l_{y}\gg a_{0}$,
the Gaussian terms (second line) effectively reduce to simple selection
rules encoded by Kronecker delta symbols:
\begin{align*}
\mathrm{e}^{-\left(\frac{1}{2}l_{x}\sum_{j}\Delta n_{j}\mathbf{e}_{x}^{T}\mathbf{b}_{j}^{\varepsilon}\right)^{2}} & \approx\delta_{-\Delta n_{1}+\Delta n_{2}+\Delta n_{3},0}\times\\
 & \phantom{\approx}\times\mathrm{e}^{-\left(\frac{1}{2}l_{x}\sum_{j}\Delta n_{j}\mathbf{e}_{x}^{T}\varepsilon\mathbf{b}_{j}\right)^{2}},\\
\mathrm{e}^{-\left(\frac{1}{2}l_{y}\sum_{j}\Delta n_{j}\mathbf{e}_{y}^{T}\mathbf{b}_{j}^{\varepsilon}\right)^{2}} & \approx\delta_{+\Delta n_{1}-\Delta n_{2}+\Delta n_{3},0}\times\\
 & \phantom{\approx}\times\mathrm{e}^{-\left(\frac{1}{2}l_{y}\sum_{j}\Delta n_{j}\mathbf{e}_{y}^{T}\varepsilon\mathbf{b}_{j}\right)^{2}}.
\end{align*}
 These selection rules greatly simplify the evaluation of the summation
in $\Delta_{\mathrm{det}}$, which yields the compact result 
\begin{equation}
\Delta_{\mathrm{det}}=\sum_{n\in\mathbb{Z}}C_{n}^{\left(2\right)}J_{n}^{\mathrm{det}}.\label{eq: Delta det longitudinal compact}
\end{equation}
Here we have introduced the family of integrals 
\begin{align}
J_{n}^{\mathrm{det}} & =\mathrm{e}^{-\left(\frac{1}{2}nG_{0,x}l_{x}\right)^{2}}\mathrm{e}^{-\left(\frac{1}{2}nG_{0,y}l_{y}\right)^{2}}\label{eq: integral J det}\\
 & \phantom{=}\times\int\mathrm{d}z\,\mathrm{e}^{-i\left(nG_{0,z}+2k_{0}\right)z}\,\bigg[U_{\mathrm{QW}}\left(z\right)+U_{F}\left(z\right)\nonumber \\
 & \phantom{\times=}+\frac{\hbar\omega_{x}}{2}\left(\frac{1}{2}-\left(\frac{nG_{0,x}l_{x}}{2}\right)^{2}\right)\nonumber \\
 & \phantom{\times=}+\frac{\hbar\omega_{y}}{2}\left(\frac{1}{2}-\left(\frac{nG_{0,y}l_{y}}{2}\right)^{2}\right)\bigg]\left|\psi_{0}\left(z\right)\right|^{2},\nonumber 
\end{align}
the coefficients 
\begin{equation}
C_{n}^{\left(2\right)}=\sum_{\mathbf{G},\mathbf{G}'}c_{+}^{*}\left(\mathbf{G}\right)c_{-}\left(\mathbf{G}'\right)\delta_{\mathbf{G}-\mathbf{G}',n\mathbf{G}_{0}}\label{eq: bandstructure coefficient C2}
\end{equation}
and the vector 
\begin{align}
\mathbf{G}_{0} & =\left(I-\varepsilon\right)\left(\mathbf{b}_{1}+\mathbf{b}_{2}\right)=\frac{4\pi}{a_{0}}\left(\begin{array}{c}
-\varepsilon_{z,x}\\
-\varepsilon_{y,z}\\
1-\varepsilon_{z,z}
\end{array}\right).\label{eq: G0 vector}
\end{align}
The vector $\mathbf{G}_{0}$ describes the separation of the two low-energy
$X$ points in the strained fcc lattice and determines both the resonance
conditions in Eq.~\eqref{eq: integral J det} as well as the selection
rule in Eq.~\eqref{eq: bandstructure coefficient C2}. We note that
shear strain components $\varepsilon_{z,x}$ and $\varepsilon_{y,z}$
lead to a Gaussian damping in Eq.~\eqref{eq: integral J det} and
are therefore expected to reduce the magnitude of the valley splitting.
The compact form of Eq.~\eqref{eq: Delta det longitudinal compact}
allows for an efficient numerical evaluation, where the double-summation
in Eq.~\eqref{eq: Delta det} was effectively replaced by a single
summation over $J_{n}^{\mathrm{det}}$ (with precomputed $C_{n}^{\left(2\right)}$)
for a few integers only.

Along the same lines, we obtain for Eq.~\eqref{eq: Delta rand abs}
\begin{align}
\langle\left|\Delta_{\mathrm{rand}}\right|^{2}\rangle & =\sum_{n\in\mathbb{Z}}C_{n}^{\left(4\right)}J_{n}^{\mathrm{rand}}\label{eq: Delta random abs compact}
\end{align}
with the integral 
\begin{align}
J_{n}^{\mathrm{rand}} & =\mathrm{e}^{-\frac{1}{8}\left(nG_{0,x}l_{x}\right)^{2}}\mathrm{e}^{-\frac{1}{8}\left(nG_{0,y}l_{y}\right)^{2}}\;\frac{\left(\Delta E_{c}\right)^{2}\Omega_{a}}{2\pi l_{x}l_{y}}\times\label{eq: integral J rand}\\
 & \phantom{=}\times\int\mathrm{d}z\,\mathrm{e}^{-inG_{0,z}z}X\left(z\right)\left(1-X\left(z\right)\right)\left|\psi_{0}\left(z\right)\right|^{4}\nonumber 
\end{align}
and a second set of coefficients governing the magnitude of the disorder-induced
contribution 
\begin{equation}
C_{n}^{\left(4\right)}=\sum_{m}C_{m-n}^{\left(2\right)*}C_{m}^{\left(2\right)}.\label{eq: coefficients C4}
\end{equation}
 Note that because of the symmetries $J_{-n}^{\mathrm{rand}}=\big(J_{n}^{\mathrm{rand}}\big)^{*}$
and $C_{-n}^{\left(4\right)}=\big(C_{n}^{\left(4\right)}\big)^{*}$,
the summation in Eq.~\eqref{eq: Delta random abs compact} can be
further reduced to 
\[
\langle\left|\Delta_{\mathrm{rand}}\right|^{2}\rangle=C_{0}^{\left(4\right)}J_{0}^{\mathrm{rand}}+2\sum_{n=1}^{\infty}\mathrm{Re}\left(C_{n}^{\left(4\right)}J_{n}^{\mathrm{rand}}\right).
\]

\section{Statistics }\label{sec: statistics}

In this section, we derive the statistical properties of the intervalley
coupling parameter $\Delta$ and the valley splitting $E_{\mathrm{VS}}$
induced by the random alloy disorder.

\subsection{Intervalley Coupling Parameter }\label{sec: statistics - Intervalley-Coupling-Parameter}

We consider the random component of the intervalley coupling parameter
given in Eq.~\eqref{eq: Delta rand} and seek a characterization
of its statistical properties in terms of its characteristic function.
The characteristic function is the Fourier transform of the probability
density function, which is given for a complex-valued variable as
\[
\varphi_{\Delta_{\mathrm{rand}}}\left(s\in\mathbb{C}\right)=\langle\mathrm{e}^{i\mathrm{Re}\left(s^{*}\Delta_{\mathrm{rand}}\right)}\rangle.
\]
Substitution of Eq.~\eqref{eq: Delta rand} and Eq.~\eqref{eq: heterostructure potential}
yields 
\[
\varphi_{\Delta_{\mathrm{rand}}}\left(s\in\mathbb{C}\right)=\langle\mathrm{e}^{i\sum_{j}\mathrm{Re}\left(s^{*}w_{j}\right)\left(N_{j}-X\left(\mathbf{R}_{j}\right)\right)}\rangle,
\]
where we have introduced the shorthand notation 
\begin{align*}
w_{j} & =\Delta E_{c}\Omega_{a}\sum_{\mathbf{G},\mathbf{G}'}c_{+}^{*}\left(\mathbf{G}\right)c_{-}\left(\mathbf{G}'\right)\times\\
 & \hphantom{\Delta E_{c}\Omega_{a}\sum_{\mathbf{G},\mathbf{G}'}}\times\mathrm{e}^{-i\left(\mathbf{G}-\mathbf{G}'+2\mathbf{k}_{0}\right)\cdot\mathbf{R}_{j}}\left|\Psi_{0}\left(\mathbf{R}_{j}\right)\right|^{2}.
\end{align*}
As the local Ge concentration at different lattice sites is statistically
independent, we arrive at 
\begin{align*}
\varphi_{\Delta_{\mathrm{rand}}}\left(s\right) & =\prod_{j}\mathrm{e}^{-i\mathrm{Re}\left(s^{*}w_{j}\right)X\left(\mathbf{R}_{j}\right)}\langle\mathrm{e}^{i\mathrm{Re}\left(s^{*}w_{j}\right)N_{j}}\rangle,
\end{align*}
where the product runs over all atomic positions in the crystal. Using
the characteristic function of a scaled Bernoulli distribution 
\[
\langle\mathrm{e}^{i\mathrm{Re}\left(s^{*}w_{j}\right)N_{j}}\rangle=1-X\left(\mathbf{R}_{j}\right)+X\left(\mathbf{R}_{j}\right)\mathrm{e}^{i\mathrm{Re}\left(s^{*}w_{j}\right)}
\]
we obtain 
\begin{align*}
\varphi_{\Delta_{\mathrm{rand}}}\left(s\right) & =\prod_{j}\mathrm{e}^{-i\mathrm{Re}\left(s^{*}w_{j}\right)X\left(\mathbf{R}_{j}\right)}\times\\
 & \hphantom{=\prod_{j}}\times\left(1-X\left(\mathbf{R}_{j}\right)+X\left(\mathbf{R}_{j}\right)\mathrm{e}^{i\mathrm{Re}\left(s^{*}w_{j}\right)}\right).
\end{align*}
Expansion to second order in $w_{j}$ (which is proportional to the
atomic volume) yields 
\begin{align*}
 & \varphi_{\Delta_{\mathrm{rand}}}\left(s\right)\approx\\
 & \quad\approx\prod_{j}\left(1-\frac{1}{2}X\left(\mathbf{R}_{j}\right)\left(1-X\left(\mathbf{R}_{j}\right)\right)\left(\mathrm{Re}\left(s^{*}w_{j}\right)\right)^{2}\right)\\
 & \quad\approx\mathrm{e}^{-\frac{1}{2}\sum_{j}X\left(\mathbf{R}_{j}\right)\left(1-X\left(\mathbf{R}_{j}\right)\right)\left(\mathrm{Re}\left(s^{*}w_{j}\right)\right)^{2}}\\
 & \quad=\mathrm{e}^{-\frac{1}{4}\left|s\right|^{2}\Gamma-\frac{1}{4}\mathrm{Re}\left(s^{*2}C\right)}.
\end{align*}
This is the characteristic function of a scalar complex-valued normal
distribution with covariance $\Gamma$ and pseudo-covariance $C$.
Hence 
\[
\Delta_{\mathrm{rand}}\sim\mathrm{ComplexNormal}\left(\mu=0,\Gamma,C\right),
\]
where the covariance reads 
\begin{equation}
\Gamma=\sum_{j}X\left(\mathbf{R}_{j}\right)\left(1-X\left(\mathbf{R}_{j}\right)\right)\left|w_{j}\right|^{2}=\langle\left|\Delta_{\mathrm{rand}}\right|^{2}\rangle\label{eq: complex normal - Gamma}
\end{equation}
and the relation parameter is 
\begin{equation}
C=\sum_{j}X\left(\mathbf{R}_{j}\right)\left(1-X\left(\mathbf{R}_{j}\right)\right)w_{j}^{2}=\langle\Delta_{\mathrm{rand}}^{2}\rangle.\label{eq: complex normal - C}
\end{equation}
From the above considerations and 
\begin{align*}
\Gamma & =\langle\left(\mathrm{Re}\left(\Delta_{\mathrm{rand}}\right)\right)^{2}\rangle+\langle\left(\mathrm{Im}\left(\Delta_{\mathrm{rand}}\right)\right)^{2}\rangle,\\
C & =\langle\left(\mathrm{Re}\left(\Delta_{\mathrm{rand}}\right)\right)^{2}\rangle-\langle\left(\mathrm{Im}\left(\Delta_{\mathrm{rand}}\right)\right)^{2}\rangle\\
 & \phantom{=}+2i\langle\mathrm{Re}\left(\Delta_{\mathrm{rand}}\right)\mathrm{Im}\left(\Delta_{\mathrm{rand}}\right)\rangle,
\end{align*}
we conclude that the real and imaginary parts of $\Delta_{\mathrm{rand}}$
are weakly correlated and have a slightly different variance:
\begin{align*}
\langle\left(\mathrm{Re}\left(\Delta_{\mathrm{rand}}\right)\right)^{2}\rangle & =\frac{1}{2}\left(\Gamma+\mathrm{Re}\left(C\right)\right),\\
\langle\left(\mathrm{Im}\left(\Delta_{\mathrm{rand}}\right)\right)^{2}\rangle & =\frac{1}{2}\left(\Gamma-\mathrm{Re}\left(C\right)\right),\\
\langle\mathrm{Re}\left(\Delta_{\mathrm{rand}}\right)\mathrm{Im}\left(\Delta_{\mathrm{rand}}\right)\rangle & =\frac{1}{2}\mathrm{Im}\left(C\right).
\end{align*}
Due to the rapidly oscillating term proportional to $4k_{0}$ in Eq.~\eqref{eq: complex normal - C},
however, the relation parameter $C$ is typically orders of magnitude
smaller than the covariance $\Gamma$, such that $C\approx0$ is a
good approximation. In this limit, the distribution of $\Delta_{\mathrm{rand}}$
becomes circularly symmetric in the complex plane. Including the deterministic contribution $\Delta_{\mathrm{det}}=\left|\Delta_{\mathrm{det}}\right|\exp{\left(i\Theta\right)}$,
we finally arrive at 
\begin{align}
\mathrm{Re}\left(\Delta\right) & \sim\mathrm{Normal}\left(\mu=\left|\Delta_{\mathrm{det}}\right|\cos{\left(\Theta\right)},\sigma^{2}=\frac{1}{2}\Gamma\right),\label{eq: normal dist Re(Delta)}\\
\mathrm{Im}\left(\Delta\right) & \sim\mathrm{Normal}\left(\mu=\left|\Delta_{\mathrm{det}}\right|\sin{\left(\Theta\right)},\sigma^{2}=\frac{1}{2}\Gamma\right),\label{eq: normal dist Im(Delta)}
\end{align}
where $\Theta$ is the deterministic valley phase.

\subsection{Valley Splitting }\label{sec: statistics - valley splitting}

The statistical distribution of the valley splitting $E_{\mathrm{VS}}$
is again obtained via computation of its characteristic function.
We consider 
\[
\varphi_{E_{\mathrm{VS}}}\left(s\right)=\langle\mathrm{e}^{iE_{\mathrm{VS}}s}\rangle=\langle\mathrm{e}^{2i\left|\Delta\right|s}\rangle.
\]
Using the probability density function implied by Eqs.~\eqref{eq: normal dist Re(Delta)}--\eqref{eq: normal dist Im(Delta)}
we obtain in polar coordinates 
\begin{align*}
\varphi_{E_{\mathrm{VS}}}\left(s\right) & =\int_{0}^{\infty}\mathrm{d}R\,\mathrm{e}^{2iRs}\frac{R}{\pi\Gamma}\mathrm{e}^{-\frac{R^{2}+\left|\Delta_{\mathrm{det}}\right|^{2}}{\Gamma}}\times\\
 & \phantom{=\phantom{=}}\times\int_{0}^{2\pi}\mathrm{d}\theta\,\mathrm{e}^{\frac{2R\left|\Delta_{\mathrm{det}}\right|}{\Gamma}\cos{\left(\theta-\Theta\right)}}\\
 & =\int_{0}^{\infty}\mathrm{d}\xi\,\mathrm{e}^{i\xi s}\frac{\xi}{2\Gamma}\mathrm{e}^{-\frac{1}{2}\frac{\xi^{2}+\left(2\left|\Delta_{\mathrm{det}}\right|\right)^{2}}{2\Gamma}}I_{0}\left(\frac{2\left|\Delta_{\mathrm{det}}\right|\xi}{2\Gamma}\right).
\end{align*}
The last line is the Fourier transform of the probability density
function of the Rice distribution. Hence, the statistical distribution
of the valley splitting reads 
\[
E_{\mathrm{VS}}\sim\mathrm{Rice}\left(\nu=2\left|\Delta_{\mathrm{det}}\right|,\sigma^{2}=2\Gamma\right).
\]


\begin{thebibliography}{81}%
\makeatletter
\providecommand \@ifxundefined [1]{%
 \@ifx{#1\undefined}
}%
\providecommand \@ifnum [1]{%
 \ifnum #1\expandafter \@firstoftwo
 \else \expandafter \@secondoftwo
 \fi
}%
\providecommand \@ifx [1]{%
 \ifx #1\expandafter \@firstoftwo
 \else \expandafter \@secondoftwo
 \fi
}%
\providecommand \natexlab [1]{#1}%
\providecommand \enquote  [1]{``#1''}%
\providecommand \bibnamefont  [1]{#1}%
\providecommand \bibfnamefont [1]{#1}%
\providecommand \citenamefont [1]{#1}%
\providecommand \href@noop [0]{\@secondoftwo}%
\providecommand \href [0]{\begingroup \@sanitize@url \@href}%
\providecommand \@href[1]{\@@startlink{#1}\@@href}%
\providecommand \@@href[1]{\endgroup#1\@@endlink}%
\providecommand \@sanitize@url [0]{\catcode `\\12\catcode `\$12\catcode
  `\&12\catcode `\#12\catcode `\^12\catcode `\_12\catcode `\%12\relax}%
\providecommand \@@startlink[1]{}%
\providecommand \@@endlink[0]{}%
\providecommand \url  [0]{\begingroup\@sanitize@url \@url }%
\providecommand \@url [1]{\endgroup\@href {#1}{\urlprefix }}%
\providecommand \urlprefix  [0]{URL }%
\providecommand \Eprint [0]{\href }%
\providecommand \doibase [0]{https://doi.org/}%
\providecommand \selectlanguage [0]{\@gobble}%
\providecommand \bibinfo  [0]{\@secondoftwo}%
\providecommand \bibfield  [0]{\@secondoftwo}%
\providecommand \translation [1]{[#1]}%
\providecommand \BibitemOpen [0]{}%
\providecommand \bibitemStop [0]{}%
\providecommand \bibitemNoStop [0]{.\EOS\space}%
\providecommand \EOS [0]{\spacefactor3000\relax}%
\providecommand \BibitemShut  [1]{\csname bibitem#1\endcsname}%
\let\auto@bib@innerbib\@empty
\bibitem [{\citenamefont {Loss}\ and\ \citenamefont
  {DiVincenzo}(1998)}]{Loss1998}%
  \BibitemOpen
  \bibfield  {author} {\bibinfo {author} {\bibfnamefont {D.}~\bibnamefont
  {Loss}}\ and\ \bibinfo {author} {\bibfnamefont {D.~P.}\ \bibnamefont
  {DiVincenzo}},\ }\href {https://doi.org/10.1103/PhysRevA.57.120} {\bibfield
  {journal} {\bibinfo  {journal} {Phys. Rev. A}\ }\textbf {\bibinfo {volume}
  {57}},\ \bibinfo {pages} {120} (\bibinfo {year} {1998})}\BibitemShut
  {NoStop}%
\bibitem [{\citenamefont {Zwanenburg}\ \emph {et~al.}(2013)\citenamefont
  {Zwanenburg}, \citenamefont {Dzurak}, \citenamefont {Morello}, \citenamefont
  {Simmons}, \citenamefont {Hollenberg}, \citenamefont {Klimeck}, \citenamefont
  {Rogge}, \citenamefont {Coppersmith},\ and\ \citenamefont
  {Eriksson}}]{Zwanenburg2013}%
  \BibitemOpen
  \bibfield  {author} {\bibinfo {author} {\bibfnamefont {F.~A.}\ \bibnamefont
  {Zwanenburg}}, \bibinfo {author} {\bibfnamefont {A.~S.}\ \bibnamefont
  {Dzurak}}, \bibinfo {author} {\bibfnamefont {A.}~\bibnamefont {Morello}},
  \bibinfo {author} {\bibfnamefont {M.~Y.}\ \bibnamefont {Simmons}}, \bibinfo
  {author} {\bibfnamefont {L.~C.~L.}\ \bibnamefont {Hollenberg}}, \bibinfo
  {author} {\bibfnamefont {G.}~\bibnamefont {Klimeck}}, \bibinfo {author}
  {\bibfnamefont {S.}~\bibnamefont {Rogge}}, \bibinfo {author} {\bibfnamefont
  {S.~N.}\ \bibnamefont {Coppersmith}},\ and\ \bibinfo {author} {\bibfnamefont
  {M.~A.}\ \bibnamefont {Eriksson}},\ }\href
  {https://doi.org/10.1103/RevModPhys.85.961} {\bibfield  {journal} {\bibinfo
  {journal} {Rev. Mod. Phys.}\ }\textbf {\bibinfo {volume} {85}},\ \bibinfo
  {pages} {961} (\bibinfo {year} {2013})}\BibitemShut {NoStop}%
\bibitem [{\citenamefont {Burkard}\ \emph {et~al.}(2023)\citenamefont
  {Burkard}, \citenamefont {Ladd}, \citenamefont {Pan}, \citenamefont
  {Nichol},\ and\ \citenamefont {Petta}}]{Burkard2023}%
  \BibitemOpen
  \bibfield  {author} {\bibinfo {author} {\bibfnamefont {G.}~\bibnamefont
  {Burkard}}, \bibinfo {author} {\bibfnamefont {T.~D.}\ \bibnamefont {Ladd}},
  \bibinfo {author} {\bibfnamefont {A.}~\bibnamefont {Pan}}, \bibinfo {author}
  {\bibfnamefont {J.~M.}\ \bibnamefont {Nichol}},\ and\ \bibinfo {author}
  {\bibfnamefont {J.~R.}\ \bibnamefont {Petta}},\ }\href
  {https://doi.org/10.1103/RevModPhys.95.025003} {\bibfield  {journal}
  {\bibinfo  {journal} {Rev. Mod. Phys.}\ }\textbf {\bibinfo {volume} {95}},\
  \bibinfo {pages} {025003} (\bibinfo {year} {2023})}\BibitemShut {NoStop}%
\bibitem [{\citenamefont {Neyens}\ \emph {et~al.}(2024)\citenamefont {Neyens},
  \citenamefont {Zietz}, \citenamefont {Watson}, \citenamefont {Luthi},
  \citenamefont {Nethwewala}, \citenamefont {George}, \citenamefont {Henry},
  \citenamefont {Islam}, \citenamefont {Wagner}, \citenamefont {Borjans},
  \citenamefont {Connors}, \citenamefont {Corrigan}, \citenamefont {Curry},
  \citenamefont {Keith}, \citenamefont {Kotlyar}, \citenamefont {Lampert},
  \citenamefont {M\k{a}dzik}, \citenamefont {Millard}, \citenamefont
  {Mohiyaddin}, \citenamefont {Pellerano}, \citenamefont {Pillarisetty},
  \citenamefont {Ramsey}, \citenamefont {Savytskyy}, \citenamefont {Schaal},
  \citenamefont {Zheng}, \citenamefont {Ziegler}, \citenamefont {Bishop},
  \citenamefont {Bojarski}, \citenamefont {Roberts},\ and\ \citenamefont
  {Clarke}}]{Neyens2024}%
  \BibitemOpen
  \bibfield  {author} {\bibinfo {author} {\bibfnamefont {S.}~\bibnamefont
  {Neyens}}, \bibinfo {author} {\bibfnamefont {O.~K.}\ \bibnamefont {Zietz}},
  \bibinfo {author} {\bibfnamefont {T.~F.}\ \bibnamefont {Watson}}, \bibinfo
  {author} {\bibfnamefont {F.}~\bibnamefont {Luthi}}, \bibinfo {author}
  {\bibfnamefont {A.}~\bibnamefont {Nethwewala}}, \bibinfo {author}
  {\bibfnamefont {H.~C.}\ \bibnamefont {George}}, \bibinfo {author}
  {\bibfnamefont {E.}~\bibnamefont {Henry}}, \bibinfo {author} {\bibfnamefont
  {M.}~\bibnamefont {Islam}}, \bibinfo {author} {\bibfnamefont {A.~J.}\
  \bibnamefont {Wagner}}, \bibinfo {author} {\bibfnamefont {F.}~\bibnamefont
  {Borjans}}, \bibinfo {author} {\bibfnamefont {E.~J.}\ \bibnamefont
  {Connors}}, \bibinfo {author} {\bibfnamefont {J.}~\bibnamefont {Corrigan}},
  \bibinfo {author} {\bibfnamefont {M.~J.}\ \bibnamefont {Curry}}, \bibinfo
  {author} {\bibfnamefont {D.}~\bibnamefont {Keith}}, \bibinfo {author}
  {\bibfnamefont {R.}~\bibnamefont {Kotlyar}}, \bibinfo {author} {\bibfnamefont
  {L.~F.}\ \bibnamefont {Lampert}}, \bibinfo {author} {\bibfnamefont {M.~T.}\
  \bibnamefont {M\k{a}dzik}}, \bibinfo {author} {\bibfnamefont
  {K.}~\bibnamefont {Millard}}, \bibinfo {author} {\bibfnamefont {F.~A.}\
  \bibnamefont {Mohiyaddin}}, \bibinfo {author} {\bibfnamefont
  {S.}~\bibnamefont {Pellerano}}, \bibinfo {author} {\bibfnamefont
  {R.}~\bibnamefont {Pillarisetty}}, \bibinfo {author} {\bibfnamefont
  {M.}~\bibnamefont {Ramsey}}, \bibinfo {author} {\bibfnamefont
  {R.}~\bibnamefont {Savytskyy}}, \bibinfo {author} {\bibfnamefont
  {S.}~\bibnamefont {Schaal}}, \bibinfo {author} {\bibfnamefont
  {G.}~\bibnamefont {Zheng}}, \bibinfo {author} {\bibfnamefont
  {J.}~\bibnamefont {Ziegler}}, \bibinfo {author} {\bibfnamefont {N.~C.}\
  \bibnamefont {Bishop}}, \bibinfo {author} {\bibfnamefont {S.}~\bibnamefont
  {Bojarski}}, \bibinfo {author} {\bibfnamefont {J.}~\bibnamefont {Roberts}},\
  and\ \bibinfo {author} {\bibfnamefont {J.~S.}\ \bibnamefont {Clarke}},\
  }\href {https://doi.org/10.1038/s41586-024-07275-6} {\bibfield  {journal}
  {\bibinfo  {journal} {Nature}\ }\textbf {\bibinfo {volume} {629}},\ \bibinfo
  {pages} {80} (\bibinfo {year} {2024})}\BibitemShut {NoStop}%
\bibitem [{\citenamefont {George}\ \emph {et~al.}(2024)\citenamefont {George},
  \citenamefont {M\k{a}dzik}, \citenamefont {Henry}, \citenamefont {Wagner},
  \citenamefont {Islam}, \citenamefont {Borjans}, \citenamefont {Connors},
  \citenamefont {Corrigan}, \citenamefont {Curry}, \citenamefont {Harper},
  \citenamefont {Keith}, \citenamefont {Lampert}, \citenamefont {Luthi},
  \citenamefont {Mohiyaddin}, \citenamefont {Murcia}, \citenamefont {Nair},
  \citenamefont {Nahm}, \citenamefont {Nethwewala}, \citenamefont {Neyens},
  \citenamefont {Raharjo}, \citenamefont {Rogan}, \citenamefont {Savytskyy},
  \citenamefont {Watson}, \citenamefont {Ziegler}, \citenamefont {Zietz},
  \citenamefont {Pillarisetty}, \citenamefont {Bishop}, \citenamefont
  {Bojarski}, \citenamefont {Roberts},\ and\ \citenamefont
  {Clarke}}]{George2024}%
  \BibitemOpen
  \bibfield  {author} {\bibinfo {author} {\bibfnamefont {H.~C.}\ \bibnamefont
  {George}}, \bibinfo {author} {\bibfnamefont {M.~T.}\ \bibnamefont
  {M\k{a}dzik}}, \bibinfo {author} {\bibfnamefont {E.~M.}\ \bibnamefont
  {Henry}}, \bibinfo {author} {\bibfnamefont {A.~J.}\ \bibnamefont {Wagner}},
  \bibinfo {author} {\bibfnamefont {M.~M.}\ \bibnamefont {Islam}}, \bibinfo
  {author} {\bibfnamefont {F.}~\bibnamefont {Borjans}}, \bibinfo {author}
  {\bibfnamefont {E.~J.}\ \bibnamefont {Connors}}, \bibinfo {author}
  {\bibfnamefont {J.}~\bibnamefont {Corrigan}}, \bibinfo {author}
  {\bibfnamefont {M.}~\bibnamefont {Curry}}, \bibinfo {author} {\bibfnamefont
  {M.~K.}\ \bibnamefont {Harper}}, \bibinfo {author} {\bibfnamefont
  {D.}~\bibnamefont {Keith}}, \bibinfo {author} {\bibfnamefont
  {L.}~\bibnamefont {Lampert}}, \bibinfo {author} {\bibfnamefont
  {F.}~\bibnamefont {Luthi}}, \bibinfo {author} {\bibfnamefont {F.~A.}\
  \bibnamefont {Mohiyaddin}}, \bibinfo {author} {\bibfnamefont
  {S.}~\bibnamefont {Murcia}}, \bibinfo {author} {\bibfnamefont
  {R.}~\bibnamefont {Nair}}, \bibinfo {author} {\bibfnamefont {R.}~\bibnamefont
  {Nahm}}, \bibinfo {author} {\bibfnamefont {A.}~\bibnamefont {Nethwewala}},
  \bibinfo {author} {\bibfnamefont {S.}~\bibnamefont {Neyens}}, \bibinfo
  {author} {\bibfnamefont {R.~D.}\ \bibnamefont {Raharjo}}, \bibinfo {author}
  {\bibfnamefont {C.}~\bibnamefont {Rogan}}, \bibinfo {author} {\bibfnamefont
  {R.}~\bibnamefont {Savytskyy}}, \bibinfo {author} {\bibfnamefont {T.~F.}\
  \bibnamefont {Watson}}, \bibinfo {author} {\bibfnamefont {J.}~\bibnamefont
  {Ziegler}}, \bibinfo {author} {\bibfnamefont {O.~K.}\ \bibnamefont {Zietz}},
  \bibinfo {author} {\bibfnamefont {R.}~\bibnamefont {Pillarisetty}}, \bibinfo
  {author} {\bibfnamefont {N.~C.}\ \bibnamefont {Bishop}}, \bibinfo {author}
  {\bibfnamefont {S.~A.}\ \bibnamefont {Bojarski}}, \bibinfo {author}
  {\bibfnamefont {J.}~\bibnamefont {Roberts}},\ and\ \bibinfo {author}
  {\bibfnamefont {J.~S.}\ \bibnamefont {Clarke}},\ }\href
  {https://doi.org/10.48550/arXiv.2410.16583} {\ ,\ \bibinfo {pages}
  {arXiv.2410.16583} (\bibinfo {year} {2024})}\BibitemShut {NoStop}%
\bibitem [{\citenamefont {Koch}\ \emph {et~al.}(2024)\citenamefont {Koch},
  \citenamefont {Godfrin}, \citenamefont {Adam}, \citenamefont {Ferrero},
  \citenamefont {Schroller}, \citenamefont {Glaeser}, \citenamefont {Kubicek},
  \citenamefont {Li}, \citenamefont {Loo}, \citenamefont {Massar},
  \citenamefont {Simion}, \citenamefont {Wan}, \citenamefont {De~Greve},\ and\
  \citenamefont {Wernsdorfer}}]{Koch2024}%
  \BibitemOpen
  \bibfield  {author} {\bibinfo {author} {\bibfnamefont {T.}~\bibnamefont
  {Koch}}, \bibinfo {author} {\bibfnamefont {C.}~\bibnamefont {Godfrin}},
  \bibinfo {author} {\bibfnamefont {V.}~\bibnamefont {Adam}}, \bibinfo {author}
  {\bibfnamefont {J.}~\bibnamefont {Ferrero}}, \bibinfo {author} {\bibfnamefont
  {D.}~\bibnamefont {Schroller}}, \bibinfo {author} {\bibfnamefont
  {N.}~\bibnamefont {Glaeser}}, \bibinfo {author} {\bibfnamefont
  {S.}~\bibnamefont {Kubicek}}, \bibinfo {author} {\bibfnamefont
  {R.}~\bibnamefont {Li}}, \bibinfo {author} {\bibfnamefont {R.}~\bibnamefont
  {Loo}}, \bibinfo {author} {\bibfnamefont {S.}~\bibnamefont {Massar}},
  \bibinfo {author} {\bibfnamefont {G.}~\bibnamefont {Simion}}, \bibinfo
  {author} {\bibfnamefont {D.}~\bibnamefont {Wan}}, \bibinfo {author}
  {\bibfnamefont {K.}~\bibnamefont {De~Greve}},\ and\ \bibinfo {author}
  {\bibfnamefont {W.}~\bibnamefont {Wernsdorfer}},\ }\href
  {https://doi.org/10.48550/arXiv.2409.12731} {\ ,\ \bibinfo {pages}
  {arXiv.2409.12731} (\bibinfo {year} {2024})}\BibitemShut {NoStop}%
\bibitem [{\citenamefont {Mills}\ \emph {et~al.}(2022)\citenamefont {Mills},
  \citenamefont {Guinn}, \citenamefont {Gullans}, \citenamefont {Sigillito},
  \citenamefont {Feldman}, \citenamefont {Nielsen},\ and\ \citenamefont
  {Petta}}]{Mills2022}%
  \BibitemOpen
  \bibfield  {author} {\bibinfo {author} {\bibfnamefont {A.~R.}\ \bibnamefont
  {Mills}}, \bibinfo {author} {\bibfnamefont {C.~R.}\ \bibnamefont {Guinn}},
  \bibinfo {author} {\bibfnamefont {M.~J.}\ \bibnamefont {Gullans}}, \bibinfo
  {author} {\bibfnamefont {A.~J.}\ \bibnamefont {Sigillito}}, \bibinfo {author}
  {\bibfnamefont {M.~M.}\ \bibnamefont {Feldman}}, \bibinfo {author}
  {\bibfnamefont {E.}~\bibnamefont {Nielsen}},\ and\ \bibinfo {author}
  {\bibfnamefont {J.~R.}\ \bibnamefont {Petta}},\ }\href
  {https://doi.org/10.1126/sciadv.abn5130} {\bibfield  {journal} {\bibinfo
  {journal} {Sci. Adv.}\ }\textbf {\bibinfo {volume} {8}},\ \bibinfo {pages}
  {eabn5130} (\bibinfo {year} {2022})}\BibitemShut {NoStop}%
\bibitem [{\citenamefont {Noiri}\ \emph {et~al.}(2022)\citenamefont {Noiri},
  \citenamefont {Takeda}, \citenamefont {Nakajima}, \citenamefont {Kobayashi},
  \citenamefont {Sammak}, \citenamefont {Scappucci},\ and\ \citenamefont
  {Tarucha}}]{Noiri2022}%
  \BibitemOpen
  \bibfield  {author} {\bibinfo {author} {\bibfnamefont {A.}~\bibnamefont
  {Noiri}}, \bibinfo {author} {\bibfnamefont {K.}~\bibnamefont {Takeda}},
  \bibinfo {author} {\bibfnamefont {T.}~\bibnamefont {Nakajima}}, \bibinfo
  {author} {\bibfnamefont {T.}~\bibnamefont {Kobayashi}}, \bibinfo {author}
  {\bibfnamefont {A.}~\bibnamefont {Sammak}}, \bibinfo {author} {\bibfnamefont
  {G.}~\bibnamefont {Scappucci}},\ and\ \bibinfo {author} {\bibfnamefont
  {S.}~\bibnamefont {Tarucha}},\ }\href
  {https://doi.org/10.1038/s41586-021-04182-y} {\bibfield  {journal} {\bibinfo
  {journal} {Nature}\ }\textbf {\bibinfo {volume} {601}},\ \bibinfo {pages}
  {338} (\bibinfo {year} {2022})}\BibitemShut {NoStop}%
\bibitem [{\citenamefont {Xue}\ \emph {et~al.}(2022)\citenamefont {Xue},
  \citenamefont {Russ}, \citenamefont {Samkharadze}, \citenamefont {Undseth},
  \citenamefont {Sammak}, \citenamefont {Scappucci},\ and\ \citenamefont
  {Vandersypen}}]{Xue2022}%
  \BibitemOpen
  \bibfield  {author} {\bibinfo {author} {\bibfnamefont {X.}~\bibnamefont
  {Xue}}, \bibinfo {author} {\bibfnamefont {M.}~\bibnamefont {Russ}}, \bibinfo
  {author} {\bibfnamefont {N.}~\bibnamefont {Samkharadze}}, \bibinfo {author}
  {\bibfnamefont {B.}~\bibnamefont {Undseth}}, \bibinfo {author} {\bibfnamefont
  {A.}~\bibnamefont {Sammak}}, \bibinfo {author} {\bibfnamefont
  {G.}~\bibnamefont {Scappucci}},\ and\ \bibinfo {author} {\bibfnamefont
  {L.~M.~K.}\ \bibnamefont {Vandersypen}},\ }\href
  {https://doi.org/10.1038/s41586-021-04273-w} {\bibfield  {journal} {\bibinfo
  {journal} {Nature}\ }\textbf {\bibinfo {volume} {601}},\ \bibinfo {pages}
  {343} (\bibinfo {year} {2022})}\BibitemShut {NoStop}%
\bibitem [{\citenamefont {Vandersypen}\ \emph {et~al.}(2017)\citenamefont
  {Vandersypen}, \citenamefont {Bluhm}, \citenamefont {Clarke}, \citenamefont
  {Dzurak}, \citenamefont {Ishihara}, \citenamefont {Morello}, \citenamefont
  {Reilly}, \citenamefont {Schreiber},\ and\ \citenamefont
  {Veldhorst}}]{Vandersypen2017}%
  \BibitemOpen
  \bibfield  {author} {\bibinfo {author} {\bibfnamefont {L.~M.~K.}\
  \bibnamefont {Vandersypen}}, \bibinfo {author} {\bibfnamefont
  {H.}~\bibnamefont {Bluhm}}, \bibinfo {author} {\bibfnamefont {J.~S.}\
  \bibnamefont {Clarke}}, \bibinfo {author} {\bibfnamefont {A.~S.}\
  \bibnamefont {Dzurak}}, \bibinfo {author} {\bibfnamefont {R.}~\bibnamefont
  {Ishihara}}, \bibinfo {author} {\bibfnamefont {A.}~\bibnamefont {Morello}},
  \bibinfo {author} {\bibfnamefont {D.~J.}\ \bibnamefont {Reilly}}, \bibinfo
  {author} {\bibfnamefont {L.~R.}\ \bibnamefont {Schreiber}},\ and\ \bibinfo
  {author} {\bibfnamefont {M.}~\bibnamefont {Veldhorst}},\ }\href
  {https://doi.org/10.1038/s41534-017-0038-y} {\bibfield  {journal} {\bibinfo
  {journal} {npj Quantum Inf.}\ }\textbf {\bibinfo {volume} {3}},\ \bibinfo
  {pages} {34} (\bibinfo {year} {2017})}\BibitemShut {NoStop}%
\bibitem [{\citenamefont {K\"{u}nne}\ \emph {et~al.}(2024)\citenamefont
  {K\"{u}nne}, \citenamefont {Willmes}, \citenamefont {Oberl\"{a}nder},
  \citenamefont {Gorjaew}, \citenamefont {Teske}, \citenamefont {Bhardwaj},
  \citenamefont {Beer}, \citenamefont {Kammerloher}, \citenamefont {Otten},
  \citenamefont {Seidler}, \citenamefont {Xue}, \citenamefont {Schreiber},\
  and\ \citenamefont {Bluhm}}]{Kuenne2024}%
  \BibitemOpen
  \bibfield  {author} {\bibinfo {author} {\bibfnamefont {M.}~\bibnamefont
  {K\"{u}nne}}, \bibinfo {author} {\bibfnamefont {A.}~\bibnamefont {Willmes}},
  \bibinfo {author} {\bibfnamefont {M.}~\bibnamefont {Oberl\"{a}nder}},
  \bibinfo {author} {\bibfnamefont {C.}~\bibnamefont {Gorjaew}}, \bibinfo
  {author} {\bibfnamefont {J.~D.}\ \bibnamefont {Teske}}, \bibinfo {author}
  {\bibfnamefont {H.}~\bibnamefont {Bhardwaj}}, \bibinfo {author}
  {\bibfnamefont {M.}~\bibnamefont {Beer}}, \bibinfo {author} {\bibfnamefont
  {E.}~\bibnamefont {Kammerloher}}, \bibinfo {author} {\bibfnamefont
  {R.}~\bibnamefont {Otten}}, \bibinfo {author} {\bibfnamefont
  {I.}~\bibnamefont {Seidler}}, \bibinfo {author} {\bibfnamefont
  {R.}~\bibnamefont {Xue}}, \bibinfo {author} {\bibfnamefont {L.~R.}\
  \bibnamefont {Schreiber}},\ and\ \bibinfo {author} {\bibfnamefont
  {H.}~\bibnamefont {Bluhm}},\ }\href
  {https://doi.org/10.1038/s41467-024-49182-4} {\bibfield  {journal} {\bibinfo
  {journal} {Nat. Commun.}\ }\textbf {\bibinfo {volume} {15}},\ \bibinfo
  {pages} {4977} (\bibinfo {year} {2024})}\BibitemShut {NoStop}%
\bibitem [{\citenamefont {Seidler}\ \emph {et~al.}(2022)\citenamefont
  {Seidler}, \citenamefont {Struck}, \citenamefont {Xue}, \citenamefont
  {Focke}, \citenamefont {Trellenkamp}, \citenamefont {Bluhm},\ and\
  \citenamefont {Schreiber}}]{Seidler2022}%
  \BibitemOpen
  \bibfield  {author} {\bibinfo {author} {\bibfnamefont {I.}~\bibnamefont
  {Seidler}}, \bibinfo {author} {\bibfnamefont {T.}~\bibnamefont {Struck}},
  \bibinfo {author} {\bibfnamefont {R.}~\bibnamefont {Xue}}, \bibinfo {author}
  {\bibfnamefont {N.}~\bibnamefont {Focke}}, \bibinfo {author} {\bibfnamefont
  {S.}~\bibnamefont {Trellenkamp}}, \bibinfo {author} {\bibfnamefont
  {H.}~\bibnamefont {Bluhm}},\ and\ \bibinfo {author} {\bibfnamefont {L.~R.}\
  \bibnamefont {Schreiber}},\ }\href
  {https://doi.org/10.1038/s41534-022-00615-2} {\bibfield  {journal} {\bibinfo
  {journal} {npj Quantum Inf.}\ }\textbf {\bibinfo {volume} {8}},\ \bibinfo
  {pages} {100} (\bibinfo {year} {2022})}\BibitemShut {NoStop}%
\bibitem [{\citenamefont {Struck}\ \emph {et~al.}(2024)\citenamefont {Struck},
  \citenamefont {Volmer}, \citenamefont {Visser}, \citenamefont {Offermann},
  \citenamefont {Xue}, \citenamefont {Tu}, \citenamefont {Trellenkamp},
  \citenamefont {Cywi{\'{n}}ski}, \citenamefont {Bluhm},\ and\ \citenamefont
  {Schreiber}}]{Struck2024}%
  \BibitemOpen
  \bibfield  {author} {\bibinfo {author} {\bibfnamefont {T.}~\bibnamefont
  {Struck}}, \bibinfo {author} {\bibfnamefont {M.}~\bibnamefont {Volmer}},
  \bibinfo {author} {\bibfnamefont {L.}~\bibnamefont {Visser}}, \bibinfo
  {author} {\bibfnamefont {T.}~\bibnamefont {Offermann}}, \bibinfo {author}
  {\bibfnamefont {R.}~\bibnamefont {Xue}}, \bibinfo {author} {\bibfnamefont
  {J.-S.}\ \bibnamefont {Tu}}, \bibinfo {author} {\bibfnamefont
  {S.}~\bibnamefont {Trellenkamp}}, \bibinfo {author} {\bibfnamefont
  {{\L}.}~\bibnamefont {Cywi{\'{n}}ski}}, \bibinfo {author} {\bibfnamefont
  {H.}~\bibnamefont {Bluhm}},\ and\ \bibinfo {author} {\bibfnamefont {L.~R.}\
  \bibnamefont {Schreiber}},\ }\href
  {https://doi.org/10.1038/s41467-024-45583-7} {\bibfield  {journal} {\bibinfo
  {journal} {Nat. Commun.}\ }\textbf {\bibinfo {volume} {15}},\ \bibinfo
  {pages} {1325} (\bibinfo {year} {2024})}\BibitemShut {NoStop}%
\bibitem [{\citenamefont {Xue}\ \emph {et~al.}(2024)\citenamefont {Xue},
  \citenamefont {Beer}, \citenamefont {Seidler}, \citenamefont {Humpohl},
  \citenamefont {Tu}, \citenamefont {Trellenkamp}, \citenamefont {Struck},
  \citenamefont {Bluhm},\ and\ \citenamefont {Schreiber}}]{Xue2024}%
  \BibitemOpen
  \bibfield  {author} {\bibinfo {author} {\bibfnamefont {R.}~\bibnamefont
  {Xue}}, \bibinfo {author} {\bibfnamefont {M.}~\bibnamefont {Beer}}, \bibinfo
  {author} {\bibfnamefont {I.}~\bibnamefont {Seidler}}, \bibinfo {author}
  {\bibfnamefont {S.}~\bibnamefont {Humpohl}}, \bibinfo {author} {\bibfnamefont
  {J.-S.}\ \bibnamefont {Tu}}, \bibinfo {author} {\bibfnamefont
  {S.}~\bibnamefont {Trellenkamp}}, \bibinfo {author} {\bibfnamefont
  {T.}~\bibnamefont {Struck}}, \bibinfo {author} {\bibfnamefont
  {H.}~\bibnamefont {Bluhm}},\ and\ \bibinfo {author} {\bibfnamefont {L.~R.}\
  \bibnamefont {Schreiber}},\ }\href
  {https://doi.org/10.1038/s41467-024-46519-x} {\bibfield  {journal} {\bibinfo
  {journal} {Nat. Commun.}\ }\textbf {\bibinfo {volume} {15}},\ \bibinfo
  {pages} {2296} (\bibinfo {year} {2024})}\BibitemShut {NoStop}%
\bibitem [{\citenamefont {Friesen}\ \emph {et~al.}(2007)\citenamefont
  {Friesen}, \citenamefont {Chutia}, \citenamefont {Tahan},\ and\ \citenamefont
  {Coppersmith}}]{Friesen2007}%
  \BibitemOpen
  \bibfield  {author} {\bibinfo {author} {\bibfnamefont {M.}~\bibnamefont
  {Friesen}}, \bibinfo {author} {\bibfnamefont {S.}~\bibnamefont {Chutia}},
  \bibinfo {author} {\bibfnamefont {C.}~\bibnamefont {Tahan}},\ and\ \bibinfo
  {author} {\bibfnamefont {S.~N.}\ \bibnamefont {Coppersmith}},\ }\href
  {https://doi.org/10.1103/PhysRevB.75.115318} {\bibfield  {journal} {\bibinfo
  {journal} {Phys. Rev. B}\ }\textbf {\bibinfo {volume} {75}},\ \bibinfo
  {pages} {115318} (\bibinfo {year} {2007})}\BibitemShut {NoStop}%
\bibitem [{\citenamefont {Saraiva}\ \emph {et~al.}(2009)\citenamefont
  {Saraiva}, \citenamefont {Calder\'{o}n}, \citenamefont {Hu}, \citenamefont
  {Das~Sarma},\ and\ \citenamefont {Koiller}}]{Saraiva2009}%
  \BibitemOpen
  \bibfield  {author} {\bibinfo {author} {\bibfnamefont {A.~L.}\ \bibnamefont
  {Saraiva}}, \bibinfo {author} {\bibfnamefont {M.~J.}\ \bibnamefont
  {Calder\'{o}n}}, \bibinfo {author} {\bibfnamefont {X.}~\bibnamefont {Hu}},
  \bibinfo {author} {\bibfnamefont {S.}~\bibnamefont {Das~Sarma}},\ and\
  \bibinfo {author} {\bibfnamefont {B.}~\bibnamefont {Koiller}},\ }\href
  {https://doi.org/10.1103/PhysRevB.80.081305} {\bibfield  {journal} {\bibinfo
  {journal} {Phys. Rev. B}\ }\textbf {\bibinfo {volume} {80}},\ \bibinfo
  {pages} {081305} (\bibinfo {year} {2009})}\BibitemShut {NoStop}%
\bibitem [{\citenamefont {Saraiva}\ \emph {et~al.}(2011)\citenamefont
  {Saraiva}, \citenamefont {Calder\'{o}n}, \citenamefont {Capaz}, \citenamefont
  {Hu}, \citenamefont {Das~Sarma},\ and\ \citenamefont
  {Koiller}}]{Saraiva2011}%
  \BibitemOpen
  \bibfield  {author} {\bibinfo {author} {\bibfnamefont {A.~L.}\ \bibnamefont
  {Saraiva}}, \bibinfo {author} {\bibfnamefont {M.~J.}\ \bibnamefont
  {Calder\'{o}n}}, \bibinfo {author} {\bibfnamefont {R.~B.}\ \bibnamefont
  {Capaz}}, \bibinfo {author} {\bibfnamefont {X.}~\bibnamefont {Hu}}, \bibinfo
  {author} {\bibfnamefont {S.}~\bibnamefont {Das~Sarma}},\ and\ \bibinfo
  {author} {\bibfnamefont {B.}~\bibnamefont {Koiller}},\ }\href
  {https://doi.org/10.1103/PhysRevB.84.155320} {\bibfield  {journal} {\bibinfo
  {journal} {Phys. Rev. B}\ }\textbf {\bibinfo {volume} {84}},\ \bibinfo
  {pages} {155320} (\bibinfo {year} {2011})}\BibitemShut {NoStop}%
\bibitem [{\citenamefont {Paquelet~Wuetz}\ \emph {et~al.}(2022)\citenamefont
  {Paquelet~Wuetz}, \citenamefont {Losert}, \citenamefont {Koelling},
  \citenamefont {Stehouwer}, \citenamefont {Zwerver}, \citenamefont {Philips},
  \citenamefont {M\k{a}dzik}, \citenamefont {Xue}, \citenamefont {Zheng},
  \citenamefont {Lodari}, \citenamefont {Amitonov}, \citenamefont
  {Samkharadze}, \citenamefont {Sammak}, \citenamefont {Vandersypen},
  \citenamefont {Rahman}, \citenamefont {Coppersmith}, \citenamefont
  {Moutanabbir}, \citenamefont {Friesen},\ and\ \citenamefont
  {Scappucci}}]{PaqueletWuetz2022}%
  \BibitemOpen
  \bibfield  {author} {\bibinfo {author} {\bibfnamefont {B.}~\bibnamefont
  {Paquelet~Wuetz}}, \bibinfo {author} {\bibfnamefont {M.~P.}\ \bibnamefont
  {Losert}}, \bibinfo {author} {\bibfnamefont {S.}~\bibnamefont {Koelling}},
  \bibinfo {author} {\bibfnamefont {L.~E.~A.}\ \bibnamefont {Stehouwer}},
  \bibinfo {author} {\bibfnamefont {A.-M.~J.}\ \bibnamefont {Zwerver}},
  \bibinfo {author} {\bibfnamefont {S.~G.~J.}\ \bibnamefont {Philips}},
  \bibinfo {author} {\bibfnamefont {M.~T.}\ \bibnamefont {M\k{a}dzik}},
  \bibinfo {author} {\bibfnamefont {X.}~\bibnamefont {Xue}}, \bibinfo {author}
  {\bibfnamefont {G.}~\bibnamefont {Zheng}}, \bibinfo {author} {\bibfnamefont
  {M.}~\bibnamefont {Lodari}}, \bibinfo {author} {\bibfnamefont {S.~V.}\
  \bibnamefont {Amitonov}}, \bibinfo {author} {\bibfnamefont {N.}~\bibnamefont
  {Samkharadze}}, \bibinfo {author} {\bibfnamefont {A.}~\bibnamefont {Sammak}},
  \bibinfo {author} {\bibfnamefont {L.~M.~K.}\ \bibnamefont {Vandersypen}},
  \bibinfo {author} {\bibfnamefont {R.}~\bibnamefont {Rahman}}, \bibinfo
  {author} {\bibfnamefont {S.~N.}\ \bibnamefont {Coppersmith}}, \bibinfo
  {author} {\bibfnamefont {O.}~\bibnamefont {Moutanabbir}}, \bibinfo {author}
  {\bibfnamefont {M.}~\bibnamefont {Friesen}},\ and\ \bibinfo {author}
  {\bibfnamefont {G.}~\bibnamefont {Scappucci}},\ }\href
  {https://doi.org/10.1038/s41467-022-35458-0} {\bibfield  {journal} {\bibinfo
  {journal} {Nat. Commun.}\ }\textbf {\bibinfo {volume} {13}},\ \bibinfo
  {pages} {7730} (\bibinfo {year} {2022})}\BibitemShut {NoStop}%
\bibitem [{\citenamefont {Losert}\ \emph {et~al.}(2023)\citenamefont {Losert},
  \citenamefont {Eriksson}, \citenamefont {Joynt}, \citenamefont {Rahman},
  \citenamefont {Scappucci}, \citenamefont {Coppersmith},\ and\ \citenamefont
  {Friesen}}]{Losert2023}%
  \BibitemOpen
  \bibfield  {author} {\bibinfo {author} {\bibfnamefont {M.~P.}\ \bibnamefont
  {Losert}}, \bibinfo {author} {\bibfnamefont {M.~A.}\ \bibnamefont
  {Eriksson}}, \bibinfo {author} {\bibfnamefont {R.}~\bibnamefont {Joynt}},
  \bibinfo {author} {\bibfnamefont {R.}~\bibnamefont {Rahman}}, \bibinfo
  {author} {\bibfnamefont {G.}~\bibnamefont {Scappucci}}, \bibinfo {author}
  {\bibfnamefont {S.~N.}\ \bibnamefont {Coppersmith}},\ and\ \bibinfo {author}
  {\bibfnamefont {M.}~\bibnamefont {Friesen}},\ }\href
  {https://doi.org/10.1103/PhysRevB.108.125405} {\bibfield  {journal} {\bibinfo
   {journal} {Phys. Rev. B}\ }\textbf {\bibinfo {volume} {108}},\ \bibinfo
  {pages} {125405} (\bibinfo {year} {2023})}\BibitemShut {NoStop}%
\bibitem [{\citenamefont {Lima}\ and\ \citenamefont
  {Burkard}(2023)}]{Lima2023}%
  \BibitemOpen
  \bibfield  {author} {\bibinfo {author} {\bibfnamefont {J.~R.~F.}\
  \bibnamefont {Lima}}\ and\ \bibinfo {author} {\bibfnamefont {G.}~\bibnamefont
  {Burkard}},\ }\href {https://doi.org/10.1088/2633-4356/acd743} {\bibfield
  {journal} {\bibinfo  {journal} {Mater. Quantum Technol.}\ }\textbf {\bibinfo
  {volume} {3}},\ \bibinfo {pages} {025004} (\bibinfo {year}
  {2023})}\BibitemShut {NoStop}%
\bibitem [{\citenamefont {Borselli}\ \emph {et~al.}(2011)\citenamefont
  {Borselli}, \citenamefont {Ross}, \citenamefont {Kiselev}, \citenamefont
  {Croke}, \citenamefont {Holabird}, \citenamefont {Deelman}, \citenamefont
  {Warren}, \citenamefont {Alvarado-Rodriguez}, \citenamefont {Milosavljevic},
  \citenamefont {Ku}, \citenamefont {Wong}, \citenamefont {Schmitz},
  \citenamefont {Sokolich}, \citenamefont {Gyure},\ and\ \citenamefont
  {Hunter}}]{Borselli2011}%
  \BibitemOpen
  \bibfield  {author} {\bibinfo {author} {\bibfnamefont {M.~G.}\ \bibnamefont
  {Borselli}}, \bibinfo {author} {\bibfnamefont {R.~S.}\ \bibnamefont {Ross}},
  \bibinfo {author} {\bibfnamefont {A.~A.}\ \bibnamefont {Kiselev}}, \bibinfo
  {author} {\bibfnamefont {E.~T.}\ \bibnamefont {Croke}}, \bibinfo {author}
  {\bibfnamefont {K.~S.}\ \bibnamefont {Holabird}}, \bibinfo {author}
  {\bibfnamefont {P.~W.}\ \bibnamefont {Deelman}}, \bibinfo {author}
  {\bibfnamefont {L.~D.}\ \bibnamefont {Warren}}, \bibinfo {author}
  {\bibfnamefont {I.}~\bibnamefont {Alvarado-Rodriguez}}, \bibinfo {author}
  {\bibfnamefont {I.}~\bibnamefont {Milosavljevic}}, \bibinfo {author}
  {\bibfnamefont {F.~C.}\ \bibnamefont {Ku}}, \bibinfo {author} {\bibfnamefont
  {W.~S.}\ \bibnamefont {Wong}}, \bibinfo {author} {\bibfnamefont {A.~E.}\
  \bibnamefont {Schmitz}}, \bibinfo {author} {\bibfnamefont {M.}~\bibnamefont
  {Sokolich}}, \bibinfo {author} {\bibfnamefont {M.~F.}\ \bibnamefont
  {Gyure}},\ and\ \bibinfo {author} {\bibfnamefont {A.~T.}\ \bibnamefont
  {Hunter}},\ }\href {https://doi.org/10.1063/1.3569717} {\bibfield  {journal}
  {\bibinfo  {journal} {Appl. Phys. Lett.}\ }\textbf {\bibinfo {volume} {98}},\
  \bibinfo {pages} {123118} (\bibinfo {year} {2011})}\BibitemShut {NoStop}%
\bibitem [{\citenamefont {Neyens}\ \emph {et~al.}(2018)\citenamefont {Neyens},
  \citenamefont {Foote}, \citenamefont {Thorgrimsson}, \citenamefont {Knapp},
  \citenamefont {McJunkin}, \citenamefont {Vandersypen}, \citenamefont {Amin},
  \citenamefont {Thomas}, \citenamefont {Clarke}, \citenamefont {Savage},
  \citenamefont {Lagally}, \citenamefont {Friesen}, \citenamefont
  {Coppersmith},\ and\ \citenamefont {Eriksson}}]{Neyens2018}%
  \BibitemOpen
  \bibfield  {author} {\bibinfo {author} {\bibfnamefont {S.~F.}\ \bibnamefont
  {Neyens}}, \bibinfo {author} {\bibfnamefont {R.~H.}\ \bibnamefont {Foote}},
  \bibinfo {author} {\bibfnamefont {B.}~\bibnamefont {Thorgrimsson}}, \bibinfo
  {author} {\bibfnamefont {T.~J.}\ \bibnamefont {Knapp}}, \bibinfo {author}
  {\bibfnamefont {T.}~\bibnamefont {McJunkin}}, \bibinfo {author}
  {\bibfnamefont {L.~M.~K.}\ \bibnamefont {Vandersypen}}, \bibinfo {author}
  {\bibfnamefont {P.}~\bibnamefont {Amin}}, \bibinfo {author} {\bibfnamefont
  {N.~K.}\ \bibnamefont {Thomas}}, \bibinfo {author} {\bibfnamefont {J.~S.}\
  \bibnamefont {Clarke}}, \bibinfo {author} {\bibfnamefont {D.~E.}\
  \bibnamefont {Savage}}, \bibinfo {author} {\bibfnamefont {M.~G.}\
  \bibnamefont {Lagally}}, \bibinfo {author} {\bibfnamefont {M.}~\bibnamefont
  {Friesen}}, \bibinfo {author} {\bibfnamefont {S.~N.}\ \bibnamefont
  {Coppersmith}},\ and\ \bibinfo {author} {\bibfnamefont {M.~A.}\ \bibnamefont
  {Eriksson}},\ }\href {https://doi.org/10.1063/1.5033447} {\bibfield
  {journal} {\bibinfo  {journal} {Appl. Phys. Lett.}\ }\textbf {\bibinfo
  {volume} {112}},\ \bibinfo {pages} {243107} (\bibinfo {year}
  {2018})}\BibitemShut {NoStop}%
\bibitem [{\citenamefont {Hollmann}\ \emph {et~al.}(2020)\citenamefont
  {Hollmann}, \citenamefont {Struck}, \citenamefont {Langrock}, \citenamefont
  {Schmidbauer}, \citenamefont {Schauer}, \citenamefont {Leonhardt},
  \citenamefont {Sawano}, \citenamefont {Riemann}, \citenamefont {Abrosimov},
  \citenamefont {Bougeard},\ and\ \citenamefont {Schreiber}}]{Hollmann2020}%
  \BibitemOpen
  \bibfield  {author} {\bibinfo {author} {\bibfnamefont {A.}~\bibnamefont
  {Hollmann}}, \bibinfo {author} {\bibfnamefont {T.}~\bibnamefont {Struck}},
  \bibinfo {author} {\bibfnamefont {V.}~\bibnamefont {Langrock}}, \bibinfo
  {author} {\bibfnamefont {A.}~\bibnamefont {Schmidbauer}}, \bibinfo {author}
  {\bibfnamefont {F.}~\bibnamefont {Schauer}}, \bibinfo {author} {\bibfnamefont
  {T.}~\bibnamefont {Leonhardt}}, \bibinfo {author} {\bibfnamefont
  {K.}~\bibnamefont {Sawano}}, \bibinfo {author} {\bibfnamefont
  {H.}~\bibnamefont {Riemann}}, \bibinfo {author} {\bibfnamefont {N.~V.}\
  \bibnamefont {Abrosimov}}, \bibinfo {author} {\bibfnamefont {D.}~\bibnamefont
  {Bougeard}},\ and\ \bibinfo {author} {\bibfnamefont {L.~R.}\ \bibnamefont
  {Schreiber}},\ }\href {https://doi.org/10.1103/PhysRevApplied.13.034068}
  {\bibfield  {journal} {\bibinfo  {journal} {Phys. Rev. Applied}\ }\textbf
  {\bibinfo {volume} {13}},\ \bibinfo {pages} {034068} (\bibinfo {year}
  {2020})}\BibitemShut {NoStop}%
\bibitem [{\citenamefont {Degli~Esposti}\ \emph {et~al.}(2024)\citenamefont
  {Degli~Esposti}, \citenamefont {Stehouwer}, \citenamefont {G\"{u}l},
  \citenamefont {Samkharadze}, \citenamefont {D\'{e}prez}, \citenamefont
  {Meyer}, \citenamefont {Meijer}, \citenamefont {Tryputen}, \citenamefont
  {Karwal}, \citenamefont {Botifoll}, \citenamefont {Arbiol}, \citenamefont
  {Amitonov}, \citenamefont {Vandersypen}, \citenamefont {Sammak},
  \citenamefont {Veldhorst},\ and\ \citenamefont
  {Scappucci}}]{DegliEsposti2024}%
  \BibitemOpen
  \bibfield  {author} {\bibinfo {author} {\bibfnamefont {D.}~\bibnamefont
  {Degli~Esposti}}, \bibinfo {author} {\bibfnamefont {L.~E.~A.}\ \bibnamefont
  {Stehouwer}}, \bibinfo {author} {\bibfnamefont {O.}~\bibnamefont {G\"{u}l}},
  \bibinfo {author} {\bibfnamefont {N.}~\bibnamefont {Samkharadze}}, \bibinfo
  {author} {\bibfnamefont {C.}~\bibnamefont {D\'{e}prez}}, \bibinfo {author}
  {\bibfnamefont {M.}~\bibnamefont {Meyer}}, \bibinfo {author} {\bibfnamefont
  {I.~N.}\ \bibnamefont {Meijer}}, \bibinfo {author} {\bibfnamefont
  {L.}~\bibnamefont {Tryputen}}, \bibinfo {author} {\bibfnamefont
  {S.}~\bibnamefont {Karwal}}, \bibinfo {author} {\bibfnamefont
  {M.}~\bibnamefont {Botifoll}}, \bibinfo {author} {\bibfnamefont
  {J.}~\bibnamefont {Arbiol}}, \bibinfo {author} {\bibfnamefont {S.~V.}\
  \bibnamefont {Amitonov}}, \bibinfo {author} {\bibfnamefont {L.~M.~K.}\
  \bibnamefont {Vandersypen}}, \bibinfo {author} {\bibfnamefont
  {A.}~\bibnamefont {Sammak}}, \bibinfo {author} {\bibfnamefont
  {M.}~\bibnamefont {Veldhorst}},\ and\ \bibinfo {author} {\bibfnamefont
  {G.}~\bibnamefont {Scappucci}},\ }\href
  {https://doi.org/10.1038/s41534-024-00826-9} {\bibfield  {journal} {\bibinfo
  {journal} {npj Quantum Inf.}\ }\textbf {\bibinfo {volume} {10}},\ \bibinfo
  {pages} {32} (\bibinfo {year} {2024})}\BibitemShut {NoStop}%
\bibitem [{\citenamefont {Yang}\ \emph {et~al.}(2013)\citenamefont {Yang},
  \citenamefont {Rossi}, \citenamefont {Ruskov}, \citenamefont {Lai},
  \citenamefont {Mohiyaddin}, \citenamefont {Lee}, \citenamefont {Tahan},
  \citenamefont {Klimeck}, \citenamefont {Morello},\ and\ \citenamefont
  {Dzurak}}]{Yang2013}%
  \BibitemOpen
  \bibfield  {author} {\bibinfo {author} {\bibfnamefont {C.~H.}\ \bibnamefont
  {Yang}}, \bibinfo {author} {\bibfnamefont {A.}~\bibnamefont {Rossi}},
  \bibinfo {author} {\bibfnamefont {R.}~\bibnamefont {Ruskov}}, \bibinfo
  {author} {\bibfnamefont {N.~S.}\ \bibnamefont {Lai}}, \bibinfo {author}
  {\bibfnamefont {F.~A.}\ \bibnamefont {Mohiyaddin}}, \bibinfo {author}
  {\bibfnamefont {S.}~\bibnamefont {Lee}}, \bibinfo {author} {\bibfnamefont
  {C.}~\bibnamefont {Tahan}}, \bibinfo {author} {\bibfnamefont
  {G.}~\bibnamefont {Klimeck}}, \bibinfo {author} {\bibfnamefont
  {A.}~\bibnamefont {Morello}},\ and\ \bibinfo {author} {\bibfnamefont {A.~S.}\
  \bibnamefont {Dzurak}},\ }\href {https://doi.org/10.1038/ncomms3069}
  {\bibfield  {journal} {\bibinfo  {journal} {Nat. Commun.}\ }\textbf {\bibinfo
  {volume} {4}},\ \bibinfo {pages} {2069} (\bibinfo {year} {2013})}\BibitemShut
  {NoStop}%
\bibitem [{\citenamefont {Borjans}\ \emph {et~al.}(2019)\citenamefont
  {Borjans}, \citenamefont {Zajac}, \citenamefont {Hazard},\ and\ \citenamefont
  {Petta}}]{Borjans2019}%
  \BibitemOpen
  \bibfield  {author} {\bibinfo {author} {\bibfnamefont {F.}~\bibnamefont
  {Borjans}}, \bibinfo {author} {\bibfnamefont {D.}~\bibnamefont {Zajac}},
  \bibinfo {author} {\bibfnamefont {T.}~\bibnamefont {Hazard}},\ and\ \bibinfo
  {author} {\bibfnamefont {J.}~\bibnamefont {Petta}},\ }\href
  {https://doi.org/10.1103/PhysRevApplied.11.044063} {\bibfield  {journal}
  {\bibinfo  {journal} {Phys. Rev. Applied}\ }\textbf {\bibinfo {volume}
  {11}},\ \bibinfo {pages} {044063} (\bibinfo {year} {2019})}\BibitemShut
  {NoStop}%
\bibitem [{\citenamefont {Losert}\ \emph {et~al.}(2024)\citenamefont {Losert},
  \citenamefont {Oberl\"ander}, \citenamefont {Teske}, \citenamefont {Volmer},
  \citenamefont {Schreiber}, \citenamefont {Bluhm}, \citenamefont
  {Coppersmith},\ and\ \citenamefont {Friesen}}]{Losert2024}%
  \BibitemOpen
  \bibfield  {author} {\bibinfo {author} {\bibfnamefont {M.~P.}\ \bibnamefont
  {Losert}}, \bibinfo {author} {\bibfnamefont {M.}~\bibnamefont
  {Oberl\"ander}}, \bibinfo {author} {\bibfnamefont {J.~D.}\ \bibnamefont
  {Teske}}, \bibinfo {author} {\bibfnamefont {M.}~\bibnamefont {Volmer}},
  \bibinfo {author} {\bibfnamefont {L.~R.}\ \bibnamefont {Schreiber}}, \bibinfo
  {author} {\bibfnamefont {H.}~\bibnamefont {Bluhm}}, \bibinfo {author}
  {\bibfnamefont {S.}~\bibnamefont {Coppersmith}},\ and\ \bibinfo {author}
  {\bibfnamefont {M.}~\bibnamefont {Friesen}},\ }\href
  {https://doi.org/10.1103/PRXQuantum.5.040322} {\bibfield  {journal} {\bibinfo
   {journal} {PRX Quantum}\ }\textbf {\bibinfo {volume} {5}},\ \bibinfo {pages}
  {040322} (\bibinfo {year} {2024})}\BibitemShut {NoStop}%
\bibitem [{\citenamefont {David}\ \emph {et~al.}(2024)\citenamefont {David},
  \citenamefont {Pazhedath}, \citenamefont {Schreiber}, \citenamefont
  {Calarco}, \citenamefont {Bluhm},\ and\ \citenamefont {Motzoi}}]{David2024}%
  \BibitemOpen
  \bibfield  {author} {\bibinfo {author} {\bibfnamefont {A.}~\bibnamefont
  {David}}, \bibinfo {author} {\bibfnamefont {A.~M.}\ \bibnamefont
  {Pazhedath}}, \bibinfo {author} {\bibfnamefont {L.~R.}\ \bibnamefont
  {Schreiber}}, \bibinfo {author} {\bibfnamefont {T.}~\bibnamefont {Calarco}},
  \bibinfo {author} {\bibfnamefont {H.}~\bibnamefont {Bluhm}},\ and\ \bibinfo
  {author} {\bibfnamefont {F.}~\bibnamefont {Motzoi}},\ }\href
  {https://doi.org/10.48550/arXiv.2409.07600} {\ ,\ \bibinfo {pages}
  {arXiv:2409.07600} (\bibinfo {year} {2024})}\BibitemShut {NoStop}%
\bibitem [{\citenamefont {Langrock}\ \emph {et~al.}(2023)\citenamefont
  {Langrock}, \citenamefont {Krzywda}, \citenamefont {Focke}, \citenamefont
  {Seidler}, \citenamefont {Schreiber},\ and\ \citenamefont
  {Cywi{\'{n}}ski}}]{Langrock2023}%
  \BibitemOpen
  \bibfield  {author} {\bibinfo {author} {\bibfnamefont {V.}~\bibnamefont
  {Langrock}}, \bibinfo {author} {\bibfnamefont {J.~A.}\ \bibnamefont
  {Krzywda}}, \bibinfo {author} {\bibfnamefont {N.}~\bibnamefont {Focke}},
  \bibinfo {author} {\bibfnamefont {I.}~\bibnamefont {Seidler}}, \bibinfo
  {author} {\bibfnamefont {L.~R.}\ \bibnamefont {Schreiber}},\ and\ \bibinfo
  {author} {\bibfnamefont {{\L}.}~\bibnamefont {Cywi{\'{n}}ski}},\ }\href
  {https://doi.org/10.1103/PRXQuantum.4.020305} {\bibfield  {journal} {\bibinfo
   {journal} {{PRX} Quantum}\ }\textbf {\bibinfo {volume} {4}},\ \bibinfo
  {pages} {020305} (\bibinfo {year} {2023})}\BibitemShut {NoStop}%
\bibitem [{\citenamefont {Volmer}\ \emph {et~al.}(2024)\citenamefont {Volmer},
  \citenamefont {Struck}, \citenamefont {Sala}, \citenamefont {Chen},
  \citenamefont {Oberl\"{a}nder}, \citenamefont {Offermann}, \citenamefont
  {Xue}, \citenamefont {Visser}, \citenamefont {Tu}, \citenamefont
  {Trellenkamp}, \citenamefont {Cywi\'{n}ski}, \citenamefont {Bluhm},\ and\
  \citenamefont {Schreiber}}]{Volmer2024}%
  \BibitemOpen
  \bibfield  {author} {\bibinfo {author} {\bibfnamefont {M.}~\bibnamefont
  {Volmer}}, \bibinfo {author} {\bibfnamefont {T.}~\bibnamefont {Struck}},
  \bibinfo {author} {\bibfnamefont {A.}~\bibnamefont {Sala}}, \bibinfo {author}
  {\bibfnamefont {B.}~\bibnamefont {Chen}}, \bibinfo {author} {\bibfnamefont
  {M.}~\bibnamefont {Oberl\"{a}nder}}, \bibinfo {author} {\bibfnamefont
  {T.}~\bibnamefont {Offermann}}, \bibinfo {author} {\bibfnamefont
  {R.}~\bibnamefont {Xue}}, \bibinfo {author} {\bibfnamefont {L.}~\bibnamefont
  {Visser}}, \bibinfo {author} {\bibfnamefont {J.-S.}\ \bibnamefont {Tu}},
  \bibinfo {author} {\bibfnamefont {S.}~\bibnamefont {Trellenkamp}}, \bibinfo
  {author} {\bibfnamefont {L.}~\bibnamefont {Cywi\'{n}ski}}, \bibinfo {author}
  {\bibfnamefont {H.}~\bibnamefont {Bluhm}},\ and\ \bibinfo {author}
  {\bibfnamefont {L.~R.}\ \bibnamefont {Schreiber}},\ }\href
  {https://doi.org/10.1038/s41534-024-00852-7} {\bibfield  {journal} {\bibinfo
  {journal} {npj Quantum Inf.}\ }\textbf {\bibinfo {volume} {10}},\ \bibinfo
  {pages} {61} (\bibinfo {year} {2024})}\BibitemShut {NoStop}%
\bibitem [{\citenamefont {Feng}\ and\ \citenamefont {Joynt}(2022)}]{Feng2022}%
  \BibitemOpen
  \bibfield  {author} {\bibinfo {author} {\bibfnamefont {Y.}~\bibnamefont
  {Feng}}\ and\ \bibinfo {author} {\bibfnamefont {R.}~\bibnamefont {Joynt}},\
  }\href {https://doi.org/10.1103/PhysRevB.106.085304} {\bibfield  {journal}
  {\bibinfo  {journal} {Phys. Rev. B}\ }\textbf {\bibinfo {volume} {106}},\
  \bibinfo {pages} {085304} (\bibinfo {year} {2022})}\BibitemShut {NoStop}%
\bibitem [{\citenamefont {Klos}\ \emph {et~al.}(2024)\citenamefont {Klos},
  \citenamefont {Tr\"{o}ger}, \citenamefont {Keutgen}, \citenamefont {Losert},
  \citenamefont {Abrosimov}, \citenamefont {Knoch}, \citenamefont {Bracht},
  \citenamefont {Coppersmith}, \citenamefont {Friesen}, \citenamefont
  {Cojocaru-Mir\'{e}din}, \citenamefont {Schreiber},\ and\ \citenamefont
  {Bougeard}}]{Klos2024}%
  \BibitemOpen
  \bibfield  {author} {\bibinfo {author} {\bibfnamefont {J.}~\bibnamefont
  {Klos}}, \bibinfo {author} {\bibfnamefont {J.}~\bibnamefont {Tr\"{o}ger}},
  \bibinfo {author} {\bibfnamefont {J.}~\bibnamefont {Keutgen}}, \bibinfo
  {author} {\bibfnamefont {M.~P.}\ \bibnamefont {Losert}}, \bibinfo {author}
  {\bibfnamefont {N.~V.}\ \bibnamefont {Abrosimov}}, \bibinfo {author}
  {\bibfnamefont {J.}~\bibnamefont {Knoch}}, \bibinfo {author} {\bibfnamefont
  {H.}~\bibnamefont {Bracht}}, \bibinfo {author} {\bibfnamefont {S.~N.}\
  \bibnamefont {Coppersmith}}, \bibinfo {author} {\bibfnamefont
  {M.}~\bibnamefont {Friesen}}, \bibinfo {author} {\bibfnamefont
  {O.}~\bibnamefont {Cojocaru-Mir\'{e}din}}, \bibinfo {author} {\bibfnamefont
  {L.~R.}\ \bibnamefont {Schreiber}},\ and\ \bibinfo {author} {\bibfnamefont
  {D.}~\bibnamefont {Bougeard}},\ }\href
  {https://doi.org/10.1002/advs.202407442} {\bibfield  {journal} {\bibinfo
  {journal} {Adv. Sci.}\ }\textbf {\bibinfo {volume} {11}},\ \bibinfo {pages}
  {2407442} (\bibinfo {year} {2024})}\BibitemShut {NoStop}%
\bibitem [{\citenamefont {Woods}\ \emph {et~al.}(2024)\citenamefont {Woods},
  \citenamefont {Soomro}, \citenamefont {Joseph}, \citenamefont {Frink},
  \citenamefont {Joynt}, \citenamefont {Eriksson},\ and\ \citenamefont
  {Friesen}}]{Woods2024}%
  \BibitemOpen
  \bibfield  {author} {\bibinfo {author} {\bibfnamefont {B.~D.}\ \bibnamefont
  {Woods}}, \bibinfo {author} {\bibfnamefont {H.}~\bibnamefont {Soomro}},
  \bibinfo {author} {\bibfnamefont {E.~S.}\ \bibnamefont {Joseph}}, \bibinfo
  {author} {\bibfnamefont {C.~C.~D.}\ \bibnamefont {Frink}}, \bibinfo {author}
  {\bibfnamefont {R.}~\bibnamefont {Joynt}}, \bibinfo {author} {\bibfnamefont
  {M.~A.}\ \bibnamefont {Eriksson}},\ and\ \bibinfo {author} {\bibfnamefont
  {M.}~\bibnamefont {Friesen}},\ }\href
  {https://doi.org/10.1038/s41534-024-00853-6} {\bibfield  {journal} {\bibinfo
  {journal} {npj Quantum Inf.}\ }\textbf {\bibinfo {volume} {10}},\ \bibinfo
  {pages} {54} (\bibinfo {year} {2024})}\BibitemShut {NoStop}%
\bibitem [{\citenamefont {McJunkin}\ \emph {et~al.}(2022)\citenamefont
  {McJunkin}, \citenamefont {Harpt}, \citenamefont {Feng}, \citenamefont
  {Losert}, \citenamefont {Rahman}, \citenamefont {Dodson}, \citenamefont
  {Wolfe}, \citenamefont {Savage}, \citenamefont {Lagally}, \citenamefont
  {Coppersmith}, \citenamefont {Friesen}, \citenamefont {Joynt},\ and\
  \citenamefont {Eriksson}}]{McJunkin2022}%
  \BibitemOpen
  \bibfield  {author} {\bibinfo {author} {\bibfnamefont {T.}~\bibnamefont
  {McJunkin}}, \bibinfo {author} {\bibfnamefont {B.}~\bibnamefont {Harpt}},
  \bibinfo {author} {\bibfnamefont {Y.}~\bibnamefont {Feng}}, \bibinfo {author}
  {\bibfnamefont {M.~P.}\ \bibnamefont {Losert}}, \bibinfo {author}
  {\bibfnamefont {R.}~\bibnamefont {Rahman}}, \bibinfo {author} {\bibfnamefont
  {J.~P.}\ \bibnamefont {Dodson}}, \bibinfo {author} {\bibfnamefont {M.~A.}\
  \bibnamefont {Wolfe}}, \bibinfo {author} {\bibfnamefont {D.~E.}\ \bibnamefont
  {Savage}}, \bibinfo {author} {\bibfnamefont {M.~G.}\ \bibnamefont {Lagally}},
  \bibinfo {author} {\bibfnamefont {S.~N.}\ \bibnamefont {Coppersmith}},
  \bibinfo {author} {\bibfnamefont {M.}~\bibnamefont {Friesen}}, \bibinfo
  {author} {\bibfnamefont {R.}~\bibnamefont {Joynt}},\ and\ \bibinfo {author}
  {\bibfnamefont {M.~A.}\ \bibnamefont {Eriksson}},\ }\href
  {https://doi.org/10.1038/s41467-022-35510-z} {\bibfield  {journal} {\bibinfo
  {journal} {Nat. Commun.}\ }\textbf {\bibinfo {volume} {13}},\ \bibinfo
  {pages} {7777} (\bibinfo {year} {2022})}\BibitemShut {NoStop}%
\bibitem [{\citenamefont {McJunkin}\ \emph {et~al.}(2021)\citenamefont
  {McJunkin}, \citenamefont {MacQuarrie}, \citenamefont {Tom}, \citenamefont
  {Neyens}, \citenamefont {Dodson}, \citenamefont {Thorgrimsson}, \citenamefont
  {Corrigan}, \citenamefont {Ercan}, \citenamefont {Savage}, \citenamefont
  {Lagally}, \citenamefont {Joynt}, \citenamefont {Coppersmith}, \citenamefont
  {Friesen},\ and\ \citenamefont {Eriksson}}]{McJunkin2021}%
  \BibitemOpen
  \bibfield  {author} {\bibinfo {author} {\bibfnamefont {T.}~\bibnamefont
  {McJunkin}}, \bibinfo {author} {\bibfnamefont {E.~R.}\ \bibnamefont
  {MacQuarrie}}, \bibinfo {author} {\bibfnamefont {L.}~\bibnamefont {Tom}},
  \bibinfo {author} {\bibfnamefont {S.~F.}\ \bibnamefont {Neyens}}, \bibinfo
  {author} {\bibfnamefont {J.~P.}\ \bibnamefont {Dodson}}, \bibinfo {author}
  {\bibfnamefont {B.}~\bibnamefont {Thorgrimsson}}, \bibinfo {author}
  {\bibfnamefont {J.}~\bibnamefont {Corrigan}}, \bibinfo {author}
  {\bibfnamefont {H.~E.}\ \bibnamefont {Ercan}}, \bibinfo {author}
  {\bibfnamefont {D.~E.}\ \bibnamefont {Savage}}, \bibinfo {author}
  {\bibfnamefont {M.~G.}\ \bibnamefont {Lagally}}, \bibinfo {author}
  {\bibfnamefont {R.}~\bibnamefont {Joynt}}, \bibinfo {author} {\bibfnamefont
  {S.~N.}\ \bibnamefont {Coppersmith}}, \bibinfo {author} {\bibfnamefont
  {M.}~\bibnamefont {Friesen}},\ and\ \bibinfo {author} {\bibfnamefont {M.~A.}\
  \bibnamefont {Eriksson}},\ }\href
  {https://doi.org/10.1103/PhysRevB.104.085406} {\bibfield  {journal} {\bibinfo
   {journal} {Phys. Rev. B}\ }\textbf {\bibinfo {volume} {104}},\ \bibinfo
  {pages} {085406} (\bibinfo {year} {2021})}\BibitemShut {NoStop}%
\bibitem [{\citenamefont {Gradwohl}\ \emph {et~al.}(2025)\citenamefont
  {Gradwohl}, \citenamefont {Cvitkovich}, \citenamefont {Lu}, \citenamefont
  {Koelling}, \citenamefont {Oezkent}, \citenamefont {Liu}, \citenamefont
  {Waldh\"{o}r}, \citenamefont {Grasser}, \citenamefont {Niquet}, \citenamefont
  {Albrecht}, \citenamefont {Richter}, \citenamefont {Moutanabbir},\ and\
  \citenamefont {Martin}}]{Gradwohl2025}%
  \BibitemOpen
  \bibfield  {author} {\bibinfo {author} {\bibfnamefont {K.-P.}\ \bibnamefont
  {Gradwohl}}, \bibinfo {author} {\bibfnamefont {L.}~\bibnamefont
  {Cvitkovich}}, \bibinfo {author} {\bibfnamefont {C.-H.}\ \bibnamefont {Lu}},
  \bibinfo {author} {\bibfnamefont {S.}~\bibnamefont {Koelling}}, \bibinfo
  {author} {\bibfnamefont {M.}~\bibnamefont {Oezkent}}, \bibinfo {author}
  {\bibfnamefont {Y.}~\bibnamefont {Liu}}, \bibinfo {author} {\bibfnamefont
  {D.}~\bibnamefont {Waldh\"{o}r}}, \bibinfo {author} {\bibfnamefont
  {T.}~\bibnamefont {Grasser}}, \bibinfo {author} {\bibfnamefont {Y.-M.}\
  \bibnamefont {Niquet}}, \bibinfo {author} {\bibfnamefont {M.}~\bibnamefont
  {Albrecht}}, \bibinfo {author} {\bibfnamefont {C.}~\bibnamefont {Richter}},
  \bibinfo {author} {\bibfnamefont {O.}~\bibnamefont {Moutanabbir}},\ and\
  \bibinfo {author} {\bibfnamefont {J.}~\bibnamefont {Martin}},\ }\href
  {https://doi.org/10.1021/acs.nanolett.4c05326} {\bibfield  {journal}
  {\bibinfo  {journal} {Nano Lett.}\ }\textbf {\bibinfo {volume} {25}},\
  \bibinfo {pages} {4204} (\bibinfo {year} {2025})}\BibitemShut {NoStop}%
\bibitem [{\citenamefont {Boykin}\ \emph {et~al.}(2004)\citenamefont {Boykin},
  \citenamefont {Klimeck}, \citenamefont {Eriksson}, \citenamefont {Friesen},
  \citenamefont {Coppersmith}, \citenamefont {von Allmen}, \citenamefont
  {Oyafuso},\ and\ \citenamefont {Lee}}]{Boykin2004}%
  \BibitemOpen
  \bibfield  {author} {\bibinfo {author} {\bibfnamefont {T.~B.}\ \bibnamefont
  {Boykin}}, \bibinfo {author} {\bibfnamefont {G.}~\bibnamefont {Klimeck}},
  \bibinfo {author} {\bibfnamefont {M.~A.}\ \bibnamefont {Eriksson}}, \bibinfo
  {author} {\bibfnamefont {M.}~\bibnamefont {Friesen}}, \bibinfo {author}
  {\bibfnamefont {S.~N.}\ \bibnamefont {Coppersmith}}, \bibinfo {author}
  {\bibfnamefont {P.}~\bibnamefont {von Allmen}}, \bibinfo {author}
  {\bibfnamefont {F.}~\bibnamefont {Oyafuso}},\ and\ \bibinfo {author}
  {\bibfnamefont {S.}~\bibnamefont {Lee}},\ }\href
  {https://doi.org/10.1063/1.1637718} {\bibfield  {journal} {\bibinfo
  {journal} {Appl. Phys. Lett.}\ }\textbf {\bibinfo {volume} {84}},\ \bibinfo
  {pages} {115} (\bibinfo {year} {2004})}\BibitemShut {NoStop}%
\bibitem [{\citenamefont {Chutia}\ \emph {et~al.}(2008)\citenamefont {Chutia},
  \citenamefont {Coppersmith},\ and\ \citenamefont {Friesen}}]{Chutia2008}%
  \BibitemOpen
  \bibfield  {author} {\bibinfo {author} {\bibfnamefont {S.}~\bibnamefont
  {Chutia}}, \bibinfo {author} {\bibfnamefont {S.~N.}\ \bibnamefont
  {Coppersmith}},\ and\ \bibinfo {author} {\bibfnamefont {M.}~\bibnamefont
  {Friesen}},\ }\href {https://doi.org/10.1103/PhysRevB.77.193311} {\bibfield
  {journal} {\bibinfo  {journal} {Phys. Rev. B}\ }\textbf {\bibinfo {volume}
  {77}},\ \bibinfo {pages} {193311} (\bibinfo {year} {2008})}\BibitemShut
  {NoStop}%
\bibitem [{\citenamefont {Zhang}\ \emph {et~al.}(2013)\citenamefont {Zhang},
  \citenamefont {Luo}, \citenamefont {Saraiva}, \citenamefont {Koiller},\ and\
  \citenamefont {Zunger}}]{Zhang2013}%
  \BibitemOpen
  \bibfield  {author} {\bibinfo {author} {\bibfnamefont {L.}~\bibnamefont
  {Zhang}}, \bibinfo {author} {\bibfnamefont {J.-W.}\ \bibnamefont {Luo}},
  \bibinfo {author} {\bibfnamefont {A.}~\bibnamefont {Saraiva}}, \bibinfo
  {author} {\bibfnamefont {B.}~\bibnamefont {Koiller}},\ and\ \bibinfo {author}
  {\bibfnamefont {A.}~\bibnamefont {Zunger}},\ }\href
  {https://doi.org/10.1038/ncomms3396} {\bibfield  {journal} {\bibinfo
  {journal} {Nat. Commun.}\ }\textbf {\bibinfo {volume} {4}},\ \bibinfo {pages}
  {2396} (\bibinfo {year} {2013})}\BibitemShut {NoStop}%
\bibitem [{\citenamefont {Cvitkovich}(2024)}]{Cvitkovich2024}%
  \BibitemOpen
  \bibfield  {author} {\bibinfo {author} {\bibfnamefont {L.}~\bibnamefont
  {Cvitkovich}},\ }\emph {\bibinfo {title} {Atomistic Modeling of {Si} Spin
  Qubits From First Principles}},\ \href
  {https://doi.org/10.34726/hss.2024.123305} {\bibinfo {type} {phdthesis}},\
  \bibinfo  {school} {TU Vienna} (\bibinfo {year} {2024})\BibitemShut {NoStop}%
\bibitem [{\citenamefont {Hosseinkhani}\ and\ \citenamefont
  {Burkard}(2020)}]{Hosseinkhani2020}%
  \BibitemOpen
  \bibfield  {author} {\bibinfo {author} {\bibfnamefont {A.}~\bibnamefont
  {Hosseinkhani}}\ and\ \bibinfo {author} {\bibfnamefont {G.}~\bibnamefont
  {Burkard}},\ }\href {https://doi.org/10.1103/PhysRevResearch.2.043180}
  {\bibfield  {journal} {\bibinfo  {journal} {Phys. Rev. Research}\ }\textbf
  {\bibinfo {volume} {2}},\ \bibinfo {pages} {043180} (\bibinfo {year}
  {2020})}\BibitemShut {NoStop}%
\bibitem [{\citenamefont {Lima}\ and\ \citenamefont
  {Burkard}(2024)}]{Lima2024}%
  \BibitemOpen
  \bibfield  {author} {\bibinfo {author} {\bibfnamefont {J.~R.~F.}\
  \bibnamefont {Lima}}\ and\ \bibinfo {author} {\bibfnamefont {G.}~\bibnamefont
  {Burkard}},\ }\href {https://doi.org/10.1103/PhysRevMaterials.8.036202}
  {\bibfield  {journal} {\bibinfo  {journal} {Phys. Rev. Materials}\ }\textbf
  {\bibinfo {volume} {8}},\ \bibinfo {pages} {036202} (\bibinfo {year}
  {2024})}\BibitemShut {NoStop}%
\bibitem [{\citenamefont {Sverdlov}\ \emph {et~al.}(2010)\citenamefont
  {Sverdlov}, \citenamefont {Baumgartner}, \citenamefont {Windbacher},\ and\
  \citenamefont {Selberherr}}]{Sverdlov2010}%
  \BibitemOpen
  \bibfield  {author} {\bibinfo {author} {\bibfnamefont {V.}~\bibnamefont
  {Sverdlov}}, \bibinfo {author} {\bibfnamefont {O.}~\bibnamefont
  {Baumgartner}}, \bibinfo {author} {\bibfnamefont {T.}~\bibnamefont
  {Windbacher}},\ and\ \bibinfo {author} {\bibfnamefont {S.}~\bibnamefont
  {Selberherr}},\ }in\ \href {https://doi.org/10.1002/9780470649343.ch24}
  {\emph {\bibinfo {booktitle} {Future Trends in Microelectronics}}},\ \bibinfo
  {editor} {edited by\ \bibinfo {editor} {\bibfnamefont {S.}~\bibnamefont
  {Luryi}}, \bibinfo {editor} {\bibfnamefont {J.}~\bibnamefont {Xu}},\ and\
  \bibinfo {editor} {\bibfnamefont {A.}~\bibnamefont {Zaslavsky}}}\ (\bibinfo
  {publisher} {Wiley},\ \bibinfo {year} {2010})\ Chap.~\bibinfo {chapter} {24},
  pp.\ \bibinfo {pages} {281--291}\BibitemShut {NoStop}%
\bibitem [{\citenamefont {Sverdlov}(2011)}]{Sverdlov2011}%
  \BibitemOpen
  \bibfield  {author} {\bibinfo {author} {\bibfnamefont {V.}~\bibnamefont
  {Sverdlov}},\ }\href {https://doi.org/10.1007/978-3-7091-0382-1} {\emph
  {\bibinfo {title} {Strain-Induced Effects in Advanced {MOSFET}s}}},\
  Computational Microelectronics\ (\bibinfo  {publisher} {Springer},\ \bibinfo
  {address} {Vienna},\ \bibinfo {year} {2011})\BibitemShut {NoStop}%
\bibitem [{\citenamefont {Sch\"{a}ffler}(1997)}]{Schaeffler1997}%
  \BibitemOpen
  \bibfield  {author} {\bibinfo {author} {\bibfnamefont {F.}~\bibnamefont
  {Sch\"{a}ffler}},\ }\href {https://doi.org/10.1088/0268-1242/12/12/001}
  {\bibfield  {journal} {\bibinfo  {journal} {Semicond. Sci. Technol.}\
  }\textbf {\bibinfo {volume} {12}},\ \bibinfo {pages} {1515} (\bibinfo {year}
  {1997})}\BibitemShut {NoStop}%
\bibitem [{Note1()}]{Note1}%
  \BibitemOpen
  \bibinfo {note} {In principle, one could also consider a (small) non-zero Ge
  concentration in the bulk potential $V\left (\protect \mathbf {r}\right )$,
  see Eq.~\protect \eqref {eq: stationary Schroedinger equation}, and treat the
  resulting Si$_{1-x}$Ge$_{x}$ alloy using the virtual crystal approximation
  (VCA). For consistency, the same Ge concentration must then be removed from
  the mean of the heterostructure potential $U\left (\protect \mathbf {r}\right
  )$. While this approach could yield some enhancements for the band structure
  coefficients and effective masses, the VCA will qualitatively preserve the
  symmetries of the diamond crystal \cite {Feng2022}, leading to a similar
  shear strain dependency. For low Ge concentrations, only minor changes in the
  quantitative results are expected since the effective mass and band structure
  parameters of the Si$_{1-x}$Ge$_{x}$-alloy (in VCA) are very close to the
  pure Si values \cite {Rieger1993,Schaeffler1997}.}\BibitemShut {Stop}%
\bibitem [{\citenamefont {Pe\~{n}a}\ \emph {et~al.}(2024)\citenamefont
  {Pe\~{n}a}, \citenamefont {Koepke}, \citenamefont {Dycus}, \citenamefont
  {Mounce}, \citenamefont {Baczewski}, \citenamefont {Jacobson},\ and\
  \citenamefont {Bussmann}}]{Pena2024}%
  \BibitemOpen
  \bibfield  {author} {\bibinfo {author} {\bibfnamefont {L.~F.}\ \bibnamefont
  {Pe\~{n}a}}, \bibinfo {author} {\bibfnamefont {J.~C.}\ \bibnamefont
  {Koepke}}, \bibinfo {author} {\bibfnamefont {J.~H.}\ \bibnamefont {Dycus}},
  \bibinfo {author} {\bibfnamefont {A.}~\bibnamefont {Mounce}}, \bibinfo
  {author} {\bibfnamefont {A.~D.}\ \bibnamefont {Baczewski}}, \bibinfo {author}
  {\bibfnamefont {N.~T.}\ \bibnamefont {Jacobson}},\ and\ \bibinfo {author}
  {\bibfnamefont {E.}~\bibnamefont {Bussmann}},\ }\href
  {https://doi.org/10.1038/s41534-024-00827-8} {\bibfield  {journal} {\bibinfo
  {journal} {npj Quantum Inf.}\ }\textbf {\bibinfo {volume} {10}},\ \bibinfo
  {pages} {33} (\bibinfo {year} {2024})}\BibitemShut {NoStop}%
\bibitem [{\citenamefont {Van~de Walle}\ and\ \citenamefont
  {Martin}(1986)}]{VandeWalle1986}%
  \BibitemOpen
  \bibfield  {author} {\bibinfo {author} {\bibfnamefont {C.~G.}\ \bibnamefont
  {Van~de Walle}}\ and\ \bibinfo {author} {\bibfnamefont {R.~M.}\ \bibnamefont
  {Martin}},\ }\href {https://doi.org/10.1103/PhysRevB.34.5621} {\bibfield
  {journal} {\bibinfo  {journal} {Phys. Rev. B}\ }\textbf {\bibinfo {volume}
  {34}},\ \bibinfo {pages} {5621} (\bibinfo {year} {1986})}\BibitemShut
  {NoStop}%
\bibitem [{\citenamefont {Cohen}\ and\ \citenamefont
  {Heine}(1970)}]{Cohen1970}%
  \BibitemOpen
  \bibfield  {author} {\bibinfo {author} {\bibfnamefont {M.~L.}\ \bibnamefont
  {Cohen}}\ and\ \bibinfo {author} {\bibfnamefont {V.}~\bibnamefont {Heine}},\
  }\href {https://doi.org/10.1016/S0081-1947(08)60070-3} {\bibfield  {journal}
  {\bibinfo  {journal} {Solid State Phys.}\ }\textbf {\bibinfo {volume} {24}},\
  \bibinfo {pages} {37} (\bibinfo {year} {1970})}\BibitemShut {NoStop}%
\bibitem [{\citenamefont {Chelikowsky}\ and\ \citenamefont
  {Cohen}(1974)}]{Chelikowsky1974}%
  \BibitemOpen
  \bibfield  {author} {\bibinfo {author} {\bibfnamefont {J.~R.}\ \bibnamefont
  {Chelikowsky}}\ and\ \bibinfo {author} {\bibfnamefont {M.~L.}\ \bibnamefont
  {Cohen}},\ }\href {https://doi.org/10.1103/PhysRevB.10.5095} {\bibfield
  {journal} {\bibinfo  {journal} {Phys. Rev. B}\ }\textbf {\bibinfo {volume}
  {10}},\ \bibinfo {pages} {5095} (\bibinfo {year} {1974})}\BibitemShut
  {NoStop}%
\bibitem [{\citenamefont {Chelikowsky}\ and\ \citenamefont
  {Cohen}(1976)}]{Chelikowsky1976}%
  \BibitemOpen
  \bibfield  {author} {\bibinfo {author} {\bibfnamefont {J.~R.}\ \bibnamefont
  {Chelikowsky}}\ and\ \bibinfo {author} {\bibfnamefont {M.~L.}\ \bibnamefont
  {Cohen}},\ }\href {https://doi.org/10.1103/PhysRevB.14.556} {\bibfield
  {journal} {\bibinfo  {journal} {Phys. Rev. B}\ }\textbf {\bibinfo {volume}
  {14}},\ \bibinfo {pages} {556} (\bibinfo {year} {1976})}\BibitemShut
  {NoStop}%
\bibitem [{\citenamefont {Fischetti}\ and\ \citenamefont
  {Higman}(1991)}]{Fischetti1991}%
  \BibitemOpen
  \bibfield  {author} {\bibinfo {author} {\bibfnamefont {M.~V.}\ \bibnamefont
  {Fischetti}}\ and\ \bibinfo {author} {\bibfnamefont {J.~M.}\ \bibnamefont
  {Higman}},\ }in\ \href {https://doi.org/10.1007/978-1-4615-4026-7_5} {\emph
  {\bibinfo {booktitle} {{Monte} {Carlo} Device Simulation}}},\ \bibinfo
  {editor} {edited by\ \bibinfo {editor} {\bibfnamefont {K.}~\bibnamefont
  {Hess}}}\ (\bibinfo  {publisher} {Springer US},\ \bibinfo {address}
  {Boston},\ \bibinfo {year} {1991})\ Chap.\ \bibinfo {chapter} {Theory and
  calculation of the deformation potential electron-phonon scattering rates in
  semiconductors}, pp.\ \bibinfo {pages} {123--160}\BibitemShut {NoStop}%
\bibitem [{\citenamefont {Fischetti}\ and\ \citenamefont
  {Laux}(1996)}]{Fischetti1996}%
  \BibitemOpen
  \bibfield  {author} {\bibinfo {author} {\bibfnamefont {M.~V.}\ \bibnamefont
  {Fischetti}}\ and\ \bibinfo {author} {\bibfnamefont {S.~E.}\ \bibnamefont
  {Laux}},\ }\href {https://doi.org/10.1063/1.363052} {\bibfield  {journal}
  {\bibinfo  {journal} {J. Appl. Phys.}\ }\textbf {\bibinfo {volume} {80}},\
  \bibinfo {pages} {2234} (\bibinfo {year} {1996})}\BibitemShut {NoStop}%
\bibitem [{\citenamefont {Rieger}\ and\ \citenamefont
  {Vogl}(1993)}]{Rieger1993}%
  \BibitemOpen
  \bibfield  {author} {\bibinfo {author} {\bibfnamefont {M.~M.}\ \bibnamefont
  {Rieger}}\ and\ \bibinfo {author} {\bibfnamefont {P.}~\bibnamefont {Vogl}},\
  }\href {https://doi.org/10.1103/PhysRevB.48.14276} {\bibfield  {journal}
  {\bibinfo  {journal} {Phys. Rev. B}\ }\textbf {\bibinfo {volume} {48}},\
  \bibinfo {pages} {14276} (\bibinfo {year} {1993})}\BibitemShut {NoStop}%
\bibitem [{\citenamefont {Ungersboeck}\ \emph {et~al.}(2007)\citenamefont
  {Ungersboeck}, \citenamefont {Dhar}, \citenamefont {Karlowatz}, \citenamefont
  {Sverdlov}, \citenamefont {Kosina},\ and\ \citenamefont
  {Selberherr}}]{Ungersboeck2007b}%
  \BibitemOpen
  \bibfield  {author} {\bibinfo {author} {\bibfnamefont {E.}~\bibnamefont
  {Ungersboeck}}, \bibinfo {author} {\bibfnamefont {S.}~\bibnamefont {Dhar}},
  \bibinfo {author} {\bibfnamefont {G.}~\bibnamefont {Karlowatz}}, \bibinfo
  {author} {\bibfnamefont {V.}~\bibnamefont {Sverdlov}}, \bibinfo {author}
  {\bibfnamefont {H.}~\bibnamefont {Kosina}},\ and\ \bibinfo {author}
  {\bibfnamefont {S.}~\bibnamefont {Selberherr}},\ }\href
  {https://doi.org/10.1109/TED.2007.902880} {\bibfield  {journal} {\bibinfo
  {journal} {IEEE Trans. Electron Devices}\ }\textbf {\bibinfo {volume} {54}},\
  \bibinfo {pages} {2183} (\bibinfo {year} {2007})}\BibitemShut {NoStop}%
\bibitem [{\citenamefont {Kim}\ and\ \citenamefont
  {Fischetti}(2010)}]{Kim2010a}%
  \BibitemOpen
  \bibfield  {author} {\bibinfo {author} {\bibfnamefont {J.}~\bibnamefont
  {Kim}}\ and\ \bibinfo {author} {\bibfnamefont {M.~V.}\ \bibnamefont
  {Fischetti}},\ }\href {https://doi.org/10.1063/1.3437655} {\bibfield
  {journal} {\bibinfo  {journal} {J. Appl. Phys.}\ }\textbf {\bibinfo {volume}
  {108}},\ \bibinfo {pages} {013710} (\bibinfo {year} {2010})}\BibitemShut
  {NoStop}%
\bibitem [{\citenamefont {Sant}\ \emph {et~al.}(2013)\citenamefont {Sant},
  \citenamefont {Lodha}, \citenamefont {Ganguly}, \citenamefont {Mahapatra},
  \citenamefont {Heinz}, \citenamefont {Smith}, \citenamefont {Moroz},\ and\
  \citenamefont {Ganguly}}]{Sant2013}%
  \BibitemOpen
  \bibfield  {author} {\bibinfo {author} {\bibfnamefont {S.}~\bibnamefont
  {Sant}}, \bibinfo {author} {\bibfnamefont {S.}~\bibnamefont {Lodha}},
  \bibinfo {author} {\bibfnamefont {U.}~\bibnamefont {Ganguly}}, \bibinfo
  {author} {\bibfnamefont {S.}~\bibnamefont {Mahapatra}}, \bibinfo {author}
  {\bibfnamefont {F.~O.}\ \bibnamefont {Heinz}}, \bibinfo {author}
  {\bibfnamefont {L.}~\bibnamefont {Smith}}, \bibinfo {author} {\bibfnamefont
  {V.}~\bibnamefont {Moroz}},\ and\ \bibinfo {author} {\bibfnamefont
  {S.}~\bibnamefont {Ganguly}},\ }\href {https://doi.org/10.1063/1.4775839}
  {\bibfield  {journal} {\bibinfo  {journal} {J. Appl. Phys.}\ }\textbf
  {\bibinfo {volume} {113}},\ \bibinfo {pages} {033708} (\bibinfo {year}
  {2013})}\BibitemShut {NoStop}%
\bibitem [{\citenamefont {Yu}\ and\ \citenamefont {Cardona}(2010)}]{Yu2010}%
  \BibitemOpen
  \bibfield  {author} {\bibinfo {author} {\bibfnamefont {P.~Y.}\ \bibnamefont
  {Yu}}\ and\ \bibinfo {author} {\bibfnamefont {M.}~\bibnamefont {Cardona}},\
  }\href {https://doi.org/10.1007/978-3-642-00710-1} {\emph {\bibinfo {title}
  {Fundamentals of Semiconductors: {Physics} and Materials Properties}}},\
  Graduate Texts in Physics\ (\bibinfo  {publisher} {Springer},\ \bibinfo
  {address} {Berlin, Heidelberg},\ \bibinfo {year} {2010})\BibitemShut
  {NoStop}%
\bibitem [{\citenamefont {Corley-Wiciak}\ \emph {et~al.}(2023)\citenamefont
  {Corley-Wiciak}, \citenamefont {Zoellner}, \citenamefont {Zaitsev},
  \citenamefont {Anand}, \citenamefont {Zatterin}, \citenamefont {Yamamoto},
  \citenamefont {Corley-Wiciak}, \citenamefont {Reichmann}, \citenamefont
  {Langheinrich}, \citenamefont {Schreiber}, \citenamefont {Manganelli},
  \citenamefont {Virgilio}, \citenamefont {Richter},\ and\ \citenamefont
  {Capellini}}]{CorleyWiciak2023}%
  \BibitemOpen
  \bibfield  {author} {\bibinfo {author} {\bibfnamefont {C.}~\bibnamefont
  {Corley-Wiciak}}, \bibinfo {author} {\bibfnamefont {M.}~\bibnamefont
  {Zoellner}}, \bibinfo {author} {\bibfnamefont {I.}~\bibnamefont {Zaitsev}},
  \bibinfo {author} {\bibfnamefont {K.}~\bibnamefont {Anand}}, \bibinfo
  {author} {\bibfnamefont {E.}~\bibnamefont {Zatterin}}, \bibinfo {author}
  {\bibfnamefont {Y.}~\bibnamefont {Yamamoto}}, \bibinfo {author}
  {\bibfnamefont {A.}~\bibnamefont {Corley-Wiciak}}, \bibinfo {author}
  {\bibfnamefont {F.}~\bibnamefont {Reichmann}}, \bibinfo {author}
  {\bibfnamefont {W.}~\bibnamefont {Langheinrich}}, \bibinfo {author}
  {\bibfnamefont {L.}~\bibnamefont {Schreiber}}, \bibinfo {author}
  {\bibfnamefont {C.}~\bibnamefont {Manganelli}}, \bibinfo {author}
  {\bibfnamefont {M.}~\bibnamefont {Virgilio}}, \bibinfo {author}
  {\bibfnamefont {C.}~\bibnamefont {Richter}},\ and\ \bibinfo {author}
  {\bibfnamefont {G.}~\bibnamefont {Capellini}},\ }\href
  {https://doi.org/10.1103/PhysRevApplied.20.024056} {\bibfield  {journal}
  {\bibinfo  {journal} {Phys. Rev. Appl.}\ }\textbf {\bibinfo {volume} {20}},\
  \bibinfo {pages} {024056} (\bibinfo {year} {2023})}\BibitemShut {NoStop}%
\bibitem [{\citenamefont {Adelsberger}\ \emph {et~al.}(2024)\citenamefont
  {Adelsberger}, \citenamefont {Bosco}, \citenamefont {Klinovaja},\ and\
  \citenamefont {Loss}}]{Adelsberger2024}%
  \BibitemOpen
  \bibfield  {author} {\bibinfo {author} {\bibfnamefont {C.}~\bibnamefont
  {Adelsberger}}, \bibinfo {author} {\bibfnamefont {S.}~\bibnamefont {Bosco}},
  \bibinfo {author} {\bibfnamefont {J.}~\bibnamefont {Klinovaja}},\ and\
  \bibinfo {author} {\bibfnamefont {D.}~\bibnamefont {Loss}},\ }\href
  {https://doi.org/10.1103/PhysRevLett.133.037001} {\bibfield  {journal}
  {\bibinfo  {journal} {Phys. Rev. Lett.}\ }\textbf {\bibinfo {volume} {133}},\
  \bibinfo {pages} {037001} (\bibinfo {year} {2024})}\BibitemShut {NoStop}%
\bibitem [{\citenamefont {Mooney}(1996)}]{Mooney1996}%
  \BibitemOpen
  \bibfield  {author} {\bibinfo {author} {\bibfnamefont {P.}~\bibnamefont
  {Mooney}},\ }\href {https://doi.org/10.1016/S0927-796X(96)00192-1} {\bibfield
   {journal} {\bibinfo  {journal} {Materials Science and Engineering: R:
  Reports}\ }\textbf {\bibinfo {volume} {17}},\ \bibinfo {pages} {105}
  (\bibinfo {year} {1996})}\BibitemShut {NoStop}%
\bibitem [{\citenamefont {Gradwohl}\ \emph {et~al.}(2023)\citenamefont
  {Gradwohl}, \citenamefont {Lu}, \citenamefont {Liu}, \citenamefont {Richter},
  \citenamefont {Boeck}, \citenamefont {Martin},\ and\ \citenamefont
  {Albrecht}}]{Gradwohl2023}%
  \BibitemOpen
  \bibfield  {author} {\bibinfo {author} {\bibfnamefont {K.-P.}\ \bibnamefont
  {Gradwohl}}, \bibinfo {author} {\bibfnamefont {C.-H.}\ \bibnamefont {Lu}},
  \bibinfo {author} {\bibfnamefont {Y.}~\bibnamefont {Liu}}, \bibinfo {author}
  {\bibfnamefont {C.}~\bibnamefont {Richter}}, \bibinfo {author} {\bibfnamefont
  {T.}~\bibnamefont {Boeck}}, \bibinfo {author} {\bibfnamefont
  {J.}~\bibnamefont {Martin}},\ and\ \bibinfo {author} {\bibfnamefont
  {M.}~\bibnamefont {Albrecht}},\ }\href
  {https://doi.org/10.1002/pssr.202200398} {\bibfield  {journal} {\bibinfo
  {journal} {Phys. Status Solidi RRL}\ }\textbf {\bibinfo {volume} {17}},\
  \bibinfo {pages} {2200398} (\bibinfo {year} {2023})}\BibitemShut {NoStop}%
\bibitem [{\citenamefont {Zaiser}\ and\ \citenamefont {Wu}(2022)}]{Zaiser2022}%
  \BibitemOpen
  \bibfield  {author} {\bibinfo {author} {\bibfnamefont {M.}~\bibnamefont
  {Zaiser}}\ and\ \bibinfo {author} {\bibfnamefont {R.}~\bibnamefont {Wu}},\
  }\href {https://doi.org/10.1186/s41313-021-00036-2} {\bibfield  {journal}
  {\bibinfo  {journal} {Materials Theory}\ }\textbf {\bibinfo {volume} {6}},\
  \bibinfo {pages} {4} (\bibinfo {year} {2022})}\BibitemShut {NoStop}%
\bibitem [{\citenamefont {Geslin}\ and\ \citenamefont
  {Rodney}(2021)}]{Geslin2021a}%
  \BibitemOpen
  \bibfield  {author} {\bibinfo {author} {\bibfnamefont {P.-A.}\ \bibnamefont
  {Geslin}}\ and\ \bibinfo {author} {\bibfnamefont {D.}~\bibnamefont
  {Rodney}},\ }\href {https://doi.org/10.1016/j.jmps.2021.104479} {\bibfield
  {journal} {\bibinfo  {journal} {J. Mech. Phys. Solids}\ }\textbf {\bibinfo
  {volume} {153}},\ \bibinfo {pages} {104479} (\bibinfo {year}
  {2021})}\BibitemShut {NoStop}%
\bibitem [{\citenamefont {Culcer}\ \emph {et~al.}(2010)\citenamefont {Culcer},
  \citenamefont {Hu},\ and\ \citenamefont {Das~Sarma}}]{Culcer2010}%
  \BibitemOpen
  \bibfield  {author} {\bibinfo {author} {\bibfnamefont {D.}~\bibnamefont
  {Culcer}}, \bibinfo {author} {\bibfnamefont {X.}~\bibnamefont {Hu}},\ and\
  \bibinfo {author} {\bibfnamefont {S.}~\bibnamefont {Das~Sarma}},\ }\href
  {https://doi.org/10.1103/PhysRevB.82.205315} {\bibfield  {journal} {\bibinfo
  {journal} {Phys. Rev. B}\ }\textbf {\bibinfo {volume} {82}},\ \bibinfo
  {pages} {205315} (\bibinfo {year} {2010})}\BibitemShut {NoStop}%
\bibitem [{\citenamefont {Woods}\ \emph {et~al.}(2023)\citenamefont {Woods},
  \citenamefont {Eriksson}, \citenamefont {Joynt},\ and\ \citenamefont
  {Friesen}}]{Woods2023}%
  \BibitemOpen
  \bibfield  {author} {\bibinfo {author} {\bibfnamefont {B.~D.}\ \bibnamefont
  {Woods}}, \bibinfo {author} {\bibfnamefont {M.~A.}\ \bibnamefont {Eriksson}},
  \bibinfo {author} {\bibfnamefont {R.}~\bibnamefont {Joynt}},\ and\ \bibinfo
  {author} {\bibfnamefont {M.}~\bibnamefont {Friesen}},\ }\href
  {https://doi.org/10.1103/PhysRevB.107.035418} {\bibfield  {journal} {\bibinfo
   {journal} {Phys. Rev. B}\ }\textbf {\bibinfo {volume} {107}},\ \bibinfo
  {pages} {035418} (\bibinfo {year} {2023})}\BibitemShut {NoStop}%
\bibitem [{\citenamefont {Frink}\ \emph {et~al.}(2023)\citenamefont {Frink},
  \citenamefont {Woods}, \citenamefont {Losert}, \citenamefont {MacQuarrie},
  \citenamefont {Eriksson},\ and\ \citenamefont {Friesen}}]{Frink2023}%
  \BibitemOpen
  \bibfield  {author} {\bibinfo {author} {\bibfnamefont {C.~C.~D.}\
  \bibnamefont {Frink}}, \bibinfo {author} {\bibfnamefont {B.~D.}\ \bibnamefont
  {Woods}}, \bibinfo {author} {\bibfnamefont {M.~P.}\ \bibnamefont {Losert}},
  \bibinfo {author} {\bibfnamefont {E.~R.}\ \bibnamefont {MacQuarrie}},
  \bibinfo {author} {\bibfnamefont {M.~A.}\ \bibnamefont {Eriksson}},\ and\
  \bibinfo {author} {\bibfnamefont {M.}~\bibnamefont {Friesen}},\ }\bibfield
  {journal} {\bibinfo  {journal} {arXiv}\ }\href
  {https://doi.org/10.48550/arXiv.2312.09235} {10.48550/arXiv.2312.09235}
  (\bibinfo {year} {2023})\BibitemShut {NoStop}%
\bibitem [{\citenamefont {Thayil}\ \emph {et~al.}()\citenamefont {Thayil},
  \citenamefont {Ermoneit},\ and\ \citenamefont {Kantner}}]{Thayil2025b}%
  \BibitemOpen
  \bibfield  {author} {\bibinfo {author} {\bibfnamefont {A.}~\bibnamefont
  {Thayil}}, \bibinfo {author} {\bibfnamefont {L.}~\bibnamefont {Ermoneit}},\
  and\ \bibinfo {author} {\bibfnamefont {M.}~\bibnamefont {Kantner}},\ }\href
  {https://github.com/kantner/valley-splitting} {\bibinfo {title}
  {Valley-splitting}},\ \bibinfo {note} {{GitHub} repository with MATLAB
  simulation code.\\
  \url{https://github.com/kantner/valley-splitting}}\BibitemShut {NoStop}%
\bibitem [{\citenamefont {Kohn}\ and\ \citenamefont
  {Luttinger}(1955)}]{Kohn1955}%
  \BibitemOpen
  \bibfield  {author} {\bibinfo {author} {\bibfnamefont {W.}~\bibnamefont
  {Kohn}}\ and\ \bibinfo {author} {\bibfnamefont {J.~M.}\ \bibnamefont
  {Luttinger}},\ }\href
  {https://doi.org/https://doi.org/10.1103/PhysRev.98.915} {\bibfield
  {journal} {\bibinfo  {journal} {Phys. Rev.}\ }\textbf {\bibinfo {volume}
  {98}},\ \bibinfo {pages} {915} (\bibinfo {year} {1955})}\BibitemShut
  {NoStop}%
\bibitem [{\citenamefont {Fritzsche}(1962)}]{Fritzsche1962}%
  \BibitemOpen
  \bibfield  {author} {\bibinfo {author} {\bibfnamefont {H.}~\bibnamefont
  {Fritzsche}},\ }\href {https://doi.org/10.1103/PhysRev.125.1560} {\bibfield
  {journal} {\bibinfo  {journal} {Phys. Rev.}\ }\textbf {\bibinfo {volume}
  {125}},\ \bibinfo {pages} {1560} (\bibinfo {year} {1962})}\BibitemShut
  {NoStop}%
\bibitem [{\citenamefont {Ning}\ and\ \citenamefont {Sah}(1971)}]{Ning1971}%
  \BibitemOpen
  \bibfield  {author} {\bibinfo {author} {\bibfnamefont {T.~H.}\ \bibnamefont
  {Ning}}\ and\ \bibinfo {author} {\bibfnamefont {C.~T.}\ \bibnamefont {Sah}},\
  }\href {https://doi.org/10.1103/PhysRevB.4.3468} {\bibfield  {journal}
  {\bibinfo  {journal} {Phys. Rev. B}\ }\textbf {\bibinfo {volume} {4}},\
  \bibinfo {pages} {3468} (\bibinfo {year} {1971})}\BibitemShut {NoStop}%
\bibitem [{\citenamefont {Shindo}\ and\ \citenamefont
  {Nara}(1976)}]{Shindo1976}%
  \BibitemOpen
  \bibfield  {author} {\bibinfo {author} {\bibfnamefont {K.}~\bibnamefont
  {Shindo}}\ and\ \bibinfo {author} {\bibfnamefont {H.}~\bibnamefont {Nara}},\
  }\href {https://doi.org/10.1143/JPSJ.40.1640} {\bibfield  {journal} {\bibinfo
   {journal} {J. Phys. Soc. Jpn.}\ }\textbf {\bibinfo {volume} {40}},\ \bibinfo
  {pages} {1640} (\bibinfo {year} {1976})}\BibitemShut {NoStop}%
\bibitem [{\citenamefont {Hui}(2013)}]{Hui2013}%
  \BibitemOpen
  \bibfield  {author} {\bibinfo {author} {\bibfnamefont {H.}~\bibnamefont
  {Hui}},\ }\href {https://doi.org/10.1016/j.ssc.2012.10.023} {\bibfield
  {journal} {\bibinfo  {journal} {Solid State Commun.}\ }\textbf {\bibinfo
  {volume} {154}},\ \bibinfo {pages} {19} (\bibinfo {year} {2013})}\BibitemShut
  {NoStop}%
\bibitem [{\citenamefont {Burt}(1988)}]{Burt1988}%
  \BibitemOpen
  \bibfield  {author} {\bibinfo {author} {\bibfnamefont {M.~G.}\ \bibnamefont
  {Burt}},\ }\href {https://doi.org/10.1088/0268-1242/3/8/003} {\bibfield
  {journal} {\bibinfo  {journal} {Semicond. Sci. Tech.}\ }\textbf {\bibinfo
  {volume} {3}},\ \bibinfo {pages} {739} (\bibinfo {year} {1988})}\BibitemShut
  {NoStop}%
\bibitem [{\citenamefont {Burt}(1992)}]{Burt1992}%
  \BibitemOpen
  \bibfield  {author} {\bibinfo {author} {\bibfnamefont {M.~G.}\ \bibnamefont
  {Burt}},\ }\href {https://doi.org/10.1088/0953-8984/4/32/003} {\bibfield
  {journal} {\bibinfo  {journal} {J. Phys. Condens. Matter}\ }\textbf {\bibinfo
  {volume} {4}},\ \bibinfo {pages} {6651} (\bibinfo {year} {1992})}\BibitemShut
  {NoStop}%
\bibitem [{\citenamefont {Foreman}(1995)}]{Foreman1995}%
  \BibitemOpen
  \bibfield  {author} {\bibinfo {author} {\bibfnamefont {B.~A.}\ \bibnamefont
  {Foreman}},\ }\href {https://doi.org/10.1103/PhysRevB.52.12241} {\bibfield
  {journal} {\bibinfo  {journal} {Phys. Rev. B}\ }\textbf {\bibinfo {volume}
  {52}},\ \bibinfo {pages} {12241} (\bibinfo {year} {1995})}\BibitemShut
  {NoStop}%
\bibitem [{\citenamefont {Foreman}(1996)}]{Foreman1996}%
  \BibitemOpen
  \bibfield  {author} {\bibinfo {author} {\bibfnamefont {B.~A.}\ \bibnamefont
  {Foreman}},\ }\href {https://doi.org/10.1103/PhysRevB.54.1909} {\bibfield
  {journal} {\bibinfo  {journal} {Phys. Rev. B}\ }\textbf {\bibinfo {volume}
  {54}},\ \bibinfo {pages} {1909} (\bibinfo {year} {1996})}\BibitemShut
  {NoStop}%
\bibitem [{\citenamefont {Klymenko}\ and\ \citenamefont
  {Remacle}(2014)}]{Klymenko2014}%
  \BibitemOpen
  \bibfield  {author} {\bibinfo {author} {\bibfnamefont {M.~V.}\ \bibnamefont
  {Klymenko}}\ and\ \bibinfo {author} {\bibfnamefont {F.}~\bibnamefont
  {Remacle}},\ }\href {https://doi.org/10.1088/0953-8984/26/6/065302}
  {\bibfield  {journal} {\bibinfo  {journal} {J. Phys. Condens. Matter}\
  }\textbf {\bibinfo {volume} {26}},\ \bibinfo {pages} {065302} (\bibinfo
  {year} {2014})}\BibitemShut {NoStop}%
\bibitem [{\citenamefont {Klymenko}\ \emph {et~al.}(2015)\citenamefont
  {Klymenko}, \citenamefont {Rogge},\ and\ \citenamefont
  {Remacle}}]{Klymenko2015}%
  \BibitemOpen
  \bibfield  {author} {\bibinfo {author} {\bibfnamefont {M.~V.}\ \bibnamefont
  {Klymenko}}, \bibinfo {author} {\bibfnamefont {S.}~\bibnamefont {Rogge}},\
  and\ \bibinfo {author} {\bibfnamefont {F.}~\bibnamefont {Remacle}},\ }\href
  {https://doi.org/10.1103/PhysRevB.92.195302} {\bibfield  {journal} {\bibinfo
  {journal} {Phys. Rev. B}\ }\textbf {\bibinfo {volume} {92}},\ \bibinfo
  {pages} {195302} (\bibinfo {year} {2015})}\BibitemShut {NoStop}%
\bibitem [{\citenamefont {Burt}(1994)}]{Burt1994}%
  \BibitemOpen
  \bibfield  {author} {\bibinfo {author} {\bibfnamefont {M.~G.}\ \bibnamefont
  {Burt}},\ }\href {https://doi.org/10.1103/PhysRevB.50.7518} {\bibfield
  {journal} {\bibinfo  {journal} {Phys. Rev. B}\ }\textbf {\bibinfo {volume}
  {50}},\ \bibinfo {pages} {7518} (\bibinfo {year} {1994})}\BibitemShut
  {NoStop}%
\bibitem [{\citenamefont {Hamaguchi}(2010)}]{Hamaguchi2010}%
  \BibitemOpen
  \bibfield  {author} {\bibinfo {author} {\bibfnamefont {C.}~\bibnamefont
  {Hamaguchi}},\ }\href {https://doi.org/10.1007/978-3-642-03303-2} {\emph
  {\bibinfo {title} {Basic Semiconductor Physics}}}\ (\bibinfo  {publisher}
  {Springer Berlin Heidelberg},\ \bibinfo {year} {2010})\BibitemShut {NoStop}%
\end{thebibliography}

%

\end{document}